\documentclass[a4paper,11pt]{article}
\pdfoutput=1
\usepackage{jheppub}
\usepackage{slashed,subcaption,amssymb,amsmath,bm}
\usepackage[utf8]{inputenc}
\usepackage{amsfonts}
\usepackage{bbold}
\usepackage{mathtools}
\usepackage{pdflscape}
\usepackage[dvipsnames]{xcolor}
\usepackage{forloop}

\usepackage{ifpdf}
\graphicspath{{./img/}}
\usepackage{hyperref}
\usepackage{cleveref}
\usepackage{tikz}
\usepackage{float}
\usepackage{cancel}
\usepackage{booktabs}
\usepackage{multirow}
\usepackage{rotating}
\newcommand{\gev}{\text{ GeV}}
\newcommand{\tev}{\text{ TeV}}

\newcommand{\Tr}[1]{\text{Tr}\left[#1\right]}
\newcommand{\brackets}[1]{\left(#1\right)}
\newcommand{\sqbrackets}[1]{\left[#1\right]}

\newcommand{\dVt}[1]{\mathrm{d}^{4} #1}

\def\figureautorefname~#1\null{Fig.\,#1\null}
\def\tableautorefname~#1\null{Tab.\,#1\null}

\def\equationautorefname~#1\null{Eq.\,(#1)\null}

\DeclareUnicodeCharacter{2212}{-}

\makeindex

\makeatletter\g@addto@macro\bfseries\boldmath
\makeatother

\title{Convergent Bayesian Global Fits of 4D Composite Higgs Models}

\author[a]{Ethan Carragher,}
\author[b,c]{Will Handley,}
\author[d]{Daniel Murnane,}
\author[e]{Peter Stangl,}
\author[a]{Wei Su,}
\author[a]{Martin White,}
\author[a]{and Anthony G. Williams}

\affiliation[a]{ARC Centre of Excellence for Dark Matter Particle Physics, Department of Physics, University of Adelaide, South Australia 5005, Australia}
\affiliation[b]{Astrophysics Group, Cavendish Laboratory, J. J. Thomson Avenue, Cambridge, CB3 0HE, UK}
\affiliation[c]{Kavli Institute for Cosmology, Madingley Road, Cambridge, CB3 0HA, UK}
\affiliation[d]{Lawrence Berkeley National Laboratory, Berkeley, CA, USA}
\affiliation[e]{Albert Einstein Center for Fundamental Physics, Institute for Theoretical Physics, University of Bern, Sidlerstrasse 5, CH-3012 Bern, Switzerland}

\emailAdd{ethan.carragher@adelaide.edu.au}
\emailAdd{wh260@cam.ac.uk}
\emailAdd{dtmurnane@lbl.gov}
\emailAdd{stangl@itp.unibe.ch}
\emailAdd{wei.su@adelaide.edu.au}
\emailAdd{martin.white@adelaide.edu.au}
\emailAdd{anthony.williams@adelaide.edu.au,}

\abstract{
Models in which the Higgs boson is a composite pseudo-Nambu-Goldstone boson offer attractive solutions to the Higgs mass naturalness problem. We consider three such models based on the minimal $SO(5) \rightarrow SO(4)$ symmetry breaking pattern, and perform convergent global fits on the models under a Bayesian framework in order to find the regions of their parameter spaces that best fit a wide range of constraints, including recent Higgs measurements. We use a novel technique to analyse the fine-tuning of the models, quantifying the tuning as the Kullback-Leibler divergence from the prior to the posterior probability on the parameter space. Each model is found to be able to satisfy all constraints at the $3\sigma$ level simultaneously.
As a by-product of the fits, we analyse the collider phenomenology of our models in these viable regions.
In two of the three models, we find that the $g g \rightarrow H \rightarrow \gamma \gamma$ cross section is less than ${\sim}90$\% that predicted by the SM, which is already in slight tension with experiment and could potentially be ruled out in the future high-luminosity run of the LHC. In addition, the lightest fermions $F$ arising from the new strong dynamics in these models are seen in general to lie above ${\sim}1.1$~TeV, with the $F \rightarrow tW^{+}$ and $F \rightarrow \bar{b}W^{+}$ decays offering particularly promising channels for probing these models in future collider searches.
}


\date{}

\begin{document}

\maketitle

\section{Introduction}

The Standard Model (SM), assuming UV completion at a high energy scale, is only able to accommodate the observed mass of the Higgs boson through an excessive fine-tuning of the model parameters on account of the large quantum loop contributions the mass receives. Composite Higgs Models (CHMs), on the other hand, contend that the Higgs boson is in fact a bound state of some new strong dynamics whose confinement scale cuts off the loop contributions at $\mathcal{O}(1)$~TeV, alleviating this issue of fine-tuning \cite{kaplan1984,kaplan1984b,kaplan1985}. In such models, the Higgs emerges as a naturally light pseudo-Nambu-Goldstone boson (pNGB) of some spontaneously broken symmetry of the new ``composite" sector to explain the absence of other composite resonances in collider searches to date. Of considerable interest is the Minimal CHM (MCHM): the simplest viable CHM, based on the symmetry breaking pattern $SO(5) \rightarrow SO(4)$ that delivers only the Higgs doublet in the NGB spectrum while also ensuring custodial symmetry~\cite{contino2006,contino2007b}, which we focus on here.

Many particular realisations of the MCHM exist, differing in their composite field content and their precise symmetry structures. It has proven particularly fruitful to consider those MCHMs that are holographically dual to certain 5D Gauge-Higgs-Unification theories, for they inherit calculability of the Higgs potential from the higher-dimensional models, and therefore possess extra predictivity in comparison to generic MCHMs~\cite{contino2003,agashe2005,contino2006,Contino:2006nn,contino2007b,giudice2007,contino2007,Panico:2007qd}. A dimensional deconstruction~\cite{ArkaniHamed:2001ca,Hill:2000mu} of the extra dimension leads to ``elementary" boundary fields and a tower of ``composite" resonances associated with the discrete points inside the extra dimension. The elementary fields mirror the non-Higgs SM fields and mix with the composite fields so that physical particles are ``partially composite" superpositions of both~\cite{kaplan1991}. We will be working with low-energy effective descriptions of such models known as Minimal 4D Composite Higgs Models (M4DCHMs), wherein the composite resonance towers are truncated to only a finite number of levels but calculability of the Higgs potential is still maintained~\cite{de2012,panico2011,MarzoccaGeneralCHMs}. See also Ref.~\cite{Panico:2015jxa} for a review of 4DCHMs.

Models of this form, which may still differ in the $SO(5)$ representations of the composite fields, are popular choices in studies of composite Higgs scenarios on account of their predictivity and relative simplicity\footnote{For studies of 4DCHMs with non-minimal symmetry structures, see for example Refs.~\cite{Bellazzini:2014yua,gripaios2009beyond,Redi:2012ha,Banerjee:2017qod}.}. Early studies mostly examined the theoretical form of the Higgs potential and its consequences on the Higgs mass and fine-tuning, at first only including composite partners for the top quark in the fundamental representation \cite{de2012}, and later also including bottom quark partners and other representations, with some simple numerical scans performed to verify the results~\cite{panico2012}. Composite lepton partners have also been considered, and have been found to give interesting fine-tuning effects~\cite{carmona2015,BarnardFT}. More comprehensive scans of M4DCHMs have since been performed, for example:
\begin{itemize}
    \item to study the Higgs phenomenology in regions of parameter space that reproduce the correct electroweak (EW) scale and accurate quark and Higgs masses, for a wide variety of models with top and bottom quark partners in different representations~\cite{Carena},
    \item to analyse points consistent with EW precision tests, Higgs physics, flavour physics, and (Run 1) LHC resonance bounds in four M4DCHMs that include all quark partners in the fundamental representation, subject to different flavour symmetries~\cite{Niehoff:2015iaa},
    \item and to investigate the fine-tuning behaviour in regions fitted to give accurate EW scales and SM masses of models with composite partners for the top quark~\cite{BarnardCC}, and also for all third generation fermions~\cite{BarnardFT}.
\end{itemize}
However, no work so far has contained rigorous, convergent statistical global fits of such models. This is primarily on account of their large parameter spaces, making exhaustive parameter scans prohibitively computationally expensive, along with the fact that the parameters all have highly non-linear effects on both the SM and composite sectors, so that even simplified analyses of the parameter spaces are difficult.

The purpose of the present work is to extend this research by providing the first full convergent global fits of several realistic M4DCHMs, mapping the regions of their parameter spaces that best fit a wide range of constraints. Given the difficulties mentioned above, we consider limiting models in which out of the SM fermions, only the top and bottom quarks couple to the composite sector, so that we may have more manageable parameter spaces. We will explore three different M4DCHMs, distinguished by the representations of the quark partners under $SO(5)$. Our fits are performed under a Bayesian framework and are facilitated by a sophisticated nested sampling algorithm that delivers convergent results where other techniques have struggled. This Bayesian approach is particularly relevant to the case of CHMs, or indeed any proposed solution to the Higgs mass naturalness problem, for it takes into account both the experimental fitness and the naturalness of the models. We use the numerical techniques of Refs.~\cite{Niehoff:2015iaa,niehoff2017electroweak} to calculate the observables at each point, and employ their same constraints (with updated values) where applicable. We also analyse the composite resonance cross sections and Higgs signal strengths predicted by the models in their viable regions to help guide future collider searches for new physics.

The structure of this paper is as follows. In \Cref{model} we give a brief overview of the M4DCHM and specify all three of our chosen fermion sectors, and in \Cref{scanning} we detail our procedure for scanning these models. We present and discuss our fit results, as well as make some comparisons between the different models based on their fits, in \Cref{results}. We analyse experimental signatures of the three models in \Cref{experimental_signatures_section}, and conclude with a summary of our findings in \Cref{sec:conclusions}.

\section{Model description}
\label{model}

Here we outline the structure of the M4DCHM, first described in Ref.~\cite{de2012}. Throughout this work we refer specifically to the \textit{two-site} M4DCHM, which contains only one level of composite resonances, for this is the minimum matter content needed for calculability of the Higgs potential. Site $0$ consists of massless elementary fields with the same quantum numbers as the non-Higgs SM fields, while Site $1$ contains the composite fields, which mix with the elementary fields both linearly and through proto-Yukawa terms to enact partial compositeness. An $SO(5) \rightarrow SO(4)$ symmetry breaking gives rise to the Higgs doublet, though the $SO(5)$ symmetry is explicitly broken by the elementary-composite mixing so that the Higgs is really a pNGB. This mixing necessitates $SU(3)$ and $U(1)_{X}$ symmetries on each site so that particles can be assigned the correct colour charge and the correct hypercharge under the definition
\begin{align}
    Y = T^3_R + X,
\end{align}
where the isospin operator $T^3_R$ is the third generator\footnote{See \Cref{SO5_appendix} for our convention with $SO(5)$ generators.} of $SU(2)_R \subset SO(5)$. The overall symmetry breaking pattern is then
\begin{align}
    \underbrace{SU(3)_{c} \times SO(5) \times U(1)_{X}}_{G} \rightarrow SU(3)_{c} \times SO(4) \times U(1)_{X}.
\end{align}
The symmetry groups $G$ on each site spontaneously break to their diagonal subgroup, leading to non-linear $\sigma$-model ``link" fields $\Omega_{i}$ ($i = 1,2$) that parameterise the resulting NGBs.
Most of them are would-be NGBs that become the longitudinal polarisations of the gauge fields on each site such that in addition to the SM bosons, there are ten massive $SO(5)$ vector bosons, eight massive $SU(3)$ resonances, and a massive abelian $U(1)_X$ resonance.
The remaining physical (p)NGBs, parameterised by the product $\Omega = \Omega_{1} \Omega_{2}$, constitute the composite Higgs field.

\subsection{Boson sector}

The boson sector of the theory is quite simple, for it is composed entirely of gauge bosons and NGBs. The low-energy dynamics of NGBs are entirely specified by their coset structure and are governed by the $\sigma$-model Lagrangian, so that the complete bosonic Lagrangian can be expressed as
\begin{align}
    \mathcal{L}_{\text{boson}} = \mathcal{L}_{\text{gauge}} + \mathcal{L}_{\sigma}.
\label{eq:boson_lagrangian}
\end{align}
The gauge sector consists of $G^0_{\mu}$, $W^0_{\mu}$, and $B^0_{\mu}$ - the elementary counterparts of the SM gauge fields - and $\rho_{G_{\mu}}$, $\rho_{\mu}$, and $\rho_{X_{\mu}}$ - the $SU(3)$, $SO(5)$, and $U(1)_X$ gauge fields of the composite sector. They have the usual kinetic terms
\begin{align}
    \mathcal{L}_{\text{gauge}} = &-\frac{1}{4} \Tr{G^0_{\mu \nu} G^{0 \mu \nu}} -\frac{1}{4} \Tr{W^0_{\mu \nu} W^{0 \mu \nu}} -\frac{1}{4} B^0_{\mu \nu} B^{0 \mu \nu} && \left.\vphantom{\frac{1}{4}}\right\rbrace \text{ elementary} \nonumber \\[0.15cm]
        &-\frac{1}{4} \Tr{\rho_{G_{\mu \nu}} \rho^{\mu \nu}_G} -\frac{1}{4} \Tr{\rho_{\mu \nu} \rho^{\mu \nu}} -\frac{1}{4} \rho_{X_{\mu \nu}} \rho^{\mu \nu}_X, && \left.\vphantom{\frac{1}{4}}\right\rbrace \text{ composite}
\end{align}
where the field strength tensors $\{G^0_{\mu \nu}, \ldots \}$ have the form
\begin{align}
    X_{\mu \nu} = \partial_{\mu} X_{\nu} - \partial_{\nu} X_{\mu} + i g_{(X)} [X_{\mu}, X_{\nu}].
\end{align}
In order to give the $\sigma$-model Lagrangian, it will be useful to recast the link fields $\Omega_i$ into independent link fields for each simple group factor in the global symmetry breakings. In this set is $\Omega_{1,X,G}$ respectively associated to the $SO(5)$, $U(1)_X$, and $SU(3)_c$ diagonal subgroup breakings between the two sites, and also $\Omega_2$ for the breaking $SO(5) \rightarrow SO(4)$. Each will carry its own NGB decay constant $f_{1,2,X,G}$ that determines the scale of the associated symmetry breaking. The link fields transform under the symmetries at each site as
\begin{align}
\begin{array}{cccc}
    SO(5)^0 \times SO(5)^1 :& \Omega_1 \rightarrow g_0 \Omega_1 g^{-1}_1, \ \ & U(1)^0_X \times U(1)^1_X :& \Omega_X \rightarrow g_0 \Omega_X g^{-1}_1,\\[0.1cm]
    SO(5)^1 \times SO(4)\hphantom{{}^1} :& \Omega_2 \rightarrow g_1 \Omega_2 h^{-1}, \ \ & SU(3)^0_c \times SU(3)^1_c :& \Omega_G \rightarrow g_0 \Omega_G g^{-1}_1,
\end{array}
\end{align}
where transformations $g_k$ come from Site $k$, and $h \in SO(4)$. Accordingly, their covariant derivatives are given by
\begin{align}
\begin{array}{lcl}
    D_{\mu} \Omega_1 & = & \partial_{\mu} \Omega_1 -i(g_0 W^{0 a}_{\mu} T^a_L + g'_{0} B^0_{\mu} T^3_R) \Omega_1 +ig_{\rho} \Omega_1 \rho_{\mu},\\[0.1cm]
    D_{\mu} \Omega_2 & = & \partial_{\mu} \Omega_2 -ig_{\rho} \rho_{\mu} \Omega_2,\\[0.1cm]
    D_{\mu} \Omega_X & = & \partial_{\mu} \Omega_X -ig'_{0} B^0_{\mu} \Omega_X +ig_{X} \Omega_X \rho_{X_{\mu}},\\[0.1cm]
    D_{\mu} \Omega_G & = & \partial_{\mu} \Omega_G -ig^{0}_{s} G^0_{\mu} \Omega_G +ig_{G} \Omega_G \rho_{G_{\mu}}.
\end{array}
\label{eq:Omega_derivatives}
\end{align}
Here there are elementary gauge couplings $g_{0}$, $g'_{0}$, and $g^{0}_{s}$ mirroring the SM couplings, as well as new composite couplings $g_{\rho}$, $g_{X}$, and $g_{G}$ of $SO(5)^{1}$, $U(1)^{1}_{X}$, and $SU(3)^{1}_{c}$. The generators $T^A = \{T^a_L, T^a_R, \hat{T}^a \}$ of $SO(5)$ have been grouped into $SU(2)_L$, $SU(2)_R$, and coset components (see \Cref{SO5_appendix}), allowing the $SO(5)$ resonances to be expressed as
\begin{align}
    \rho_{\mu} = \rho^A_{\mu} T^A = \rho^a_{L_{\mu}} T^a_L + \rho^a_{R_{\mu}} T^a_R + \mathfrak{a}^a_{\mu} \hat{T}^a.
\end{align}
Note that since $SO(5)$ is spontaneously broken to $SO(4)$, there is a vacuum vector $\Phi_0$ left invariant under $SO(4)$, so the product $\Phi := \Omega_2 \Phi_0$ transforms as a fundamental of $SO(5)^1$. We will use a basis where $\Phi_0 = (0,0,0,0,1)^\intercal$. With these, the leading-order $\sigma$-model Lagrangian is
\begin{align}
    \mathcal{L}_{\sigma} = \sum_{i = 1,X,G} \frac{f^2_i}{4} \Tr{(D_{\mu} \Omega_i)^{\dagger}(D^{\mu} \Omega_i)} + \frac{f^2_2}{2} (D_{\mu} \Omega_2 \Phi_0)^{\intercal}(D^{\mu} \Omega_2 \Phi_0).
\label{eq:sigma_Lagrangian}
\end{align}
We now go to the Site $0$ holographic gauge to remove the unphysical NGBs, setting
\begin{align}
    \Omega_2 = \Omega_X = \Omega_G = \mathbb{1}, \quad \Omega_1 = e^{i \frac{\sqrt{2}}{f_1} h^a \hat{T}^a},
\end{align}
where $h^a$ are the Higgs doublet components. From here on, unless otherwise stated all equations should be understood to be in the SM unitary gauge wherein $h^a = (0,0,0,h)^\intercal$. With our convention for the $SO(5)$ generators given in \Cref{SO5_appendix}, this makes the NGB matrix come out to
\begin{align}
\Omega \equiv \Omega_1 \Omega_2 = \Omega_1 = \left( \begin{matrix}
\ 1\ & & & & \\
   &\ 1\ & & & \\
   & &\ 1\ & & \\
   & & & \cos\frac{h}{f_1} & \sin\frac{h}{f_1} \\[2pt]
   & & & -\sin\frac{h}{f_1} & \cos\frac{h}{f_1}
\end{matrix}\right).\label{eq:explicit_goldstone_matrix}
\end{align}
Now the gauge boson mixing terms in \Cref{eq:sigma_Lagrangian} become readily apparent:
\begin{align}
    \mathcal{L}_{\sigma} = &\frac{f^2_1}{4} \Tr{(D_{\mu} \Omega_1)^{\dagger}(D^{\mu} \Omega_1)} + \frac{f^2_2}{2} (D_{\mu} \Omega_2 \Phi_0)^{\intercal}(D^{\mu} \Omega_2 \Phi_0) \nonumber \\
    & + \frac{f^2_X}{4} (g^{\prime}_0 B^0_{\mu} - g_{X} \rho_{X_{\mu}})^2 + \frac{f^2_G}{4} (g_{0_{s}} G^0_{\mu} - g_{G} \rho_{G_{\mu}})^2.
\label{eq:sigma_mixing_Lagrangian}
\end{align}
Hidden inside the first line of \Cref{eq:sigma_mixing_Lagrangian} is mixing between $\mathfrak{a}^{4}_{\mu}$ and the Higgs:
\begin{align}
    \mathcal{L}_{\sigma} \supset \frac{1}{2} (\partial_{\mu} h)(\partial^{\mu} h) + \frac{f_1}{\sqrt{2}} g_{\rho} \mathfrak{a}^{4}_{\mu} \partial^{\mu} h + \frac{g^2_{\rho}}{4} (f^2_1 + f^2_2) \mathfrak{a}^{4}_{\mu} \mathfrak{a}^{4 \mu}.
\end{align}
This mixing is removed, keeping the Higgs kinetic term conventionally normalised, with the field redefinitions
\begin{align}
    \mathfrak{a}^{4}_{\mu} \rightarrow \mathfrak{a}^{4}_{\mu} - \frac{\sqrt{2}}{g_{\rho}} \frac{f}{f^2_2} \partial_{\mu} h, \quad h \rightarrow \frac{f_1}{f} h, \quad \text{where} \quad \frac{1}{f^2} \equiv \frac{1}{f^2_1} + \frac{1}{f^2_2}.
\label{field_redefinitions}
\end{align}
Note that the Higgs interactions now depend only on the quantity
\begin{align}
    s_{h} := \sin \brackets{\frac{h}{f}}.
\end{align}
Hence, masses can only depend on the Higgs field through the ``vacuum misalignment" parameter $s_{\langle h \rangle}$. We will return to this point in \Cref{particle_content_section}.

\subsection{Fermion sector}
\label{fermion_sector}

At this point, the remaining freedom in specifying a particular model lies in establishing the fermionic content of the composite site and the representations of $SO(5) \times U(1)_X$ under which it transforms. We will be restricting to the case where the composite sector mixes only with the third generation elementary quarks in the interests of reducing the parameter space of the theory. This is a reasonable limiting case, for the lighter fermions must mix with the composite sector only weakly if their left- and right-handed chiralities mix with roughly equal strengths.

\begin{figure}
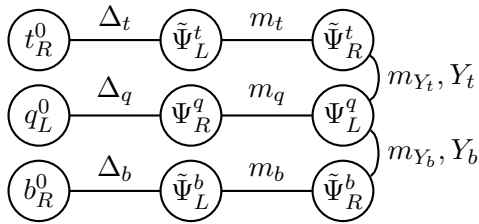

\centering\tikz[scale=1, every node/.style={transform shape}]{
\draw [thick] (2,1) circle [radius=0.4] node {$t^0_R$};
\draw [thick] (2,0) circle [radius=0.4] node {$q^0_L$};
\draw [thick] (2,-1) circle [radius=0.4] node {$b^0_R$};
\draw [thick] (2.4,1) -- (3.6,1);
\draw [thick] (2.4,0) -- (3.6,0);
\draw [thick] (2.4,-1) -- (3.6,-1);
\node [above] at (3,1) {$\Delta_t$};
\node [above] at (3,0) {$\Delta_q$};
\node [above] at (3,-1) {$\Delta_b$};
\draw [thick] (4,1) circle [radius=0.4] node {$\tilde{\Psi}^t_L$};
\draw [thick] (4,0) circle [radius=0.4] node {$\Psi^q_R$};
\draw [thick] (4,-1) circle [radius=0.4] node {$\tilde{\Psi}^b_L$};
\draw [thick] (4.4,1) -- (5.6,1);
\draw [thick] (4.4,0) -- (5.6,0);
\draw [thick] (4.4,-1) -- (5.6,-1);
\node [above] at (5,1) {$m_t$};
\node [above] at (5,0) {$m_q$};
\node [above] at (5,-1) {$m_b$};
\draw [thick] (6,1) circle [radius=0.4] node {$\tilde{\Psi}^t_R$};
\draw [thick] (6,0) circle [radius=0.4] node {$\Psi^q_L$};
\draw [thick] (6,-1) circle [radius=0.4] node {$\tilde{\Psi}^b_R$};
\draw [thick] plot [smooth, tension=1] coordinates {(6.346,0.8) (6.45,0.65) (6.45,0.35) (6.346,0.2)};
\node [right] at (6.45,0.5) {$m_{Y_t}, Y_t$};
\draw [thick] plot [smooth, tension=1] coordinates {(6.364,-0.2) (6.45,-0.35) (6.45,-0.65) (6.364,-0.8)};
\node [right] at (6.45,-0.5) {$m_{Y_b}, Y_b$};
}
\caption{Structure of couplings between elementary and composite resonances. For the M4DCHM$^{5-5-5}$, the left-handed elementary quark doublet $q^{0}_{L}$ actually couples to two different composite multiplets $\Psi^{t}$ and $\Psi^{b}$, which respectively couple to $\tilde{\Psi}^{t}$ and $\tilde{\Psi}^{b}$ with independent interaction strengths. \label{fig:interaction_structure}}
\end{figure}

We will be considering three different models in this work: the M4DCHM$^{5-5-5}$, the $\text{M4DCHM}^{14-14-10}$, and the M4DCHM$^{14-1-10}$, all of which have been outlined in Ref.~\cite{Carena}. The labels M4DCHM$^{q-t-b}$ specify the representations $(\mathbf{q}, \mathbf{t}, \mathbf{b})$ of $SO(5)$ in which the respective composite partners $(\Psi^q, \tilde{\Psi}^t, \tilde{\Psi}^b)$ of the elementary $( q^0_L = (t^0_L, b^0_L)^\intercal, t^0_R, b^0_R )$ transform. The structure of the elementary-composite interactions is depicted graphically in \Cref{fig:interaction_structure}.

These models are chosen primarily because they provide custodial protection for the $Z b_L \bar{b}_L$ coupling from tree-level corrections \cite{contino2006}, and because of their different relationships with fine-tuning. The fine-tuning of a model is commonly estimated as $f^2/v^2$, where $v \approx 246 \gev$ is the EWSB scale, though the M4DCHM$^{5-5-5}$ is subject to a parametrically higher ``double tuning" because it requires precise cancellations of leading terms in the Higgs potential to achieve a non-trivial minimum and to produce the correct EWSB scale \cite{panico2012,matsedonskyi2012}. The other models need not feature such extra tunings on account of extra leading-order invariants possible in the Higgs potential with the $\mathbf{14}$ representation. The M4DCHM$^{14-14-10}$ and M4DCHM$^{14-1-10}$ each also offer interesting prospects, with the former having symmetries that allow for two independent proto-Yukawa couplings for the top quark, and the latter being able to feature an entirely composite right-handed top quark (though we do not take it as composite in this work). These are not the only possible models that fit our criteria, but it has been found in Ref.~\cite{Carena} that the various other representation combinations either tend to give experimentally unfavourable predictions, or are qualitatively similar to the three we consider here.

In specifying the Lagrangian for the fermion sector of each model, we split up the contributions as
\begin{align}
    \mathcal{L}^{q-t-b}_{\text{fermion}} = \mathcal{L}^{q-t-b}_{\text{comp. quark}} + \mathcal{L}_{\text{elem. quark}} + \mathcal{L}_{\text{lepton}}.
\end{align}
The high-energy Lagrangian for the (partially) composite quarks, consisting of the third generation elementary quarks and their composite partners $\Psi$ and $\tilde{\Psi}$, will be given below for each model. Elementary fields will be embedded into incomplete multiplets $\psi$ in the same representation as their partners for convenience in writing gauge-invariant interactions\footnote{A more general approach is to have the SM-like elementary fields embedded in \textit{complete} representations, with mass terms given to the supplementary fields in the elementary multiplets to explicitly break the $SO(5)$ symmetry~\cite{Blasi:2019jqc,Blasi:2020ktl}. Our formulation is obtained in the limit where these mass couplings go to infinity. Along with the assumption of maximal symmetry~\cite{Csaki:2017cep}, such general models alleviate issues with light quark partners and promise favourable levels of fine-tuning. It would be interesting to fit these models in future work.}. In all cases the covariant derivatives of the fields are given by
\begin{align}
\label{eq:covariant_derivatives}
D_\mu \psi_R &= \left(\partial_\mu  - i g'_0 B^0_\mu Y - i g^{0}_{s} G^{0 a}_\mu \frac{\lambda^a}{2} \right) \psi_R, \nonumber \\
D_\mu \psi_L &= \left(\partial_\mu  - i g_0 W^{0 a}_\mu T_L^a - i g'_0 B^0_\mu Y - i g^{0}_{s} G^{0 a}_\mu \frac{\lambda^a}{2} \right) \psi_L, \\
D_\mu \Psi &= \left(\partial_\mu  - i g_\rho \rho^A_\mu T^A - i g_X \rho_X X  - i g_G \rho^a_{G_{\mu}} \frac{\lambda^a}{2} \right) \Psi, \nonumber
\end{align}
where the hypercharges are $(Y_{q_L}, Y_{t_R}, Y_{b_R}, Y_{l_L}, Y_{e_R}) = (\frac{1}{6}, \frac{2}{3}, \frac{-1}{3}, \frac{-1}{2}, -1)$, and $\lambda^a$ are the $SU(3)$ generators, taken so that each elementary or composite field couples with the same strength to the $SU(3)$ gauge bosons. It should be kept in mind that a multiplet $\Psi$ in the symmetric $\mathbf{14}$ or antisymmetric $\mathbf{10}$ of $SO(5)$, when expressed as a matrix, is acted upon by the generators as
\begin{align}
    T^A \Psi = [T^A, \Psi].
\end{align}
Refer to \Cref{SO5_appendix} for the explicit embeddings of fields into representations of $SO(5)$.

\subsubsection{M4DCHM$^{5-5-5}$}

Unlike the other models we will be exploring, the M4DCHM$^{5-5-5}$ requires \emph{two} composite multiplets $\Psi^{t,b}$ to couple to $q^0_L$, simply because a single partner multiplet does not provide sufficient couplings for both the top and the bottom quark to have mass. The partner content of this model consists of $\Psi^t$, $\tilde{\Psi}^t$ in the $\mathbf{5}_{+\frac{2}{3}}$ representation of $SO(5) \times U(1)_X$, and  $\Psi^b$, $\tilde{\Psi}^b \sim \mathbf{5}_{-\frac{1}{3}}$. Their interactions are specified by the quark Lagrangian
\begin{align}
\mathcal{L}^{5-5-5}_\text{comp. quark} =\  &\bar{q}^0_L i \slashed{D} q^0_L + \bar{t}^0_R i \slashed{D} t^0_R + \bar{b}^0_R i \slashed{D} b^0_R && \left.\vphantom{\bar{b}^0_R \slashed{D}}\right\rbrace \text{ elementary} \nonumber \\[0.1cm]
& + \bar{\Psi}^t\left(i \slashed{D} - m_{t} \right)\Psi^t + \bar{\tilde{\Psi}}^t \left( i\slashed{D} - m_{\tilde{t}} \right) \tilde{\Psi}^t && \left.\vphantom{\bar{\tilde{\Psi}}^u}\right\rbrace \text{ composite} \nonumber\\[0.1cm]
& +  \Delta_{tL} \bar{\psi}^t_L \Omega_1 \Psi^t_R +  \Delta_{tR} \bar{\psi}^t_R \Omega_1 \tilde{\Psi}^t_L && \left.\vphantom{\bar{\psi}^u_R}\right\rbrace \text{ link} \nonumber\\[0.1cm]
& - m_{Y_t} \bar{\Psi}^t_L \tilde{\Psi}^t_R - Y_t \bar{\Psi}^t_L \Phi \Phi^\dagger \tilde{\Psi}^t_R && \left.\vphantom{\bar{\psi}^u_R}\right\rbrace \text{ Yukawa}\nonumber \\[0.1cm]
& + (t \rightarrow b) + \text{h.c.} \label{eq:fundamental_fermion_lagrangian_5-5-5}
\end{align}
Here the third generation elementary quarks are embedded into the incomplete $SO(5)$ fundamentals
\begin{align}
\psi^t_L = \frac{1}{\sqrt{2}} \left( \begin{matrix}
    b^0_L \\[1pt]
    -ib^0_L \\[1pt]
    t^0_L \\[1pt]
    it^0_L \\[1pt]
    0
\end{matrix} \right), \quad \psi^t_R  = \left( \begin{matrix}
        {} \\[1pt]
         \vec{0}  \\[1pt]
        {} \\[1pt]
        {} \\[1pt]
        \hline
        t^0_R
    \end{matrix} \right), \quad \psi^b_L = \frac{1}{\sqrt{2}} \left( \begin{matrix}
    t^0_L\\[1pt]
    i t^0_L\\[1pt]
    - b^0_L\\[1pt]
    i b^0_L\\[1pt]
    0
\end{matrix} \right), \quad \psi^b_R  = \left( \begin{matrix}
        {} \\[1pt]
         \vec{0}  \\[1pt]
        {} \\[1pt]
        {} \\[1pt]
        \hline
        b^0_R
    \end{matrix} \right).
\end{align}
Note that this is not the most general Lagrangian obeying the symmetries of the theory. Terms such as $\psi_L \Omega_1 \tilde{\Psi}_R$ or $\bar{\tilde{\Psi}}_L \Psi_R$, if included, would reduce the model to the two-site Discrete CHM and lead to a logarithmic divergence in the Higgs potential. Similar considerations exist for the other representations.

\subsubsection{M4DCHM$^{14-14-10}$}
Here we have two composite multiplets, $\Psi^q$ and $\tilde{\Psi}^t$, in the traceless symmetric $\mathbf{14}$ of $SO(5)$, and one multiplet $\tilde{\Psi}^b$ in the antisymmetric $\mathbf{10}$. Each composite multiplet carries an $X$ charge of $+\frac{2}{3}$. All of them transform adjointly,
\begin{align}
    \Psi \xrightarrow{g \in SO(5)} g \Psi g^{-1},
\end{align}
so invariant terms may be found by taking the traces of various field combinations. Writing out the invariant interactions yields the quark Lagrangian
\begin{align}
\mathcal{L}^{14-14-10}_\text{comp. quark} =\  &\bar{q}^0_L i \slashed{D} q^0_L + \bar{t}^0_R i \slashed{D} t^0_R + \bar{b}^0_R i \slashed{D} b^0_R && \left.\vphantom{\bar{b}^0_R \slashed{D}}\right\rbrace \text{ elementary} \nonumber \\[0.1cm]
& + \Tr{\bar{\Psi}^q \left(i \slashed{D} - m_q \right) \Psi^q} + \Tr{\bar{\tilde{\Psi}}^t \left(i \slashed{D} - m_t \right) \tilde{\Psi}^t} && \left.\vphantom{\Tr{\bar{\tilde{\Psi}}^u}}\right\rbrace \text{ composite} \nonumber\\[0.1cm]
& +  \Delta_{q} \Tr{\bar{\psi}^q_L \Omega_1 \Psi^q_R \Omega^\dagger_1} +  \Delta_{t} \Tr{\bar{\psi}^t_R \Omega_1 \tilde{\Psi}^t_L \Omega^\dagger_1} && \left.\vphantom{\Tr{\tilde{\Psi}^u}}\right\rbrace \text{ link} \nonumber\\[0.1cm]
&- m_{Y_t} \Tr{\bar{\Psi}^q_L \tilde{\Psi}^t_R} - Y_t \Phi^\dagger \bar{\Psi}^q_L \tilde{\Psi}^t_R \Phi - \tilde{Y}_t \Phi^\dagger \bar{\Psi}^q_L \Phi \Phi^\dagger \tilde{\Psi}^t_R \Phi && \left.\vphantom{\tilde{\Psi}^u_R}\right\rbrace \text{ Yukawa} \nonumber\\[0.1cm]
& + (t \rightarrow b) + \text{h.c.}
\label{eq:fundamental_fermion_lagrangian_14-14-10}
\end{align}
The elementary quark fields are embedded into incomplete representations as
\begin{align}
\psi^q_L &= \frac{1}{2} \left( \begin{array}{c|c}
        {} & i b^0_L\\[1.5pt]
        0_{4{\times}4} &   b^0_L\\[1.5pt]
        {} & i t^0_L\\[1.5pt]
        {} & - t^0_L\\[1.5pt]
        \hline
        i b^0_L \ \ b^0_L \ \ i t^0_L \ - t^0_L & 0\\
    \end{array} \right), \quad \psi^b_R  = \frac{b^0_R}{\sqrt{8}}  \left( \begin{matrix}
    0 & 0 & i  & -1 & 0\\[1pt]
    0 & 0 & 1 & i  & 0\\[1pt]
    -i  & -1 & 0 & 0 & 0\\[1pt]
    1    & -i  & 0 & 0 & 0\\[1pt]
    0 & 0 & 0 & 0 & 0
\end{matrix} \right), \nonumber\\[8pt]
\psi^t_R  &= - \frac{t^0_R}{\sqrt{20}} \text{ diag}(1,1,1,1,-4).
\label{eq:14-14-10_incomplete}
\end{align}

Notice that by the symmetries of this model there is an extra allowed Yukawa-like term, having a coupling strength $\tilde{Y}_{t}$. The analogous terms proportional to $m_{Y_b}$ and $\tilde{Y}_b$ vanish due to the symmetry (antisymmetry) of $\Psi^q$ ($\tilde{\Psi}^b$).

\subsubsection{M4DCHM$^{14-1-10}$}
The M4DCHM$^{14-1-10}$ has a similar structure to the M4DCHM$^{14-14-10}$, except with $\tilde{\Psi}^t$ now being a single field in the $\mathbf{1}_{+\frac{2}{3}}$ of $SO(5) \times U(1)_X$ instead of a multiplet. The quark Lagrangian is given by
\begin{align}
\mathcal{L}^{14-1-10}_\text{comp. quark} =\ &\bar{q}^0_L i \slashed{D} q^0_L + \bar{t}^0_R i \slashed{D} t^0_R + \bar{b}^0_R i \slashed{D} b^0_R \nonumber \\[0.1cm]
& + \Tr{\bar{\Psi}^q \left(i \slashed{D} - m_q \right) \Psi^q} + \bar{\tilde{\Psi}}^t \left(i \slashed{D} - m_t \right) \tilde{\Psi}^t + \Tr{\bar{\tilde{\Psi}}^b \left(i \slashed{D} - m_b \right) \tilde{\Psi}^b} \nonumber\\[0.1cm]
& +  \Delta_{q} \Tr{\bar{\psi}^q_L \Omega_1 \Psi^q_R \Omega^\dagger_1} +  \Delta_{t} \bar{\psi}^t_R \tilde{\Psi}^t_L  +  \Delta_{b} \Tr{\bar{\psi}^b_R \Omega_1 \tilde{\Psi}^b_L \Omega^\dagger_1} \nonumber\\[0.1cm]
& - Y_t \Phi^\dagger \bar{\Psi}^q_L \Phi \tilde{\Psi}^t_R  - Y_b \Phi^\dagger \bar{\Psi}^q_L \tilde{\Psi}^b_R \Phi \nonumber \\[0.1cm]
& + \text{h.c.}
\label{eq:fundamental_fermion_lagrangian_14-1-10}
\end{align}
The elementary $\psi^q_L$ and $\psi^b_R$ multiplets are the same here as in \Cref{eq:14-14-10_incomplete}, but now $\psi^t_R = t^0_R$. Notice that the symmetries do not allow any off-diagonal mass terms $m_{Y}$ in this model.

\subsection{Elementary fermions}
In a truly consistent CHM, all of the lighter fermions would be incorporated through partial compositeness in a similar manner to the above. However, in our work, we treat them as entirely elementary for simplicity. We reproduce the SM lepton sector with bilinear couplings to the low-energy Higgs pNGB:
\begin{equation}
    \mathcal{L}_\text{lepton} = \sum_{\text{generations}} \left( i \bar{l}_L \slashed{D} l_L + i \bar{\ell}_R \slashed{D} \ell_R - \frac{m_\text{SM}}{v} \bar{l}_L  \left(\begin{matrix}
                0\\
                h
        \end{matrix} \right) \ell_R + \text{h.c.} \right).
\end{equation}
The covariant derivatives here couple the leptons to the elementary gauge fields, with SM quantum numbers as in \Cref{eq:covariant_derivatives}. The elementary quarks, on the other hand, are given simple Dirac mass terms
\begin{equation}
    \mathcal{L}_\text{elem. quark} = \sum_{q = u,d,c,s} \bar{q} (i \slashed{D} - m_q) q
\end{equation}
with no Higgs couplings, and again coupling to the elementary gauge fields. Admittedly there is no reason to treat the mass terms of the elementary quarks and leptons differently. However, the difference is largely inconsequential due to the small couplings to the Higgs involved, and because there is little distinction between a mass term and a Higgs coupling below the EW scale.

\subsection{Higgs potential}
\label{Higgs_potential_section}

Mixing between the elementary and composite sites explicitly breaks the global symmetries of the composite sector, and therefore leads to a quantum effective potential for the (now pseudo-) NGB Higgs. At the low energies that we are capable of probing, it is useful to integrate out the composite degrees of freedom and work with the effective (composite quark) Lagrangians that, in momentum space, will be of the form
\begin{align}
    \mathcal{L}_{\text{eff}} = \sum_{\psi=t,b} \left[ \bar{\psi}^{0}_{L} \slashed{p} \brackets{1+\Pi_{\psi_{L}}(p^{2})} \psi^{0}_{L} + \bar{\psi}^{0}_{R} \slashed{p} \brackets{1+\Pi_{\psi_{R}}(p^{2})} \psi^{0}_{R} + \bar{\psi}^{0}_{L} M_{\psi}(p^{2}) \psi^{0}_{R} + \text{h.c.}\right]
\label{eqn:form_factor_Lagrangian}
\end{align}
for some model-dependent correlators $\Pi_{\psi}$ and $M_{\psi}$ provided in \Cref{appendix_correlators}. With these, the fermion contributions to the effective potential, with only partially composite third generation quarks, can be shown to be
\begin{align}
    V^{\text{fermion}}_{\text{eff}}(h) &= 2 i N_{c} \sum_{\psi=t,b} \int\frac{\dVt{p}}{(2 \pi)^{4}} \ln\sqbrackets{\brackets{1 + \Pi_{\psi_{L}}(p^{2})}\brackets{1 + \Pi_{\psi_{R}}(p^{2})} + \frac{|M_{\psi}(p^{2})|^{2}}{p^{2}}}, \nonumber\\
    &= -2 N_{c} \sum_{\psi=t,b} \int\frac{\mathrm{d} p^{2}_{E}}{16 \pi^{2}} \ln \left[\brackets{1 + \Pi_{\psi_{L}}(-p^{2}_{E})}\brackets{1 + \Pi_{\psi_{R}}(-p^{2}_{E})} - \frac{|M_{\psi}(-p^{2}_{E})|^{2}}{p^{2}_{E}} \right],
\label{eqn:Veff}
\end{align}
where $N_{c}=3$ is the number of colours and a Wick rotation to Euclidean space was used. Notice that scaling the mass parameters does not change the stationary points of the potential. The location of its minimum - the Higgs VEV $\langle h \rangle$ - can be found by expanding the potential in powers of $s_{h}$ as $V_{\text{eff}}(h) =: - \gamma s_{h}^{2} + \beta s_{h}^{4}$ to obtain the coefficients $\gamma$ and $\beta$ that dictate the VEV satisfies
\begin{align}
    s_{\langle h \rangle} = \frac{\gamma}{2\beta},
\end{align}
and which can further be used to calculate the Higgs mass as
\begin{align}
    m_{H} = \sqrt{8 \beta \brackets{1-s^2_{\langle h \rangle}}} \frac{s_{\langle h \rangle}}{f}.
\end{align}

We neglect detailing the incorporation of the gauge boson contributions to the potential in this framework, since they are of secondary importance. Suffice to say that the gauge boson contribution to $\gamma$ can be calculated analytically, with the result that \cite{BarnardFT}
\begin{align}
    \gamma_{\text{gauge}} = - \frac{9 m^{4}_{\rho} \brackets{m^{2}_{a} - m^{2}_{\rho}} t_{\theta}}{64 \pi^{2} \brackets{m^{2}_{a} - \brackets{1 + t_{\theta}} m^{2}_{\rho}}} \ln\sqbrackets{\frac{m^{2}_{a}}{\brackets{1 + t_{\theta}} m^{2}_{\rho}}}
\end{align}
to first order in $t_{\theta} := g_{0}/g_{\rho}$. Here we have the approximate masses of the lightest composite gauge bosons (see \Cref{particle_content_section})
\begin{align}
    m_{\rho}^{2} := \frac{1}{2} g_{\rho}^{2} f_{1}^{2}, \qquad m_{a}^{2} := \frac{1}{2} g_{\rho}^{2} (f_{1}^{2} + f_{2}^{2}).
\label{eq:gauge_boson_mass_parameters_definitions}
\end{align}
We use this formula in our scans, and neglect the contributions to $\beta$ from the gauge bosons.

\subsection{Particle content}
\label{particle_content_section}

One interesting signature of composite Higgs models is the existence of new resonances in the few-TeV mass range. In this section we give an overview of the particle content in each of the models we are considering, including some analysis of the particles' masses.

\subsubsection*{Boson sector}
Each model shares the same boson sector. It contains, in addition to the SM bosons, the heavy gluons, five other neutral bosons, and three bosons of unit electric charge. Their mass mixing matrices are given as functions of the Lagrangian parameters in \Cref{appendix_mass_matrices}. Singular values of the matrices are the squared (tree-level) masses of the resonances, though not all have useful analytic forms. The bosons with easily calculable masses are the heavy gluons and the neutral $\mathfrak{a}^4$ resonance, which have squared masses of
\begin{align}
    m^2_G = \frac{1}{2} f^2_G ((g^{0}_{s})^{2} + g^{2}_{G}), \quad m^2_{\mathfrak{a}^4} = \frac{1}{2} \frac{f^4_1}{f^2_1 - f^2} g^2_{\rho},
\end{align}
and the charged bosons, whose mass spectrum is given by $m^2_{\text{charged}} = \left\{ \frac{1}{2} f^2_1 g^2_{\rho}, m^2_1, m^2_2, m^2_3 \right\}$,
where the last three masses are the solutions to
\begin{align}
     m^2 \left( m^2 -\frac{1}{2} (g^2_0 + g^2_{\rho}) f^2_1 \right) \left( m^2 - \frac{g^2_{\rho} f^2_1}{2} \frac{f^2_1}{f^2_1-f^2} \right) = \frac{1}{16} g^2_0 g^4_{\rho} \frac{f^6_1 f^2}{f^2_1 - f^2} s^{2}_{\langle h \rangle}.
\end{align}
The W boson mass, for example, distinguished as being $\mathcal{O}(s_{\langle h \rangle})$, is given by
\begin{align}
    m_{W} = \frac{1}{2} \frac{g_{0} g_{\rho}}{\sqrt{g^{2}_{0} +  g^{2}_{\rho}}} f s_{\langle h \rangle} + \mathcal{O}(s^{3}_{\langle h \rangle}).
\label{eq:W_mass}
\end{align}
Neglecting the higher-order terms, we can match to the SM expression $m_{W} = \frac{1}{2}gv$ to find the relations
\begin{align}
    g = \frac{g_{0} g_{\rho}}{\sqrt{g^{2}_{0} +  g^{2}_{\rho}}}, \qquad s_{\langle h \rangle} = \frac{v}{f}.
\label{eq:matching_conditions}
\end{align}
Similar matchings can also be done to yield relations for the remaining SM gauge couplings in terms of the elementary and composite couplings that we use throughout this work:
\begin{align}
    \frac{1}{g^{\prime 2}} = \frac{1}{g^{\prime 2}_0} + \frac{1}{g^2_{\rho}} + \frac{1}{g^2_{X}}, \quad \text{and} \quad \frac{1}{g^2_s} = \frac{1}{(g^{0}_{s})^{2}} + \frac{1}{g^2_{G}}.
\label{eq:SM_gauge_couplings}
\end{align}

The same sort of analysis can be done for the other resonances, but the contributions to their masses from the Higgs field are only relatively minor. We simply give their masses in the $s_{\langle h \rangle} \rightarrow 0$ limit, which will be approximately satisfied by realistic points and can be used as a zeroth order estimate. In this limit, excluding the electroweak bosons (which will be massless), the charged bosons have masses
\begin{align}
    \lim\limits_{s_{\langle h \rangle} \rightarrow 0} m^2_{\text{charged}} = \left\{ \frac{1}{2} f^2_1 g^2_{\rho},\ \frac{1}{2} f^2_1 \left(g^2_0 + g^2_{\rho} \right),\ \frac{1}{2} \frac{f^4_1}{f^2_1 - f^2} g^2_{\rho} \right\},
\end{align}
and the other neutral bosons have masses
\begin{align}
    \lim\limits_{s_{\langle h \rangle} \rightarrow 0} m^2_{\text{neutral}} = \left\{\frac{1}{2} f^2_1 \left(g^2_0 + g^2_{\rho} \right),\ \frac{1}{2} \frac{f^4_1}{f^2_1 - f^2} g^2_{\rho},\ \frac{1}{2} M^2_{\pm} \left(f_1 g_{\rho}, f_X g_{X}, f_X g_{\rho}, f_1 g^{\prime}_0 \right) \right\}.
\end{align}
Here we have defined the functions
\begin{align}
    M^2_{\pm}(x_1, x_2, x_3, x_4) :=  \frac{|\vec{x}|^2}{2} \pm \sqrt{\frac{|\vec{x}|^4}{4} - \left(x_1^2 x_2^2 + x_2^2 x_4^2 + x_3^2 x_4^2 \right)},
\label{eq:mass_function}
\end{align}
which will also be useful in expressing the fermion masses.

Note there is some degeneracy among the masses of the charged and neutral bosons in the $s_{\langle h \rangle} \rightarrow 0$ limit. In fact, numerical calculation of the masses for realistic parameter points with small non-zero $s_{\langle h \rangle}$ suggests that two charged bosons will always separately be very close in mass to two neutral bosons, and the last charged boson will be very close in mass to two other neutral bosons (``very close" here meaning within ${\sim}0.1\%$).

\subsubsection*{Fermion sector}

\begin{table}[t]
\begin{center}
\begin{tabular}{l|ccccc}
Model & U & D & Q$_{4/3}$ & Q$_{5/3}$ & Q$_{8/3}$\\[1pt]
\hline
\\[-1em]
M4DCHM$^{5-5-5}$ & 8 & 8 & 2 & 2 & 0\\
M4DCHM$^{14-14-10}$ & 16 & 9 & 2 & 9 & 2\\
M4DCHM$^{14-1-10}$ & 11 & 6 & 1 & 6 & 1\\
\end{tabular}
\caption{Number of non-SM particles of a given type in each model. There are up-type (U) and down-type (D) particles, as well as particles Q$_x$ that have exotic electric charge $x$.
\label{tab:particle_content}}
\end{center}
\end{table}

The fermion sector of each model can be analysed using the explicit embeddings of fields in the different representations of $SO(5)$ given in \Cref{SO5_appendix}. We group the particles according to their electric charge, which is determined through the relation
\begin{align}
    Q = T^3_L + Y = T^3_L + T^3_R + X.
\end{align}
From the fields' $SU(2)_L \times SU(2)_R$ quantum numbers and the $U(1)_X$ charges of the multiplets specified in \Cref{fermion_sector}, the models are seen to deliver some up-type (U) and down-type (D) resonances, along with particles of exotic charge Q$_{4/3}$, Q$_{5/3}$, and Q$_{8/3}$. The number of new particles in each model is listed in \Cref{tab:particle_content}.

\begin{figure}[t]
\centering
\includegraphics[width=1\linewidth]{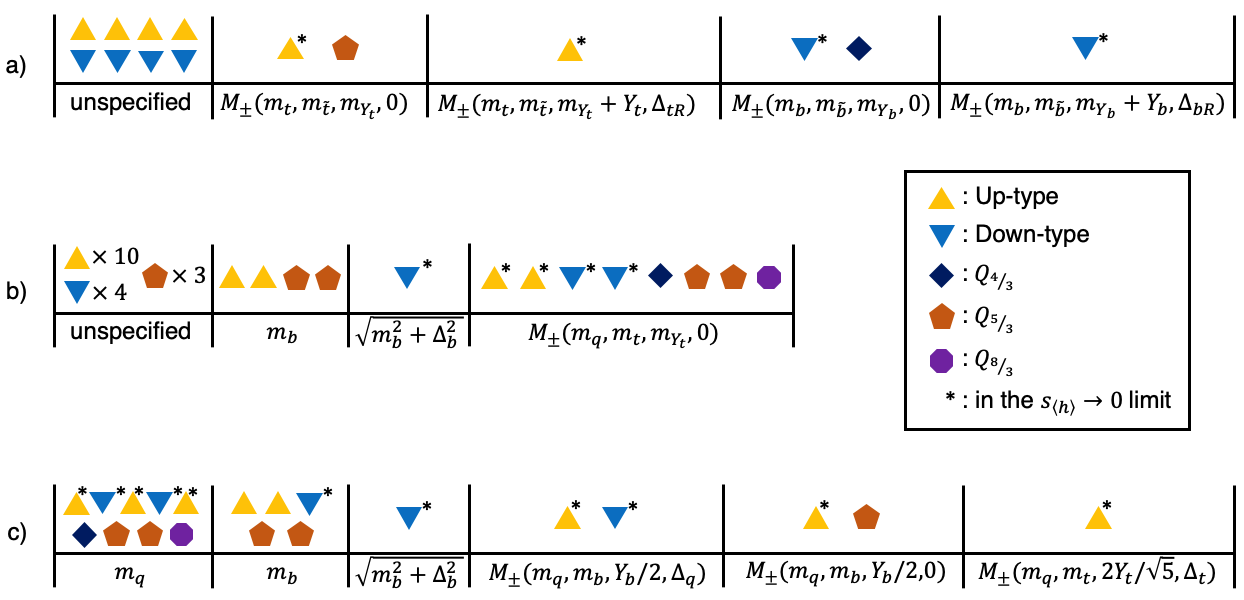}
\caption{Tree-level masses of the non-SM fermions (a) in the M4DCHM$^{5-5-5}$, (b) in the M4DCHM$^{14-14-10}$, and (c) in the M4DCHM$^{14-1-10}$. Each icon corresponds to one particle, or to a pair of particles for masses given in terms of the functions $M_{\pm}$ defined by \Cref{eq:mass_function}. Some masses cannot be expressed in a reasonably understandable form, in which case they are left unspecified.}
\label{fig:fermion_masses}
\end{figure}

The (tree-level) masses of the fermions are found as the singular values of the fermion mass matrices in \Cref{appendix_mass_matrices}. For convenience, the non-SM fermion masses in each model are provided in \Cref{fig:fermion_masses}. Similarly to the bosons, many fermion masses are not easily expressible analytically, and we resort to giving some masses in the $s_{\langle h \rangle} \rightarrow 0$ limit. Note the many degeneracies and approximate degeneracies among the particle masses in each model.

\section{Scan procedure}
\label{scanning}

In this section we provide details regarding our global fits of the models specified in \Cref{model}. A description of the scanning algorithm used for the fits is given in \Cref{scanning_algorithm}, and our treatment of the scan parameters in \Cref{scan_parameters}. The experimental constraints employed to determine the validity of a given parameter point are outlined in \Cref{constraints}.

\subsection{Scanning algorithm}
\label{scanning_algorithm}

The purpose of our fits is twofold: firstly, to obtain estimates of the parameters of each model given current experimental data, and secondly to facilitate model comparisons under a Bayesian framework through the calculation of their Bayesian evidences. Both are made possible by scanning the parameter spaces with the nested sampling algorithm \cite{skilling2006}, which has proven effective at sampling difficult spaces in many physical theories, including CHMs. For our fits we make use of the recently developed nested sampling package \texttt{PolyChord}, which uses a sophisticated version of \textit{slice sampling} to generate points efficiently \cite{Handley:2015fda,Handley_2015}.

Nested sampling algorithms explore a model's parameter space using a certain number $n_{\text{live}}$ of ``live" parameter points $\mathbf{p}$ that are iteratively constrained into progressively smaller regions in which the predictions of the model better match observed data. The experimental constraints are enacted through a likelihood function $\mathcal{L}(\mathbf{p})$ that the algorithm tries to minimise. We take the likelihood to be Gaussian in the observables:
\begin{align}
    P(\mathcal{O}|\mathbf{p})\propto\mathcal{L}(\mathbf{p}) = e^{-\frac{1}{2}\chi^2(\mathbf{p})}, \qquad \chi^2(\mathbf{p}) = ( \mathbf{\mathcal{O}}^{\text{theo}}(\mathbf{p}) - \mathbf{\mathcal{O}}^{\text{exp}} )^\intercal C^{-1} ( \mathbf{\mathcal{O}}^{\text{theo}}(\mathbf{p}) - \mathbf{\mathcal{O}}^{\text{exp}} ),
\label{eq:likelihood_chi2_definition}
\end{align}
where $\mathbf{\mathcal{O}}^{\text{theo}}(\mathbf{p})$ is a ``vector" of the predicted values of the observables that have experimental values $\mathbf{\mathcal{O}}^{\text{exp}}$, and $C$ is the covariance matrix that takes into account theoretical and experimental uncertainties, and also correlations between observables. Details about the specific data we use are given in \Cref{constraints}. The live points are sampled according to a prior distribution $\pi$ imposed on the space, which together with the likelihood $\mathcal{L}$ gives the posterior probability $\mathcal{P}$ of each point
\begin{align}
    P(\mathbf{p}|\mathcal{O}) = \mathcal{P}(\mathbf{p}) = \frac{\mathcal{L}(\mathbf{p}) \pi(\mathbf{p})}{\mathcal{Z}},
\label{eqn:Bayes}
\end{align}
where
\begin{align}
    \mathcal{Z} := \int \mathrm{d}^n p\  \mathcal{L}(\mathbf{p}) \pi(\mathbf{p})
\label{eqn:evidence_definition}
\end{align}
is the Bayesian evidence of the model in question. Throughout nested sampling exploration, the Bayesian evidence is able to be estimated from the prior-weighted volumes of the live points at each iteration and their likelihoods. We declare the algorithm ``converged" when the evidence carried by the remaining live points falls below $10^{-3}$ times that contributed from the previous iterations.

The posterior distribution and the evidence both weigh the fitness of a point or region of space against its naturalness, and so are well suited for judging the viability of composite Higgs models. A similarly helpful quantity that may be defined is the Kullback-Leibler (KL) divergence between the posterior and the prior,
\begin{align}
    D_{\text{KL}} := \int \mathrm{d}^n p\  \mathcal{P}(\mathbf{p} )  \ln\brackets{\frac{\mathcal{P}(\mathbf{p})}{\pi(\mathbf{p})}},
\label{eqn:KL_divergence}
\end{align}
which measures, in a sense, the extra information gained when going from the prior to the posterior \cite{kullback1951}, and therefore is an indirect measure of the fine-tuning of a model. Using \Cref{eqn:Bayes}, this may be rewritten in terms of the average loglikelihood over the posterior as~\cite{2020arXiv200715632H,hergt}
\begin{align}
     \langle \ln(\mathcal{L}) \rangle_{\mathcal{P}} = D_{\text{KL}} + \ln(\mathcal{Z}).
\end{align}
Once the evidence and KL divergence are known, this is a useful way to calculate the posterior-averaged log-likelihood as a quick check of how well the model fits the data.

\subsection{Scan parameters}
\label{scan_parameters}

\begin{table}[t]
\begin{center}
\begin{tabular}{l|l|l|l}
M4DCHM & \multicolumn{1}{c}{\(\textbf{5}-\textbf{5}-\textbf{5}\)} & \multicolumn{1}{|c|}{\(\textbf{14}-\textbf{14}-\textbf{10}\)} & \multicolumn{1}{c}{\(\textbf{14}-\textbf{1}-\textbf{10}\)} \\
\midrule
Decay constants & $f$, $f_1$, $f_X$, $f_G$ & $f$, $f_1$, $f_X$, $f_G$ & $f$, $f_1$, $f_X$, $f_G$ \\[2pt]
Gauge couplings & $g_\rho$, $g_X$, $g_G$ & $g_\rho$, $g_X$, $g_G$ & $g_\rho$, $g_X$, $g_G$ \\[2pt]
Link couplings & $\Delta_{t_L}$, $\Delta_{t_R}$, $\Delta_{b_L}$, $\Delta_{b_R}$ &  $\Delta_{q}$, $\Delta_{t}$, $\Delta_{b}$ & $\Delta_{q}$, $\Delta_{t}$, $\Delta_{b}$ \\[2pt]
On-diagonal masses & $m_{t}$, $m_{\tilde{t}}$, $m_{b}$, $m_{\tilde{b}}$ & $m_{q}$, $m_{t}$, $m_{b}$ & $m_{q}$, $m_{t}$, $m_{b}$ \\[2pt]
Off-diagonal masses & $m_{Y_t}$, $m_{Y_b}$ & $m_{Y_t}$ & {} \\[2pt]
Proto-Yukawa couplings & $Y_t$, $Y_b$ & $Y_t$, $Y_b$, $\tilde{Y}_t$ & $Y_t$, $Y_b$ \\
\midrule
Dimensionality & \multicolumn{1}{c}{19} & \multicolumn{1}{|c|}{17} & \multicolumn{1}{c}{15}
\end{tabular}
\end{center}
\caption{Parameters present in each model.}
\label{tab:model_parameters}
\end{table}

A full list of the parameters for each of our models is given in \Cref{tab:model_parameters}. We take the approach of Refs.~\cite{BarnardCC,BarnardFT,Carena} and do not scan over these parameters as presented, but instead scan over the mass-dimension parameters normalised by the scale $f$. This choice is convenient because the Higgs potential can be minimised\footnote{Recall from \Cref{eqn:Veff} that the Higgs potential can be calculated using any units of mass without altering the location of its minimum.} to find the misalignment $s_{\langle h \rangle}$, and the mass scale only set later by \textit{defining}
\begin{align}
    f \equiv \frac{v}{s_{\langle h \rangle}} = \frac{246}{s_{\langle h \rangle}} \gev,
\end{align}
in accordance with \Cref{eq:matching_conditions} to automatically reproduce the correct EWSB scale, and by extension the experimental W and Z boson masses. Our treatment of the parameters in our scans is summarised in \Cref{tab:parameter_bounds}, and explained in greater detail below.

\begin{table}[ht]
\begin{center}
\begin{tabular}{ @{}lllc @{} }
\toprule
{} & Parameters & Scan Range & Prior\\
\midrule
\multirow{4}{*}{All Models} & $m_{\rho}/f,\ m_{a}/f$ & $[1/\sqrt{2}, 4 \pi]$ & \multirow{4}{*}{Uniform}\\[8pt]
 & $f_{X}/f,\ f_{G}/f$ & $[0.5, 2 \sqrt{3}]$\\[4pt]
 & $g_{\rho},\ g_{X},\ g_{G}$ & $[1.0, 4\pi]$\\[4pt]
\midrule
\multirow{12}{*}{M4DCHM$^{5-5-5}$} & $\Delta_{t_{L}}/f$ & $[e^{-0.25}, e^{1.5}]$ & \multirow{12}{*}{Logarithmic}\\[4pt]
 & $\Delta_{t_{R}}/f$ & $[e^{-0.75}, 4 \pi]$\\[4pt]
 & $\Delta_{b_{L}}/f$ & $[e^{-5.0}, e^{-3.0}]$\\[4pt]
 & $\Delta_{b_{R}}/f$ & $[e^{-0.5}, 4 \pi]$\\[4pt]
 & $m_{t}/f,\ m_{\tilde{b}}/f$ & $[e^{-0.5}, e^{1.5}]$\\[4pt]
 & $m_{\tilde{t}}/f$ & $[e^{-1.0}, 4 \pi]$\\[4pt]
 & $m_{b}/f$ & $[e^{-1.0}, e^{1.5}]$\\[4pt]
 & $m_{Y_{t}}/f$ & $[e^{-8.5}, 4 \pi]$\\[4pt]
 & $m_{Y_{b}}/f$ & $[e^{-0.25}, 4 \pi]$\\[4pt]
 & $(m_{Y_{t}} + Y_{t})/f$ & $[e^{-0.5}, 8 \pi]$\\[4pt]
 & $(m_{Y_{b}} + Y_{b})/f$ & $[e^{-8.5}, e^{-0.5}]$\\[4pt]
\midrule
\multirow{11}{*}{M4DCHM$^{14-14-10}$} & $\Delta_{q}/f$ & $[e^{-1.0}, e^{2.0}]$ & \multirow{11}{*}{Logarithmic}\\[4pt]
 & $\Delta_{t}/f$ & $[e^{-2.5}, e^{2.0}]$\\[4pt]
 & $\Delta_{b}/f$ & $[e^{-4.0}, e^{2.0}]$\\[4pt]
 & $m_{q}/f$ & $[e^{-1.0}, 4 \pi]$\\[4pt]
 & $m_{t}/f$ & $[e^{-2.5}, 4 \pi]$\\[4pt]
 & $m_{b}/f$ & $[e^{-2.0}, 4 \pi]$\\[4pt]
 & $m_{Y_{t}}/f$ & $[e^{-8.5}, 4 \pi]$\\[4pt]
 & $(m_{Y_{t}} + \frac{1}{2}Y_{t})/f$ & $[e^{-8.5}, 1.0]$\\[4pt]
 & $(m_{Y_{t}} + \frac{4}{5}(Y_{t} + \tilde{Y}_{t}))/f$ & $[e^{-3.0}, 2.6 \times 4 \pi]$\\[4pt]
 & $Y_{b}/f$ & $[e^{-4.0}, 4 \pi]$\\[4pt]
\midrule
\multirow{6}{*}{M4DCHM$^{14-1-10}$} & $\Delta_{q}/f,\ \Delta_{t}/f,\ Y_{t}/f$ & $[e^{-5.0}, 4 \pi]$ & \multirow{6}{*}{Logarithmic}\\[4pt]
 & $\Delta_{b}/f$ & $[e^{-7.0}, 4 \pi]$\\[4pt]
 & $m_{q}/f,\ m_{b}/f$ & $[e^{-3.0}, 4 \pi]$\\[4pt]
 & $m_{t}/f$ & $[e^{-9.0}, 4 \pi]$\\[4pt]
 & $Y_{b}/f$ & $[e^{-6.0}, 4 \pi]$\\[4pt]
\bottomrule
\end{tabular}
\caption{Ranges and priors used for the parameters in our scans. The masses $m_{\rho}$ and $m_{a}$, defined in \Cref{eq:gauge_boson_mass_parameters_definitions}, reparameterise the decay constants into a form more suitable for calculating the gauge boson contribution to the Higgs potential. The normalisation factor $f$ is determined after the potential is minimised. Some parameters are subject to further constraints specified in the main text.}
\label{tab:parameter_bounds}
\end{center}
\end{table}

\subsection*{NGB decay constants}

The Higgs decay constant $f$ is one of the more consequential parameters in the theory, defining the approximate mass scale $m_{*} \sim g_{\rho} f$ as well as the na{\"i}ve energy cutoff $\Lambda_f = 4 \pi f$. The ranges of other parameters are limited by the value of $f$. For example, $f_1$ must be greater than $f$ by virtue of \Cref{field_redefinitions}, and less than $\sqrt{3} f$ to maintain partial unitarisation of NGB scattering \cite{MarzoccaGeneralCHMs}. The other decay constants are constrained by
\begin{align}
    \frac{f_1}{2} \leq f_{X,G} \leq 2 f_1
\end{align}
to avoid decoupling any resonances. We use these constraints to define the bounds for these parameters, with the added condition that all are greater than $0.5$~TeV. Points that are generated within the bounds but that do not satisfy these consistency conditions are immediately discarded - a process that simply amounts to a modification of the prior such that only consistent points are admitted.

\subsection*{Gauge couplings}

Given that the composite sector is strongly coupled and we can only perform calculations in the semi-perturbative regime, we must take the gauge couplings in \Cref{tab:model_parameters} to be between $1$ and $4\pi$. In addition to these couplings, we also vary the SM gauge couplings within their experimental limits (at the scale of the top mass). The elementary gauge couplings are then determined by the relations in \Cref{eq:matching_conditions,eq:SM_gauge_couplings}. For $g^{0}_{s}$ to be real we require that $g_G > g_s$. We also impose the restrictions
\begin{align}
    \frac{1}{\sqrt{2}} f_{1} g_{\rho} < \Lambda_f, \quad \frac{1}{\sqrt{2}} f_{X} g_{X} < \Lambda_f, \quad \frac{1}{\sqrt{2}} f_{G} g_{G} < \Lambda_f,
\end{align}
to avoid vector resonance masses above the cutoff $\Lambda_f$.

\subsection*{Link, mass, and proto-Yukawa couplings}

The remaining parameters in \Cref{tab:model_parameters} have dimensions of mass and so are given upper bounds of $\Lambda_f$. Instead of directly scanning over the off-diagonal masses and proto-Yukawa couplings in \Cref{tab:model_parameters}, however, it turns out to be more convenient to scan over particular linear combinations that appear in the correlators and mass matrices. Specifically, we scan over
\begin{align}
\begin{array}{lllll}
    \text{M4DCHM}^{5-5-5} &:\ m_{Y_u},\ & m_{Y_d},\ & m_{Y_u} + Y_u,\ & m_{Y_d} + Y_d, \\[2pt]
    \text{M4DCHM}^{14-14-10} &:\ m_{Y_u},\ & Y_d,\ & m_{Y_u} + \frac{1}{2}Y_u,\ & m_{Y_u} + \frac{4}{5}(Y_u + \tilde{Y}_u), \\[2pt]
    \text{M4DCHM}^{14-1-10} &:\ Y_u,\ & Y_d,
\end{array}
\end{align}
normalised by $f$. All of these parameters that we scan over are taken to be positive through field redefinitions, and in our initial test scans they were given arbitrarily chosen lower bounds of $e^{-8.5}$. These ranges were further restricted into those listed in \Cref{tab:parameter_bounds} based on the results of these tests. It was also apparent that the experimental constraints were drawing many of these parameters towards lower values, so scanning them with logarithmic priors was found to be advantageous.

\clearpage
\subsection{Constraints}
\label{constraints}

We employ a wide range of observables to constrain the models, consisting of SM masses, EW radiative corrections, Z boson decay ratios, and Higgs signal strengths. Bounds on the production of new heavy resonances from direct collider searches are also considered for realistic points found in the scans, though we chose not to include these constraints in the scans because of the increase in computing time they would require. We use software developed for Ref.~\cite{Niehoff:2015iaa} (and the subsequent modifications for Ref.~\cite{niehoff2017electroweak}) to calculate the observables and likelihood for each point, so we only give a brief overview of each constraint below and refer the reader to Section~3.1 of Ref.~\cite{Niehoff:2015iaa} for details on their implementation and further  discussion. Only those constraints applicable to our third-generation-partially-composite-only models are included. The experimental values have been updated for this work and are given in \Cref{tab:experimental_values}.

\begin{table}[h]
\begin{center}
\begin{tabular}{clll}
\toprule
\multicolumn{2}{c}{Observable} & \multicolumn{1}{c}{Value(s)} & Ref.\\[1.5pt]
\midrule
    {} & {      } $m_b$ & $ 4.18(4) \gev$ & \cite{PhysRevD.98.030001}\\[1.5pt]
    {} & {      } $m_t$ & $ (173.0(4) \pm 1.0^{\text{theory}}) \gev$ & \cite{PhysRevD.98.030001, Hoang:2014oea}\\[1.5pt]
    {} & {      } $m_H$  & $ 125.18(16) \gev$ & \cite{PhysRevD.98.030001}\\[1.5pt]
\midrule
    {} & {      } $S$ & $0.02(10)$ & \cite{PhysRevD.98.030001}\\[1.5pt]
    {} & {      } $T$ & $0.07(12)$ & \cite{PhysRevD.98.030001}\\[1.5pt]
\midrule
    {} & {      } $R_e$ & $20.804(50)$ & \cite{ALEPH:2005ab} \\[1.5pt]
    {} & {      } $R_\mu$ & $20.785(33)$ & \cite{ALEPH:2005ab} \\[1.5pt]
    {} & {      } $R_\tau$ & $20.764(45)$ & \cite{ALEPH:2005ab} \\[1.5pt]
    {} & {      } $R_b$ & $0.21629(66)$ & \cite{ALEPH:2005ab} \\[1.5pt]
\midrule
    {} & {      } $\mu^{gg}_{\tau \tau}$ & $1.0(6),\ \quad 1.05(50),\ 0.97(56)$ & \cite{Khachatryan:2016vau, Sirunyan:2018koj, ATLAS-CONF-2018-031} \\[1.5pt]
    {} & {      } $\mu^{gg}_{WW}$ & $0.84(17),\ 1.35(20),\ 1.20(20)$ & \cite{Khachatryan:2016vau, Sirunyan:2018koj, ATLAS-CONF-2018-031} \\[1.5pt]
    {} & {      } $\mu^{gg}_{ZZ}$ & $1.13(33),\ 1.22(22),\ 1.03(16)$ & \cite{Khachatryan:2016vau, Sirunyan:2018koj, ATLAS-CONF-2018-031} \\[1.5pt]
    {} & {      } $\mu^{gg}_{\gamma \gamma}$ & $1.10(23),\ 1.15(15),\ 0.97(15)$ & \cite{Khachatryan:2016vau, CMS:1900lgv, ATLAS-CONF-2018-031} \\[1.5pt]
\bottomrule
\end{tabular}
\caption{Experimental values used for constraints in our scans. The observables are grouped (in order) into SM masses, oblique parameters, Z decay ratios, and Higgs signal strengths. The latter employ multiple independent measurements. Note the values for $S$ and $T$ have a correlation coefficient of $+0.92$ that is taken into account in the $\chi^2$ calculation.}
\label{tab:experimental_values}
\end{center}
\end{table}

\subsubsection*{SM masses}

All of the masses of SM particles that are predictions of the theory - namely, the top and bottom quark masses and the Higgs mass - are used as constraints. Once the Higgs mass is found by the method outlined in \Cref{Higgs_potential_section}, the other masses are found at tree level by diagonalising the mass matrices in \Cref{appendix_mass_matrices} at the point $h=\langle h \rangle$. The top and bottom quarks are identified as the third lightest up- and down-type particles in the theory.

\subsubsection*{Oblique parameters}

Important constraints for CHMs come from EW precision observables, which restrict the non-linear dynamics of the pNGB Higgs and place lower bounds on composite vector resonance masses. Such observables are conveniently parameterised by the Peskin-Takeuchi $S$ and $T$ parameters \cite{Peskin:1991sw, Barbieri:2004qk}, which we take as constraints.
The $S$ parameter is calculated at tree level and mostly constrains the vector resonance masses.
The $T$ parameter arises at one-loop and is given in terms of the vacuum polarisation of the electroweak bosons.
We consider only the dominant contributions from fermion loops, including in particular the effects of composite fermion resonances.
Absolute theoretical uncertainties of $0.05$ and $0.10$ are assigned to the respective theory predictions of $S$ and $T$ and they are assumed to be uncorrelated in contrast to the experimental uncertainties.

\subsubsection*{Z decays}

The compositeness of the third generation quarks modifies the $Z$ boson couplings, and in particular $Z b_L \bar{b}_L$.
These couplings are tightly constrained by the $Z$ boson decay widths and we consider their ratios
\begin{align}
    R_b := \frac{\Gamma(Z \rightarrow b \bar{b})}{\Gamma_{\text{had}}}, \qquad R_\ell := \frac{\Gamma_{\text{had}}}{\Gamma(Z \rightarrow \ell \bar{\ell})},
\end{align}
(for $\ell = e, \mu, \tau$), where
\begin{align}
\Gamma_{\text{had}} = \sum\limits_{q = u,d,c,s,b} \Gamma(Z \rightarrow q \bar{q})
\end{align}
is the total Z hadronic width.

\subsubsection*{Higgs signal strengths}

We use Higgs signal strengths from gluon fusion production to further constrain the non-linear dynamics of the pNGB Higgs. The signal strength $\mu^{gg}_X$ for the decay into a final state $X$ is defined as the ratio of the measured cross section to the predicted SM value:
\begin{align}
    \mu^{gg}_X := \frac{\sqbrackets{\sigma (gg \rightarrow h) \text{BR} (h \rightarrow X)}_{\text{exp}}}{\sqbrackets{\sigma (gg \rightarrow h) \text{BR} (h \rightarrow X)}_{|\text{SM}}}.
\end{align}
We include as observables $\mu^{gg}_X$ for $X = \tau \tau$, $WW$, $ZZ$, and $\gamma \gamma$. The partial widths for decays into massive particles are calculated at tree level, and those for decays into massless particles are calculated at one-loop order including all fermionic and bosonic contributions. For each of these observables we use three independent measurements: one using combined ATLAS and CMS data from Run 1 of the LHC, and the others separately from ATLAS and CMS using Run 2 data. The effects of these constraints on the parameter spaces of M4DCHMs and similar models have recently been analysed in detail in Refs.~\cite{Banerjee:2017wmg,Banerjee:2020tqc}.

\subsubsection*{Collider searches}

Searches for heavy resonances in colliders place useful upper bounds on the production cross section times branching ratio of composite resonances for various decay modes. As mentioned above, we do not include these as constraints in the scans themselves, and only analyse them for realistic points found in the scans after convergence. All of the experimental analyses we use are featured in \Cref{tab:V_l,tab:V_h,tab:V_b,tab:F_1,tab:F_2}, in \Cref{collider_search_appendix}. This builds on the list of analyses used in Ref.~\cite{niehoff2017electroweak} with an additional 40 LHC searches at $\sqrt{s} = 13\tev$.

Note that since the searches do not provide measured values and uncertainties but upper bounds, there is a slight departure from \Cref{eq:likelihood_chi2_definition} in calculating the $\chi^{2}$ contribution of each decay.
In this case, the 95\% CL upper bound on the cross section times the branching ratio, denoted by $X_{\rm bound}$, is taken into account in such a way that a theory prediction $X_{\rm theo}$ of the same value should correspond to a $\chi^2=1.96^2$, i.e. the $\chi^2$ contribution of a particle with $X_{\rm theo}$ in a given decay channel is
\begin{align}\label{eq:chi2_coll_simpl}
    \chi^{2} = \frac{X_{\text{theo}}^2 (\mathbf{p}) }{\sigma^2_X},
\end{align}
where $\sigma_X=X_{\rm bound}/1.96$.
While this is applicable for the case that the expected bound $X_{\rm exp}$ equals the observed bound $X_{\rm obs}$, we correct this relation in the case that both are different by using $\sigma_X=X_{\rm exp}/1.96$ and defining
\begin{align}\label{eq:chi2_coll_full}
    \chi^{2} = \frac{(X_{\text{theo}} (\mathbf{p}) - x)^2 - x^2}{\sigma^2_X},
\end{align}
where $x = X_{\rm obs} - \kappa(X_{\rm obs}/X_{\rm exp})\,X_{\rm exp}$. The function $\kappa$ corrects the $\chi^{2}$ contribution if $X_{\rm obs}/X_{\rm exp}\neq1$ while for $X_{\rm obs}/X_{\rm exp}=1$, we have $\kappa(1)=1$, i.e.\ in this case we have $x=0$ and \Cref{eq:chi2_coll_full} reduces to \Cref{eq:chi2_coll_simpl}.

In calculating the $\chi^2$ for each decay, we only use the maximum $\chi^2$ contribution from all the analyses restricting the decay. While a tighter bound would be given by combining the contributions from independent analyses, we have found that the improvement in the excluded cross section at the $3\sigma$ level when doing this is often smaller than the error introduced by the narrow width approximation used in our calculations, and so implementing this would not be worthwhile.

\subsection*{Mass cuts}

As a final measure, we impose harsh penalties for any new fermionic resonances below $500\gev$ because such a resonance would likely have already been discovered. Ruling out points using hard mass cuts below $500\gev$ makes it prohibitively time consuming to find an initial population of viable points, so instead we assign a steep one-sided Gaussian likelihood to the fermionic masses. This makes initial populations of points easier to find, and they will evolve towards points that give more reasonable masses as the scans proceed.

\section{Results}
\label{results}

The results of our global fits of the M4DCHM$^{5-5-5}$, the M4DCHM$^{14-14-10}$, and the M4DCHM$^{14-1-10}$ are presented below in \Cref{5-5-5_results_section,14-14-10_results_section,14-1-10_results_section}, and are collectively discussed in \Cref{model_comparison_section}. As a by-product of the fits, we analyse the experimental signatures of the three models in \Cref{experimental_signatures_section} in the hope of guiding the search for evidence of these models at the LHC.

For each of the models, we performed multiple scans to verify the reproducibility of the results. This was a lengthy process, with each of our production-ready scans taking at least one week to converge running on $64$ cores, or equivalent. The levels of agreement between the scans are shown in \Cref{scan_agreement_appendix}.
Broadly, results for the M4DCHM$^{14-14-10}$ and the M4DCHM$^{14-1-10}$ are highly reproducible, while there is somewhat less agreement in the results for the M4DCHM$^{5-5-5}$. We analyse the combined samples from all scans for each model\footnote{Nested sampling is trivially parallelisable, meaning that any collection of nested sampling runs can be combined by simply merging their samples and ordering them according to their likelihoods, and the result is equivalent to a single run that uses a number of live points equal to the sum of the live points from each individual run.} using \texttt{anesthetic} \cite{anesthetic} in order to maximise the robustness of our conclusions. In the following discussion we will focus on only those features of the results that are seen across multiple scans.

\subsection{M4DCHM$^{5-5-5}$}
\label{5-5-5_results_section}

\begin{figure}
\centering
  \includegraphics[width=1\linewidth]{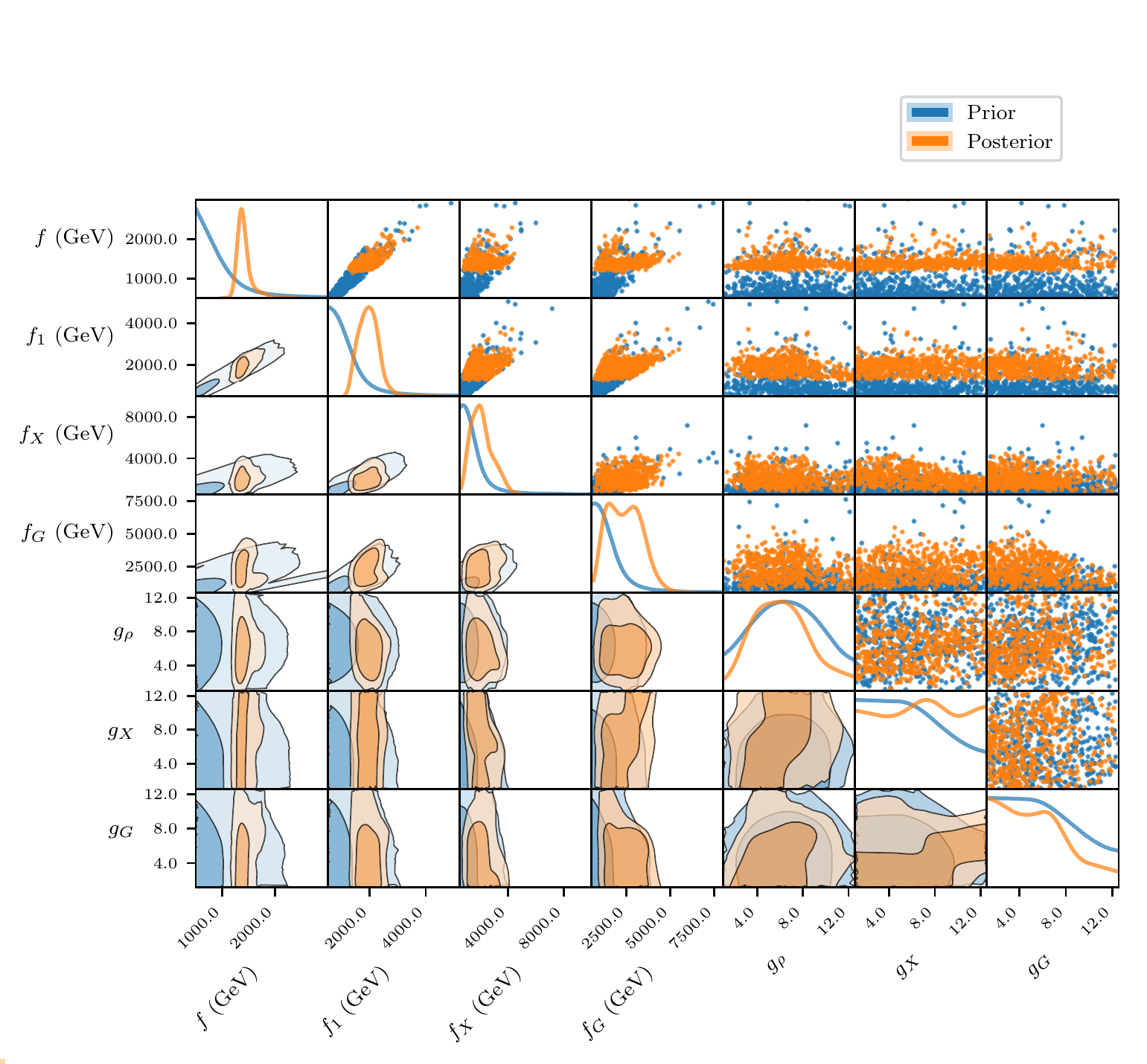}
\caption{1D and 2D marginalised priors and posteriors for the gauge sector parameters in the M4DCHM$^{5-5-5}$.}
\label{fig:5-5-5_gauge_posterior_prior}
\end{figure}

\begin{figure}
\centering
  \includegraphics[width=1\linewidth]{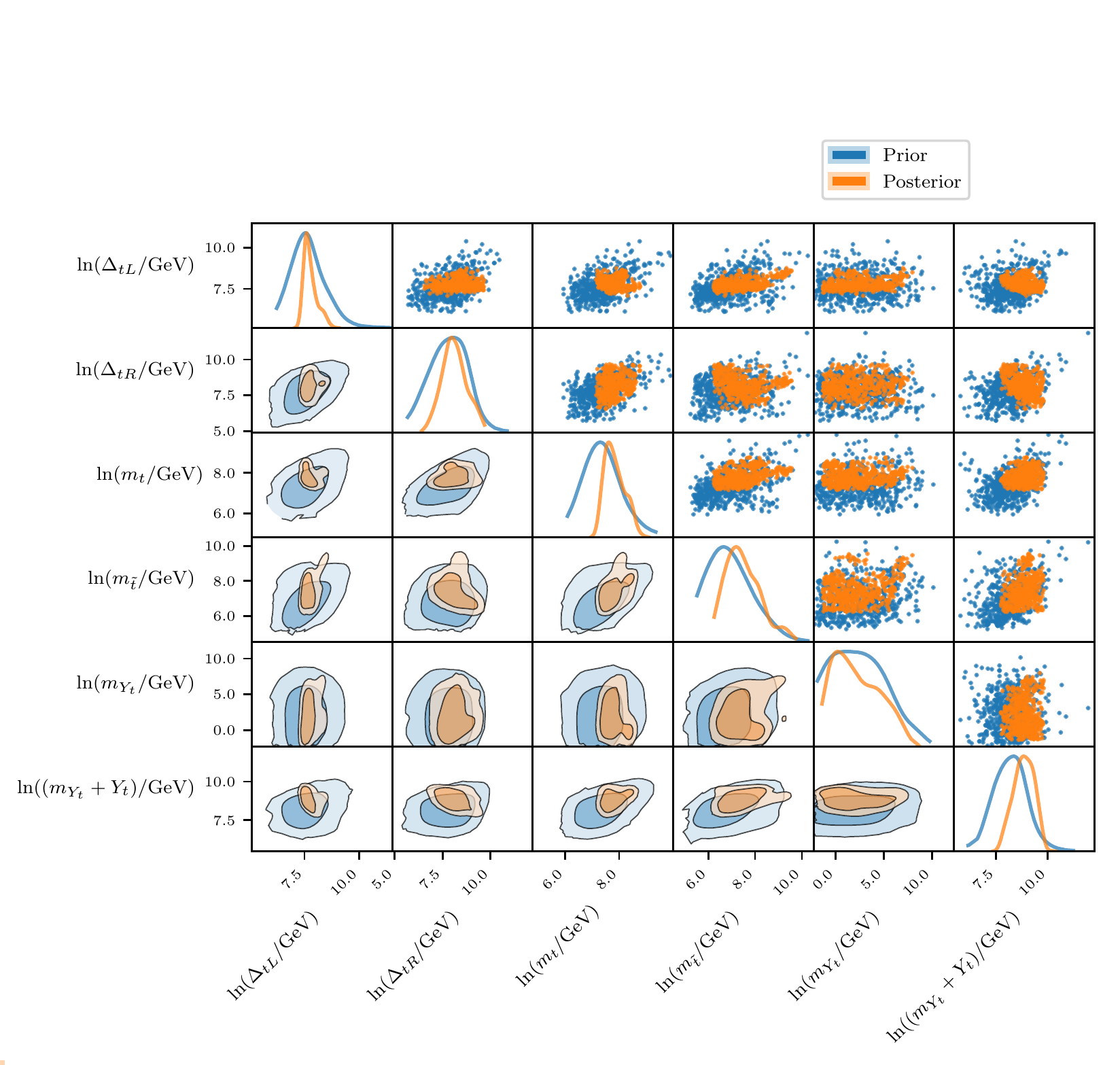}
\caption{1D and 2D marginalised priors and posteriors for the top partner parameters in the M4DCHM$^{5-5-5}$.}
\label{fig:5-5-5_T_posterior_prior}
\end{figure}

\begin{figure}
\centering
  \includegraphics[width=1\linewidth]{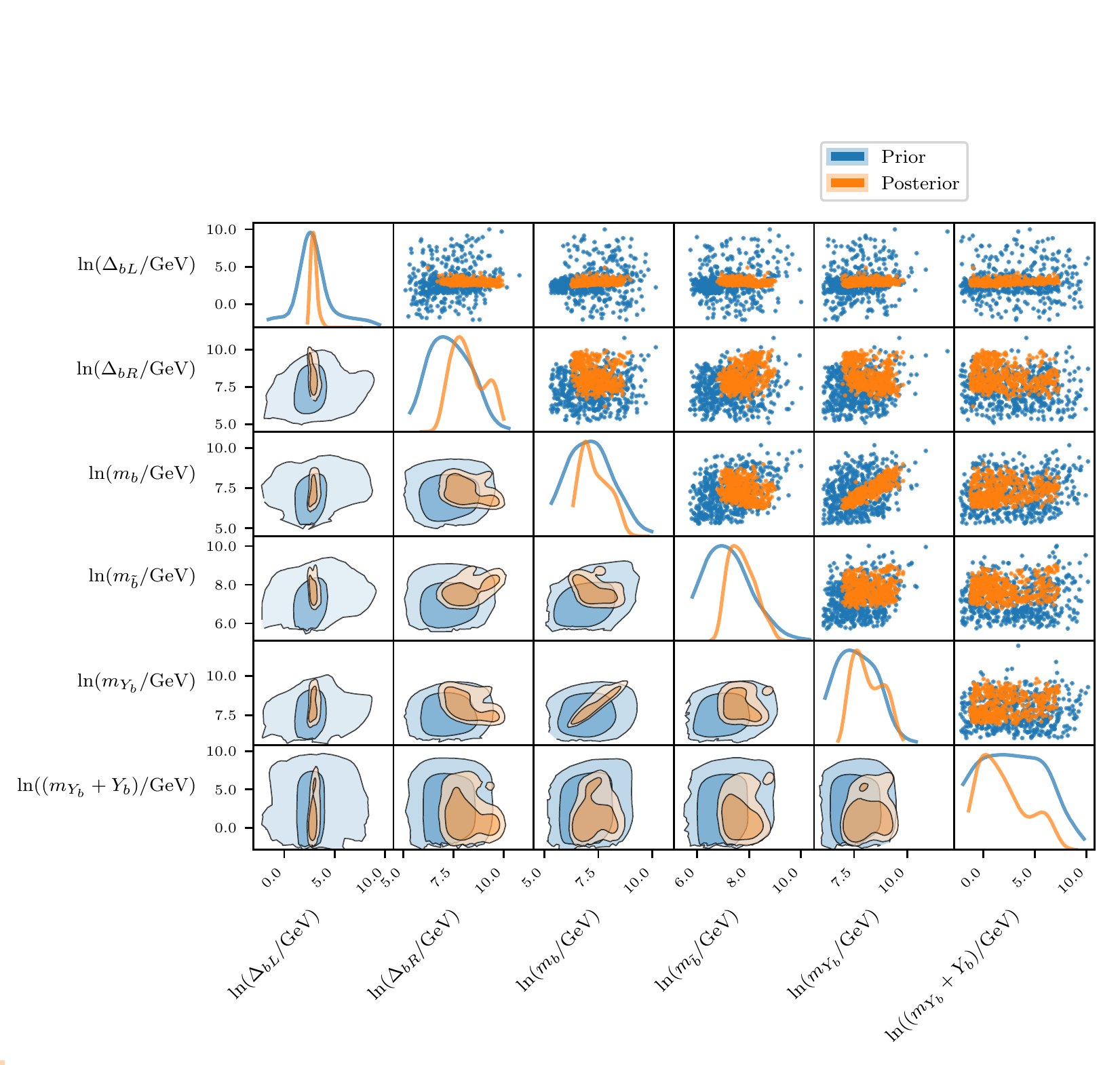}
\caption{1D and 2D marginalised priors and posteriors for the bottom partner parameters in the M4DCHM$^{5-5-5}$.}
\label{fig:5-5-5_B_posterior_prior}
\end{figure}

Our main results for this model, the marginalised prior and posterior distributions of the parameters from the combined nested sampling runs, are shown in \Cref{fig:5-5-5_gauge_posterior_prior,fig:5-5-5_T_posterior_prior,fig:5-5-5_B_posterior_prior}. In the type of figure shown here, the diagonal plots show the 1D marginalised priors and posteriors of the parameters, with the 2D marginalised distributions of the various parameter pairings filling the off-diagonal plots. The upper-right plots show samples drawn directly from the prior and posterior, while the lower-left plots are kernel density estimates of these distributions, with iso-likelihood contours containing  $66$\% and $95$\% of the probability mass.

We look first at the results for the gauge parameters, in \Cref{fig:5-5-5_gauge_posterior_prior}. There is a clear pattern for the NGB decay constants: their priors all favour lower values of the constants, successfully encoding the notion that smaller decay constants are more natural (\textit{cf}. discussion in \Cref{fermion_sector}), while their posteriors occupy the higher values, indicating the smaller decay constants are experimentally disfavoured as is already known from precision EW tests \cite{agashe2005}. The posterior for $f$, for example, lies mostly between $1.2$~TeV and $1.9$~TeV, with a maximum at ${\sim}1.4$~TeV. The upper bound here comes from suppression from the prior, while the lower bound is due to a variety of constraints; namely, the oblique constraints, the Higgs signal strength constraints, and the Z decay constraints. It is not \textit{im}possible for such constraints to be individually well-satisfied at low $f$; it is simply that many points in this range do not satisfy the constraints well, and this large low-likelihood region suppresses the posterior over this range because of the marginalisation process\footnote{It is also possible that satisfying one constraint well necessitates a violation in another constraint, which would lead to a large \textit{total} $\chi^{2}$ and a direct suppression of the posterior from the likelihood, rather than from a volume effect.}.

There does not seem to be any structure in the posteriors of the other decay constants in \Cref{fig:5-5-5_gauge_posterior_prior} beyond the consistency conditions outlined in \Cref{scan_parameters}. That is, the $SO(5)$ decay constant $f_{1}$ has no particular preference for any values between its theoretical bounds of $f$ and $\sqrt{3}f$, having a posterior peaking around $2$~TeV, and likewise for the $U(1)_{X}$ and $SU(3)_{c}$ decay constants $f_{X}$ and $f_{G}$, which have similar posterior distributions between ${\sim}1$~TeV and ${\sim}4$~TeV, though the results for $f_{G}$ are not particularly consistent. These parameters are constrained by the same factors as $f$ mentioned above, and this is true across all models.

As for the gauge couplings, the posterior for the $SO(5)^{1}$ coupling $g_{\rho}$ extends across the entire range $[1,4\pi]$ with a preference for values between ${\sim}2$ and ${\sim}8$, driven largely by the prior with some additional experimental factors at play, while no definitive results were found for the $U(1)^{1}_{X}$ and $SU(3)^{1}_{3}$ couplings $g_{X}$ and $g_{G}$. The Higgs signal strength, SM mass, and oblique constraints all seem to slightly disfavour values for $g_{\rho}$ at the extreme ends of the range, while the Z decay constraint $R_{b}$ disfavours values below ${\sim}2$, in opposition to $R_{e}$ and $R_{\mu}$, which favour those values. It is unsurprising that there is little structure in the posteriors of $f_{G}$ and $g_{G}$ since we do not include any constraints on heavy gluon decays or flavour physics in our scans.

For the top sector parameters in \Cref{fig:5-5-5_T_posterior_prior}, the posteriors are significantly localised in comparison to the priors, demonstrating the effects of the constraints on the space. However, the posteriors all peak around the same regions as the priors, so it seems that satisfying the constraints does not require excessive fine-tuning. Indeed, the elementary-composite mixing and composite mass parameters all tend to be constrained around $\Delta_{tL} \sim \Delta_{tR} \sim m_{t} \sim m_{\tilde{t}} \sim 1.8$~TeV, which, from the findings of Ref.~\cite{panico2012}, indicates that double tuning is not so much of an issue. All of these parameters are primarily constrained by the SM masses, except for the lower bound of $0.5$~TeV on $m_{\tilde{t}}$ that comes from the resonance mass cutoff\footnote{We could not impose this bound at the prior level because we were imposing the prior on the dimensionless variable $m_{\tilde{t}}/f$ for an unspecified scale $f$.}. Not particularly well constrained is the off-diagonal mass $m_{Y_{t}}$, which may take any value below ${\sim}2$~TeV. The last top sector parameter, $m_{Y_{t}} + Y_{t}$, clearly prefers larger values from around $2$~TeV to $20$~TeV, and this is due largely to the SM mass and oblique constraints.

Finally, we discuss the bottom sector parameters in \Cref{fig:5-5-5_B_posterior_prior}. There is not as much uniformity among these as for the top parameters, presumably because these play a subdominant role in the Higgs potential. Interestingly, $\Delta_{bL}$ is very narrowly constrained, between around $10$~GeV and $30$~GeV. Much of this is due to the prior, which suppresses values above ${\sim}150$~GeV, but again there are other factors present. The lower bound on its posterior is easily identified as coming from the bottom quark mass constraint, while the upper bound comes more subtly from a large region of points with $\Delta_{bL} \gtrsim 50$~GeV that significantly violate the Z decay constraints. The right-handed bottom quark coupling $\Delta_{bR}$, on the other hand, is quite large, ranging between roughly $1$~TeV and $10$~TeV, with the lower bound coming from the bottom quark and Higgs mass constraints. The bottom mass constrains $m_{\tilde{b}}$ to lie above ${\sim}1$~TeV, and the prior below ${\sim}5$~TeV. The final parameter, $m_{Y_{b}} + Y_{b}$, lies mostly below $2$~TeV, with a preference towards the lower values stemming from the SM mass and Z decay constraints. Results for $m_{b}$ and $m_{Y_{b}}$ are not particularly consistent, but both generally occupy large values, with the former ranging up to ${\sim}4.5$~TeV, and the latter up to ${\sim}10$~TeV.

One interesting feature to note here is that the posteriors of $m_{b}$ and $m_{Y_{b}}$ display a strong correlation that is not present in the prior, and therefore must be a result of one or more of the experimental constraints, though it is difficult to pinpoint the exact cause. Other structures to note in \Cref{fig:5-5-5_T_posterior_prior,fig:5-5-5_B_posterior_prior} are slight correlations in the priors and posteriors of the pairs $(m_{t},\ m_{Y_{t}} + Y_{t}$) and $(m_{t},\ m_{\tilde{t}}$), a mild correlation in the posteriors of $(\Delta_{bR},\ m_{\tilde{b}}$), and posteriors that seem to avoid small values for both parameters in the pairs $(\Delta_{tL},\ m_{t}$), $(\Delta_{tL},\ m_{Y_{t}} + Y_{t}$), and $(\Delta_{bR},\ m_{Y_{b}}$). More will be said about these results when comparing models in \Cref{model_comparison_section}.

\subsection{M4DCHM$^{14-14-10}$}
\label{14-14-10_results_section}

\begin{figure}
\centering
  \includegraphics[width=1\linewidth]{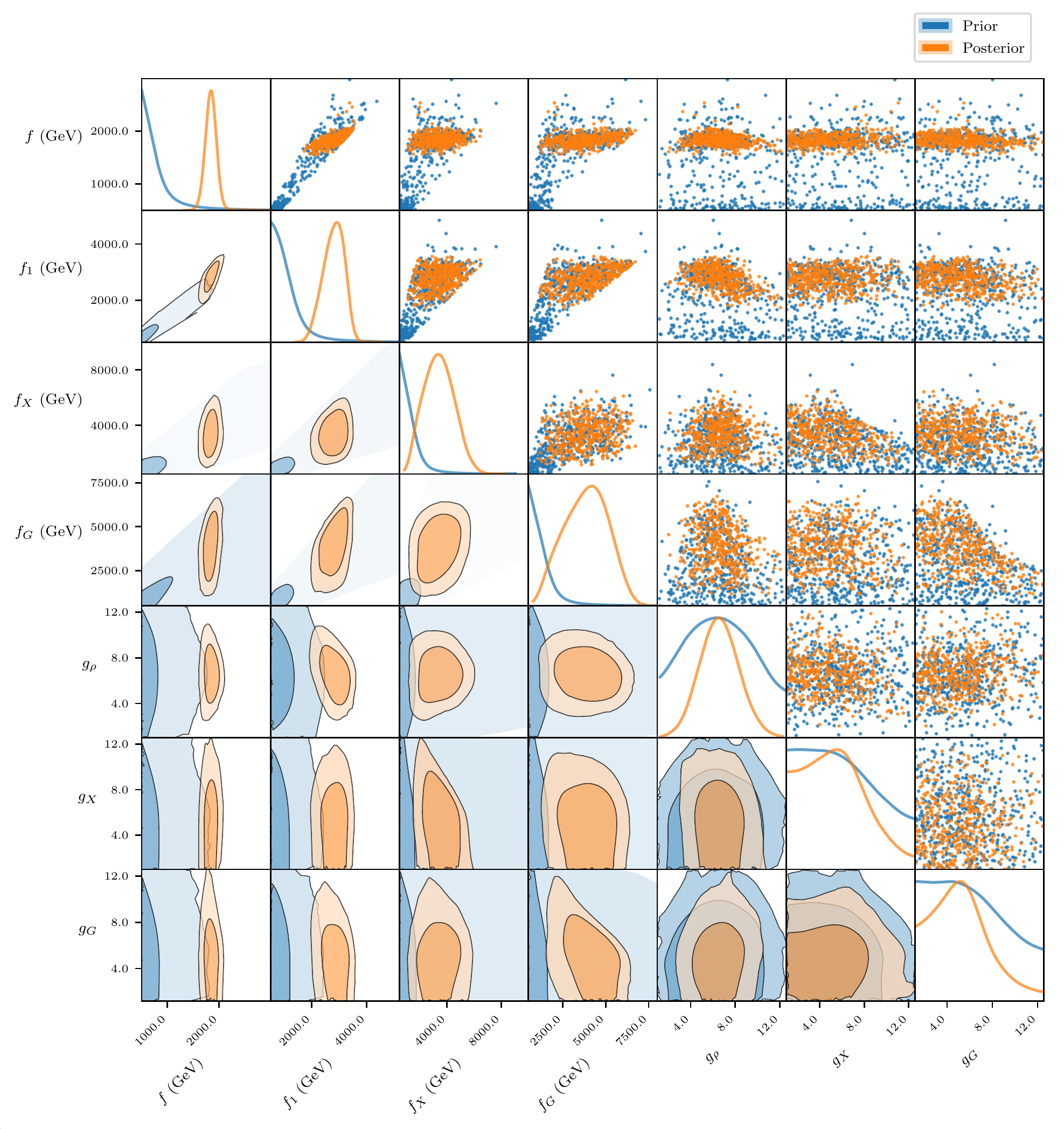}
\caption{1D and 2D marginalised priors and posteriors for the gauge sector parameters in the M4DCHM$^{14-14-10}$.}
\label{fig:14-14-10_gauge_posterior_prior}
\end{figure}

\begin{figure}
\centering
  \includegraphics[width=1\linewidth]{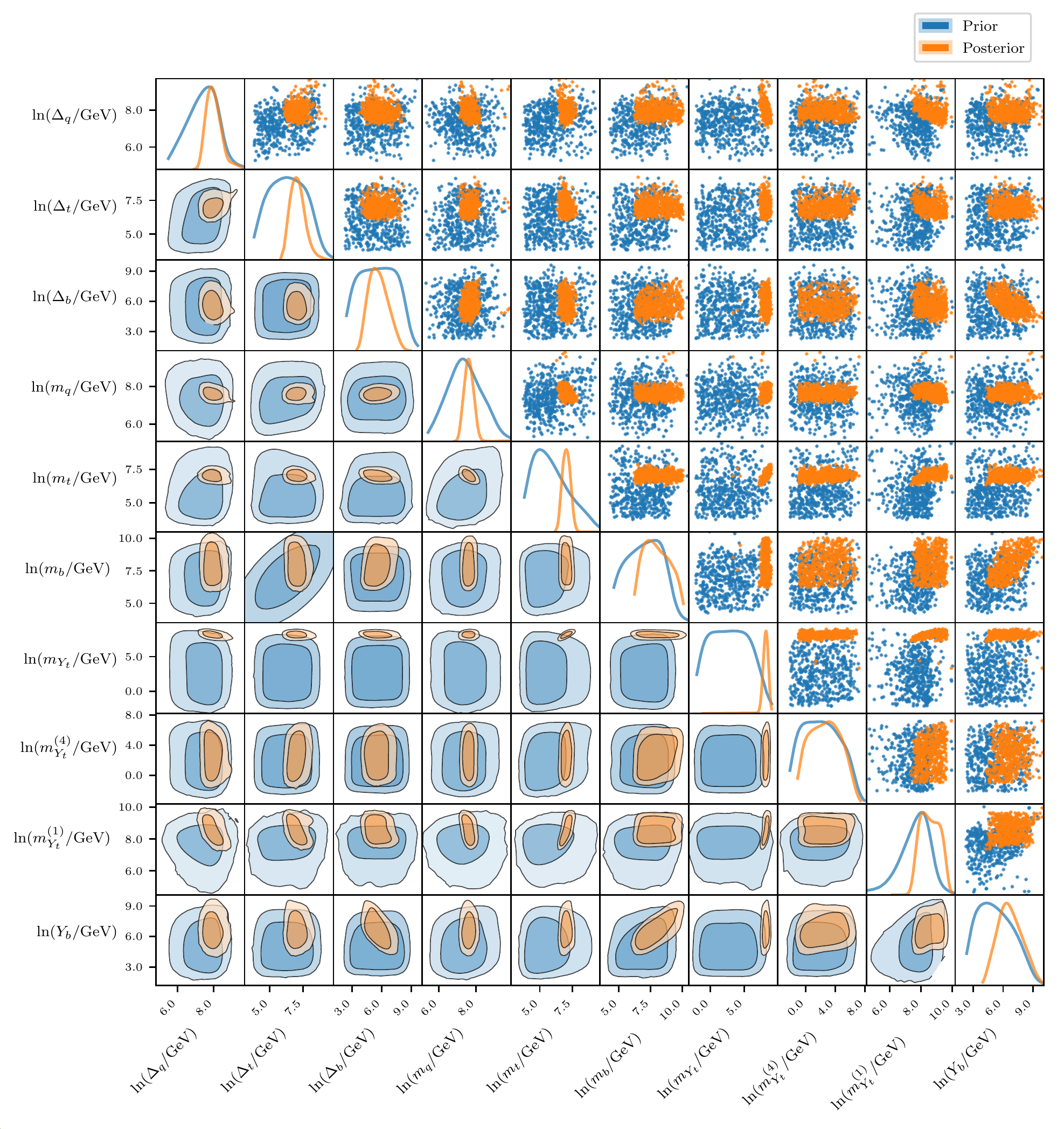}
\caption{1D and 2D marginalised priors and posteriors for the fermion sector parameters in the M4DCHM$^{14-14-10}$. The parameters $m^{(4)}_{Y_{t}}$ and $m^{(1)}_{Y_{t}}$ are defined by \Cref{eqn:mYt4,eqn:mYt1}.}
\label{fig:14-14-10_fermion_posterior_prior}
\end{figure}

Results for this model are given in \Cref{fig:14-14-10_gauge_posterior_prior,fig:14-14-10_fermion_posterior_prior}. The priors and posteriors for the gauge sector parameters, in \Cref{fig:14-14-10_gauge_posterior_prior}, look much the same as in the M4DCHM$^{5-5-5}$, with the NGB decay constant posteriors concentrated well above the small values preferred by the prior. The posterior for $f$ enjoys a slightly higher range than before, from around $1.5$~TeV to $2.2$~TeV. Tending towards the higher end of its range is $f_{1}$, spanning between ${\sim}1.5$~TeV and ${\sim}3.8$~TeV, while the remaining decay constants are constrained as $1.2\tev \lesssim f_{X} \lesssim 7.5\tev$ and $1.1\tev \lesssim f_{G} \lesssim 6.5\tev$. The gauge coupling $g_{\rho}$ has a similar distribution to before, preferring mid-range values, again seemingly driven by the prior in combination with the Higgs signal strength and Z decay constraints. As before, $g_{X}$ and $g_{G}$ were not sufficiently well constrained to give consistent results. However, it seems that stronger $g_{G}$ couplings are disfavoured as they appear to entail undesirably small values of $f_{G}$.

The M4DCHM$^{14-14-10}$ fermion parameters are seen from \Cref{fig:14-14-10_fermion_posterior_prior} to be significantly more constrained than those in the M4DCHM$^{5-5-5}$, with posteriors concentrated on considerably smaller regions in comparison to the priors - a feature emblematic of higher fine-tuning. This is particularly evident for the on-diagonal mass parameter $m_{t}$, whose prior allows it to take effectively any value up to $10$~TeV but whose posterior is localised between ${\sim}0.6$~TeV and ${\sim}2$~TeV. Mass constraints are the main factor here, with the Z decay and Higgs signal strength constraints also contributing to the upper bound somewhat, and the same is true for all of the other on-diagonal mass and mixing parameters as well. The left-handed quark partner couplings are constrained as $1.3\tev \lesssim \Delta_{q} \lesssim 8.5\tev$ and $1.1\tev \lesssim m_{q} \lesssim 3.0\tev$, while those couplings for the right-handed quarks have ranges $0.5\tev \lesssim \Delta_{t} \lesssim 4.4\tev$ and $50\gev \lesssim \Delta_{b} \lesssim 2.5\tev$. The final mass parameter, $m_{b}$, can range up to ${\sim}22$~TeV. Evidently, the right-handed bottom quark must possess a small compositeness $\Delta_{b}/m_{b}$ in this model, as is to be reasonably expected from its relatively small mass.

Despite its prior, $m_{Y_{t}}$ tends towards large values with a posterior between ${\sim}1.1$~TeV and ${\sim}6.6$~TeV, while
\begin{align}
    m_{Y_{t}}^{(4)} := m_{Y_{t}} + \frac{1}{2}Y_{t}
\label{eqn:mYt4}
\end{align}
has a posterior constrained to lower values $\lesssim 0.8$~TeV. In other words, it must be that $Y_{t} \approx -2 m_{Y_{t}}$. It is not clear exactly which experiments are to blame, but it seems that the posterior is suppressed at smaller values of $m_{Y_{t}}$ and at greater values of $m_{Y_{t}}^{(4)}$ by large low-likelihood regions stemming from the Higgs signal strength constraints. Z decays also slightly favour the smaller values of $m_{Y_{t}}^{(4)}$. Meanwhile, both the prior and posterior for
\begin{align}
    m^{(1)}_{Y_{t}} := m_{Y_{t}} + 4Y_{t}/5 + 4\tilde{Y}_{t}/5
\label{eqn:mYt1}
\end{align}
favour large values, from around $1.7$~TeV to $16$~TeV, with smaller values again perhaps being suppressed in the posterior due to a volume effect from the Higgs signal strengths. From the previous observations, this implies $\tilde{Y}_{t}$ must be greater than around $1.4$~TeV. Finally, SM mass constraints force $Y_{b}$ to be larger than $0.1$~TeV, even though smaller values are favoured by the prior.

There are some interesting correlations among these fermion parameters - specifically, strong positive correlations between $m_{b}$ and $Y_{b}$, and between $m_{t}$ and $m_{Y_{t}}$, a negative correlation between $\Delta_{b}$ and $Y_{b}$, and mild positive correlations between $m_{t}$ and $m^{(1)}_{Y_{t}}$, and between $m_{Y_{t}}$ and $m^{(1)}_{Y_{t}}$.

\subsection{M4DCHM$^{14-1-10}$}
\label{14-1-10_results_section}

\begin{figure}
\centering
  \includegraphics[width=1\linewidth]{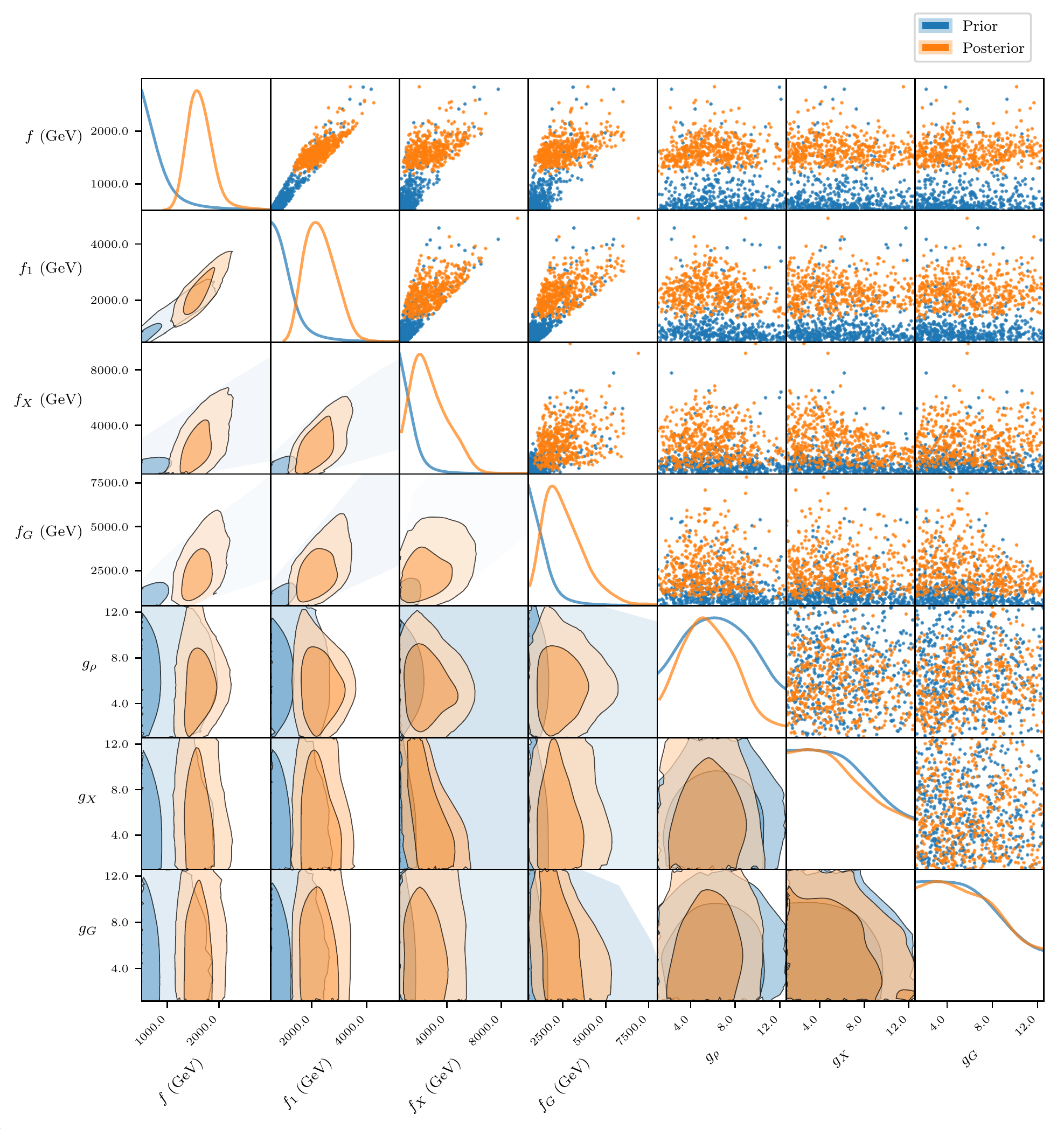}
\caption{1D and 2D marginalised priors and posteriors for the gauge sector parameters in the M4DCHM$^{14-1-10}$.}
\label{fig:14-1-10_gauge_posterior_prior}
\end{figure}

\begin{figure}
\centering
  \includegraphics[width=1\linewidth]{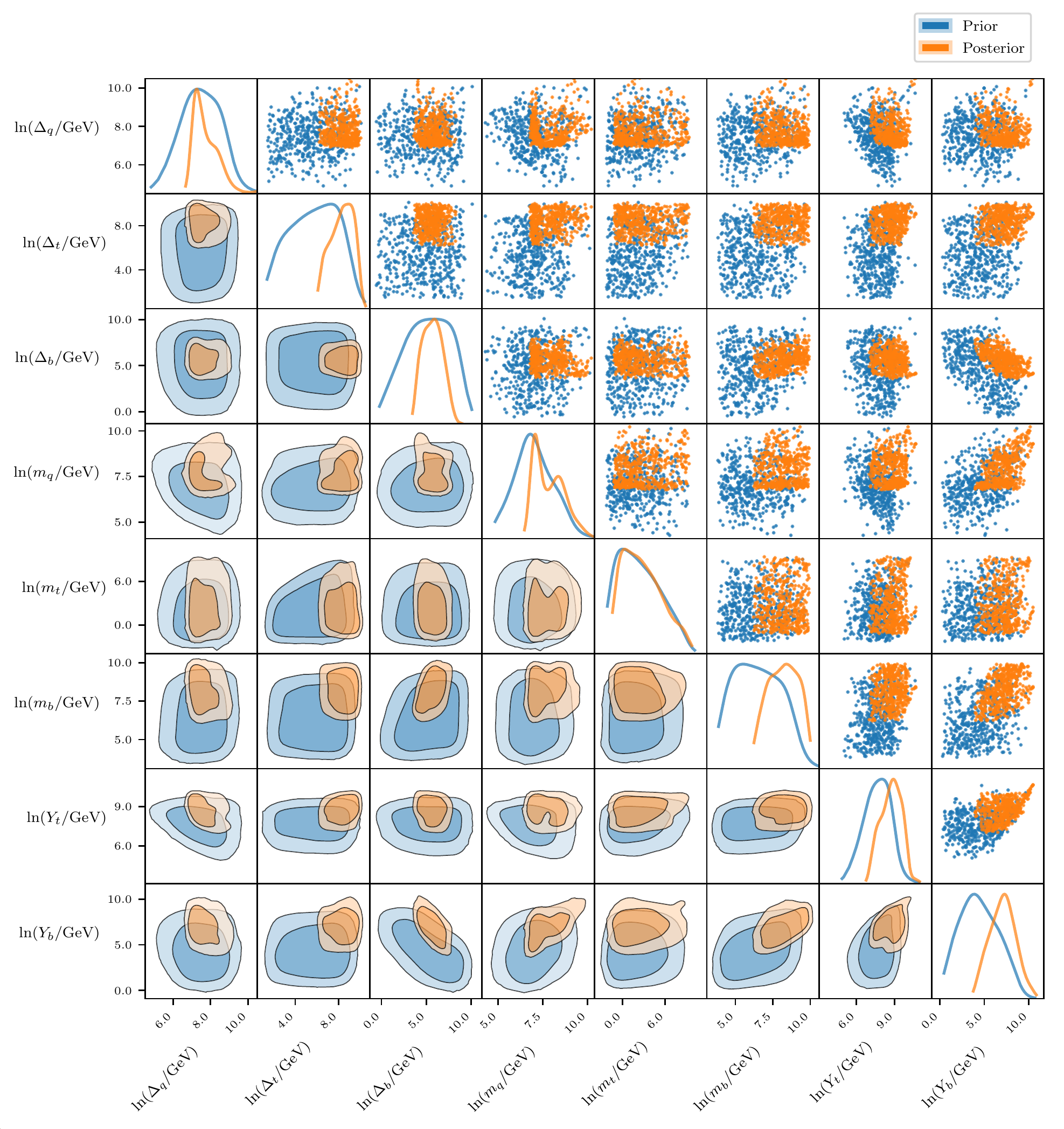}
\caption{1D and 2D marginalised priors and posteriors for the fermion sector parameters in the M4DCHM$^{14-1-10}$.}
\label{fig:14-1-10_fermion_posterior_prior}
\end{figure}

Our results for the M4DCHM$^{14-1-10}$ are given in \Cref{fig:14-1-10_gauge_posterior_prior,fig:14-1-10_fermion_posterior_prior}. There is little qualitative difference between the priors and posteriors for the gauge sector couplings in this model and those in the previous models. Here the posterior for $f$ is mostly contained between ${\sim}1.3$~TeV and ${\sim}1.8$~TeV, though it possesses a tail reaching upwards of about $2.2$~TeV. Like in the M4DCHM$^{14-14-10}$, $f_{1}$ tends to favour the upper end of its range, spanning between roughly $1.3$~TeV and $3.4$~TeV. The $U(1)_{X}$ and $SU(3)_{c}$ decay constants both extend down to their lower bounds of $0.5$~TeV and range up to around $5.5$~TeV, both with an affinity for the lower values. The gauge coupling $g_{\rho}$ behaves in the same way as in the other models, both in its preference for values from ${\sim}2$ to ${\sim}8$, and in its relationships to the various constraints. However, the same is not true for the other gauge couplings. Surprisingly, in this model the posteriors for $g_{X}$ and $g_{G}$ almost exactly align with the priors, and indeed there does not seem to be any significant dependence on these parameters for any of the observables.

In contrast to the gauge parameters, there is a clear difference in the posteriors of the fermion parameters between the M4DCHM$^{14-1-10}$ and the M4DCHM$^{14-14-10}$: those in the M4DCHM$^{14-1-10}$, shown in \Cref{fig:14-1-10_fermion_posterior_prior}, cover a markedly larger fraction of volume of the total parameter space, so these parameters require less fine-tuning to satisfy the constraints than those in the M4DCHM$^{14-14-10}$. This is taken to the extreme by the mass parameter $m_{t}$ being almost entirely unconstrained in this model: it is able to take values up to ${\sim}10$~TeV but with great preference towards values $\lesssim 0.5$~TeV purely stemming from the prior. Other parameters that require little fine-tuning, having posteriors well within the regions favoured by their priors, are $\Delta_{q}$, $\Delta_{b}$, and $m_{q}$. The posteriors of the first two of these respectively span from ${\sim}0.9$~TeV to ${\sim}5.0$~TeV, and from ${\sim}70$~GeV to ${\sim}1.8$~TeV, again reflecting the right-handed bottom quark being mostly elementary. The last of these, $m_{q}$, along with $m_{b}$, has a hard lower bound of $0.5$~TeV from the resonance mass cutoffs, with the former ranging upwards of $8$~TeV and the latter being drawn up to ${\sim}21$~TeV by the constraints. If the LHC constraints were included in this fit, we might expect these lower bounds to be increased and larger parameter values to be favoured. Unlike the other models, here the right-handed top quark can mix very strongly with the composite sector, with the posterior for $\Delta_{t}$ going between ${\sim}0.5$~TeV and ${\sim}17$~TeV, being maximised at ${\sim}7.5$~TeV despite the prior giving preferable weight to lower values. Similarly, the proto-Yukawa couplings $Y_{t}$ and $Y_{b}$ are drawn to unnaturally large values by the constraints, with posteriors ranging from ${\sim}1.6$~TeV to ${\sim}22$~TeV, and from ${\sim}50$~GeV to ${\sim}63$~TeV respectively. The posteriors for all of these parameters are shaped primarily by the SM mass constraints, along with some contributions from the Z decay constraints that slightly disfavour very large values of the couplings.

There is a slight positive correlation between the posteriors of $m_{b}$ and $Y_{b}$, and a negative correlation in both the priors and posteriors of $\Delta_{b}$ and $Y_{b}$. There is also a notable effect wherein large values of $Y_{t}$ are favoured in both the prior and posterior when $Y_{b}$ is also large. The prior tends to disfavour both $\Delta_{q}$ and $Y_{t}$ having small values, and to a lesser extent the same is true for both $\Delta_{q}$ and $m_{q}$, and the posteriors for these pairs seem to be constrained to regions for which $\Delta_{q} Y_{t}$ and $\Delta_{q} m_{q}$ lie above some respective fixed values. Unfortunately, analysing the causes of these features would be too involved a process to be viable for us.

As a final note, our result that $\Delta_{t}$ is usually much greater than $m_{t}$ indicates that the right-handed top quark must be almost completely composite in this model given the constraints. It might therefore do well to consider a model with the same $\mathbf{14}-\mathbf{1}-\mathbf{10}$ symmetry structure that incorporates $t_{R}$ as an entirely composite state as a limiting case of the M4DCHM$^{14-1-10}$.
\widowpenalty = 10000

\subsection{Discussion and model comparisons}
\label{model_comparison_section}

Having seen the fit results for all of our models, we are now in a position to discuss the similarities and differences between the models, and compare them based on their fits. The Bayesian evidences $\mathcal{Z}$ calculated from the combined samples from the scans of each model are given in \Cref{tab:Bayesian_statistics}, along with other quantities of interest such as the KL divergence $D_{\text{KL}}$, the posterior-averaged log-likelihood, and the maximum likelihoods found in the scans.

\begin{table}
\begin{center}
\begin{tabular}{l|ccccc}
{} & $\ln(\mathcal{Z}$) & $D_{\text{KL}}$ & $ \langle \ln(\mathcal{L}) \rangle_{\mathcal{P}}$ & $\max \ln(\mathcal{L}$) & BMD \\[2pt]
\midrule
M4DCHM$^{5-5-5}$ & $-28.62 \pm 0.04$ & $15.71$ & $-12.92$ & $-8.62$ & $7.23$ \\[2pt]
M4DCHM$^{14-14-10}$ & $-37.72 \pm 0.05$ & $22.58$ & $-15.14$ & $-9.29$ & $12.15$ \\[2pt]
M4DCHM$^{14-1-10}$ & $-37.58 \pm 0.04$ & $15.83$ & $-21.75$ & $-16.66$ & $7.52$ \\[2pt]
\end{tabular}
\end{center}
\caption{Statistics from the combined Bayesian scans of each model, using the imposed priors.}
\label{tab:Bayesian_statistics}
\end{table}

\begin{table}
\begin{center}
\begin{tabular}{l|ccc}
{} & $\ln(\mathcal{Z}$) & $D_{\text{KL}}$ & $ \langle \ln(\mathcal{L}) \rangle_{\mathcal{P}}$\\[2pt]
\midrule
M4DCHM$^{5-5-5}$ & $-42.1$ & $28.0$ & $-14.1$ \\[2pt]
M4DCHM$^{14-14-10}$ & $-49.8$ & $33.8$ & $-16.0$ \\[2pt]
M4DCHM$^{14-1-10}$ & $-45.3$ & $23.3$ & $-22.1$ \\[2pt]
\end{tabular}
\end{center}
\caption{Statistics from the combined Bayesian scans of each model, re-weighted as if the priors on the fermion parameters were uniform.}
\label{tab:Bayesian_statistics_uniform}
\end{table}

The most obvious point to note from \Cref{tab:Bayesian_statistics} is that the M4DCHM$^{5-5-5}$ has by far the greatest evidence of the three models, being ${\sim}7800$ times greater than the evidence for the M4DCHM$^{14-1-10}$, itself ${\sim}15$\% greater than that for the M4DCHM$^{14-14-10}$. Granted we assign an equal prior likelihood to each of our models, the M4DCHM$^{5-5-5}$ would then be decisively preferred over the others, while there is no conclusive preference between the M4DCHM$^{14-1-10}$ and the M4DCHM$^{14-14-10}$. But note that these evidences are prior-dependent; the choice of the prior distribution as well as the lower and upper bounds of the prior parameter ranges affect these results. A benefit of our sampling method is that it is possible to quantify the degree to which the evidences and other quantities in \Cref{tab:Bayesian_statistics} are dependent on the prior distribution (but not parameter bounds) by re-weighting our samples as if a different prior distribution was chosen. Uniform priors on the fermion parameters, for example, lead to \Cref{tab:Bayesian_statistics_uniform}, where it is seen the evidences are considerably different in comparison with the logarithmic priors. While in this case the M4DCHM$^{5-5-5}$ still seems to be superior compared to the other models, this, however, also depends on the parameter bounds that have been determined in the initial test scans and which are listed in \Cref{tab:parameter_bounds}.
In particular, for the left-handed composite-elementary mixings $\Delta_{t_L}$ and $\Delta_{b_L}$ in the M4DCHM$^{5-5-5}$, we find that the bounds on $\Delta_{t_L}$ are larger than those on $\Delta_{b_L}$ by a factor of roughly the size of the top to bottom mass ratio.
This means that in our priors for the M4DCHM$^{5-5-5}$, the hierarchy among the third generation quark masses is present in the bounds of left-handed composite-elementary mixings, while this is not possible in the other two models, which have only a single left-handed mixing parameter $\Delta_q$.
The fact that different lower bounds have a much smaller effect in the case of uniform priors might be a reason why the evidences in \Cref{tab:Bayesian_statistics_uniform} do not show such large differences between models as those in \Cref{tab:Bayesian_statistics}.
However, also in this case, the prior parameter ranges are favourable for the M4DCHM$^{5-5-5}$.
Therefore, a fully conclusive model comparison would require a detailed analysis of the impact of prior distributions and bounds on the evidences and the other statistics shown in \Cref{tab:Bayesian_statistics}, and it would be interesting to perform such an analysis in future work.

\begin{figure}[t]
\centering
\includegraphics[width=1\linewidth]{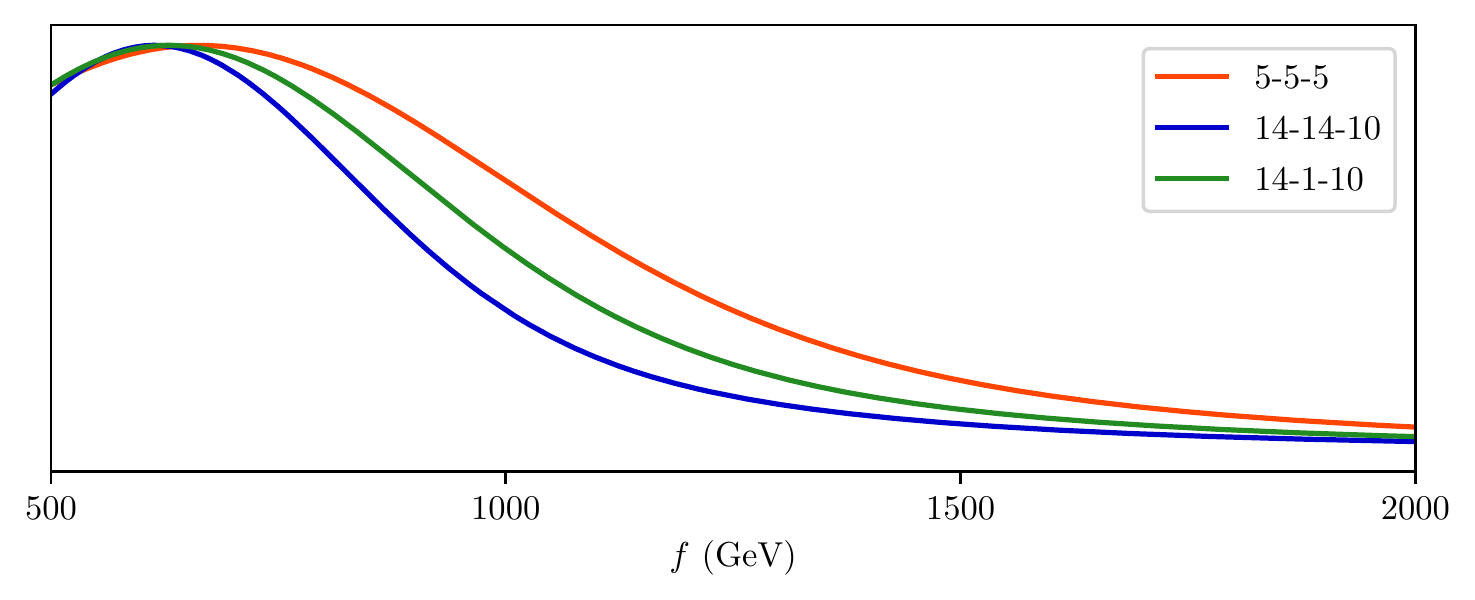}
\caption{Priors for the Higgs decay constant $f$ in each model. These priors are not put in by hand, but are the result of fixing $f$ to give the correct EWSB scale for each point.}
\label{fig:f_priors}
\end{figure}

Apart from the prior dependence, there are also other reasons for the hierarchy of preferences that can be ascertained from \Cref{tab:Bayesian_statistics}.
First, the evidence for the M4DCHM$^{14-14-10}$ is relatively low primarily because of a large KL divergence, \textit{i.e.} large fine-tuning. We have already recognised this tuning by way of the narrowly constrained posteriors in \Cref{fig:14-14-10_gauge_posterior_prior,fig:14-14-10_fermion_posterior_prior}. On the other hand, the M4DCHM$^{14-1-10}$ is disfavoured mostly because it has a much smaller posterior-averaged log-likelihood, \textit{i.e.} because it has a more difficult time satisfying the experimental constraints. We find the Higgs signal strength constraints to be the main factor here.
The M4DCHM$^{5-5-5}$ is the only model with both a low KL divergence (tuning) and relatively large average likelihood, which combine to give the greatest evidence. But here too, fine-tuning is wholly dependent on the prior distribution, as this defines the ``natural" distribution of the parameters. The worsened tuning of the M4DCHM$^{5-5-5}$ under the uniform priors in \Cref{tab:Bayesian_statistics_uniform} makes it clear that this model benefits greatly from the logarithmic priors on the fermion parameters in comparison to the other models. Note that fine-tuning has previously only been quantified in the literature of CHMs through the sensitivity of the observables with respect to the input parameters, either to first order with the Barbieri-Giudice (BG) measure \cite{panico2012,Carena,Niehoff:2015iaa,niehoff2017electroweak}, or with a higher-order measure \cite{BarnardFT,BarnardCC,Murnane:2018ynd}, and our approach of using the KL divergence as a fine-tuning measure has not been taken before.

It is of interest to compare the above findings to the na{\"i}ve fine-tuning estimates suggested by the distributions of the $f$ parameter in each model. Recall the BG tuning required to reproduce the correct EWSB scale is ${\sim}f^{2}/v^{2}$, so the posterior distributions we have found for $f$ correspond to relatively reasonable tunings of ${\sim}25{-}60$ in the M4DCHM$^{5-5-5}$, ${\sim}40{-}80$ in the M4DCHM$^{14-14-10}$, and ${\sim}30{-}80$ in the M4DCHM$^{14-1-10}$. Note that despite its double tuning, the M4DCHM$^{5-5-5}$ has both the lowest KL divergence and the lowest na{\"i}ve tuning of the three models, corroborating the findings of Ref.~\cite{Carena}. A further advantage of the M4DCHM$^{5-5-5}$ is apparent in the priors on $f$, displayed in \Cref{fig:f_priors}. Here it is seen that $f$ is generated at higher values more often in the M4DCHM$^{5-5-5}$ than in the other models, so the M4DCHM$^{5-5-5}$ can accommodate larger values of $f$ with lower levels of tuning.

\begin{figure}[t]
\centering
\includegraphics[width=1\textwidth]{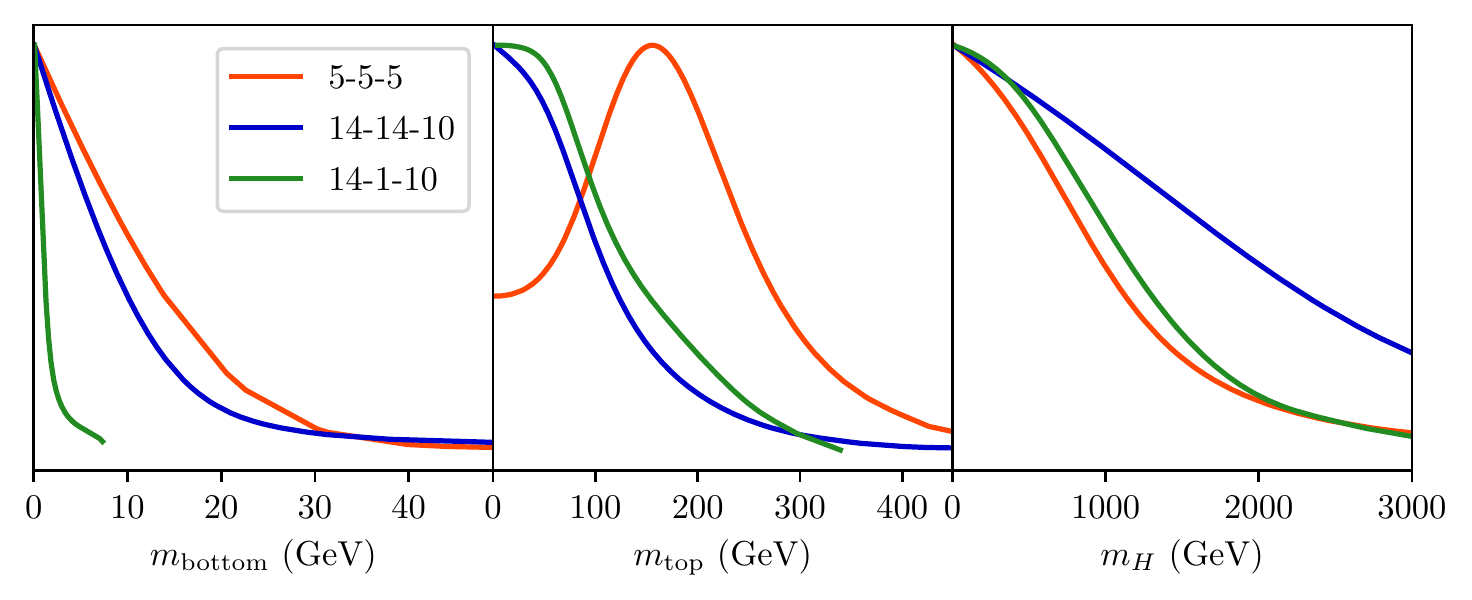}
\caption{Priors for the bottom, top, and Higgs masses in all models.
}
\label{fig:mass_prior_posterior_all}
\end{figure}

It also proves useful to analyse the priors for observables taken as constraints in each model, to see whether any model gives realistic predictions more naturally than others. We give the priors for the masses of the SM particles in \Cref{fig:mass_prior_posterior_all}. As expected, the Higgs, being a pNGB, is generated most often with a low mass. But it also lies above $1$~TeV relatively often, especially in the M4DCHM$^{14-14-10}$, and to a lesser extent in the M4DCHM$^{14-1-10}$, both of which have been noted in other works as having a tendency for producing too large a Higgs mass \cite{panico2012,Carena}. The top and bottom quark masses, on the other hand, all tend to be generated relatively close to their experimental values\footnote{Those in the M4DCHM$^{14-1-10}$ span significantly shorter ranges than in the other models because we applied hard mass cuts $30\gev \leqslant m_{\text{top}} \leqslant 350\gev$ and $0.1\gev \leqslant m_{\text{bottom}} \leqslant 8\gev$ in this model, but not in the others, on account of the computational expense required.}. The M4DCHM$^{14-14-10}$ gives quite an unfavourable prior for the top mass, with very light values $\lesssim 80$~GeV being preferentially generated. Remarkably, the prior for the top mass in the M4DCHM$^{5-5-5}$ has a maximum almost exactly at the experimental value. It should be stressed that this is not a feature of the model itself, but rather a consequence of the bounds imposed on the top sector parameters. Indeed, if the priors for the bottom sector and top sector parameters were switched in this model, then so too would the priors for the bottom quark and top quark masses switch. This is another example of the prior dependence discussed above which makes it easier for the M4DCHM$^{5-5-5}$ to reproduce the top mass, leading to a lower KL divergence and a larger evidence.

We conclude this section with some general comments on the parameter spaces of our models. During our scans, we noticed that the slice sampling steps take $\mathcal{O}(100$) likelihood evaluations in each slice instead of the more typical $\mathcal{O}(5$) evaluations, indicating that the likelihood is quite ``sheety": it is only non-negligible across thin hypersurfaces in the parameter spaces. This notion can be quantified with the Bayesian model dimensionality (BMD) of each model, defined similarly to the KL divergence \Cref{eqn:KL_divergence},
\begin{align}
    \text{BMD} := 2 \int \mathrm{d}^n p\  \mathcal{P}(\mathbf{p} | \mathcal{O})  \brackets{ \ln\brackets{\frac{\mathcal{P}(\mathbf{p} | \mathcal{O})}{\pi(\mathbf{p})}} - D_{\text{KL}}}^{2},
\end{align}
which measures the \textit{effective} dimensionality of the region of parameter space that the posterior occupies \cite{Handley:2019pqx}. This definition is used because it matches the \textit{actual} dimensionality when the posterior is Gaussian. Broader posterior distributions will have lower BMDs, and narrower distributions higher BMDs. It can be difficult to intuit the BMD of a model based on the marginalised posteriors such as those given in \Cref{5-5-5_results_section,14-14-10_results_section,14-1-10_results_section}, since it is most often the case that the constraints restrict only certain non-trivial combinations of the input parameters, and these correlations would be obscured in a 2D plot by the marginalisation. In any case, these effective dimensionalities can be calculated with \texttt{anesthetic} and are listed for our models in \Cref{tab:Bayesian_statistics}.

Clearly, the three models are indeed very sheety, with effective dimensionalities of ${\sim}7$ in the M4DCHM$^{5-5-5}$ and M4DCHM$^{14-1-10}$, and ${\sim}12$ in the M4DCHM$^{14-14-10}$, in contrast to their respective true dimensionalities of $22$, $18$, and $20$ (including the SM gauge couplings). Positive constraints that restrict the space to reproduce particular values for the observables are evidently not sufficient to localise the parameters completely, and leave these ${\sim}7-12$ degrees of freedom within the spaces. These effective dimensionalities should be expected to be quite small on account of the fine-tuning required for EWSB and the effect of the constraints on the space, but it is difficult to say whether ${\sim}7-12$ is a reasonable number of dimensions on purely theoretical grounds.

It is further possible to calculate the principal directions in parameter space that point along these hypersurfaces with \texttt{anesthetic}. If these models were to be explored in future scans, it would be beneficial to use these directions to cleverly parameterise the spaces and scan along these optimal surfaces directly, to achieve greatly improved scanning efficiencies.

\section{Experimental signatures}
\label{experimental_signatures_section}

In our fits, we have found a number of points in each model that satisfy all constraints and direct collider search bounds at the $3\sigma$ level individually, which we regard as \textit{valid} points. We have collected, in total, ${\sim}14{,}000$ valid points in the M4DCHM$^{5-5-5}$, ${\sim}77{,}000$ in the M4DCHM$^{14-14-10}$, and ${\sim}280{,}000$ in the M4DCHM$^{14-1-10}$. Here we analyse the phenomenology of these experimentally viable points so that we may anticipate possible signatures of each model in future collider experiments.

\subsection{Composite resonances}

\begin{figure}[t]
\centering
\begin{subfigure}{.49\linewidth}
  \centering
  \includegraphics[width=1\textwidth]{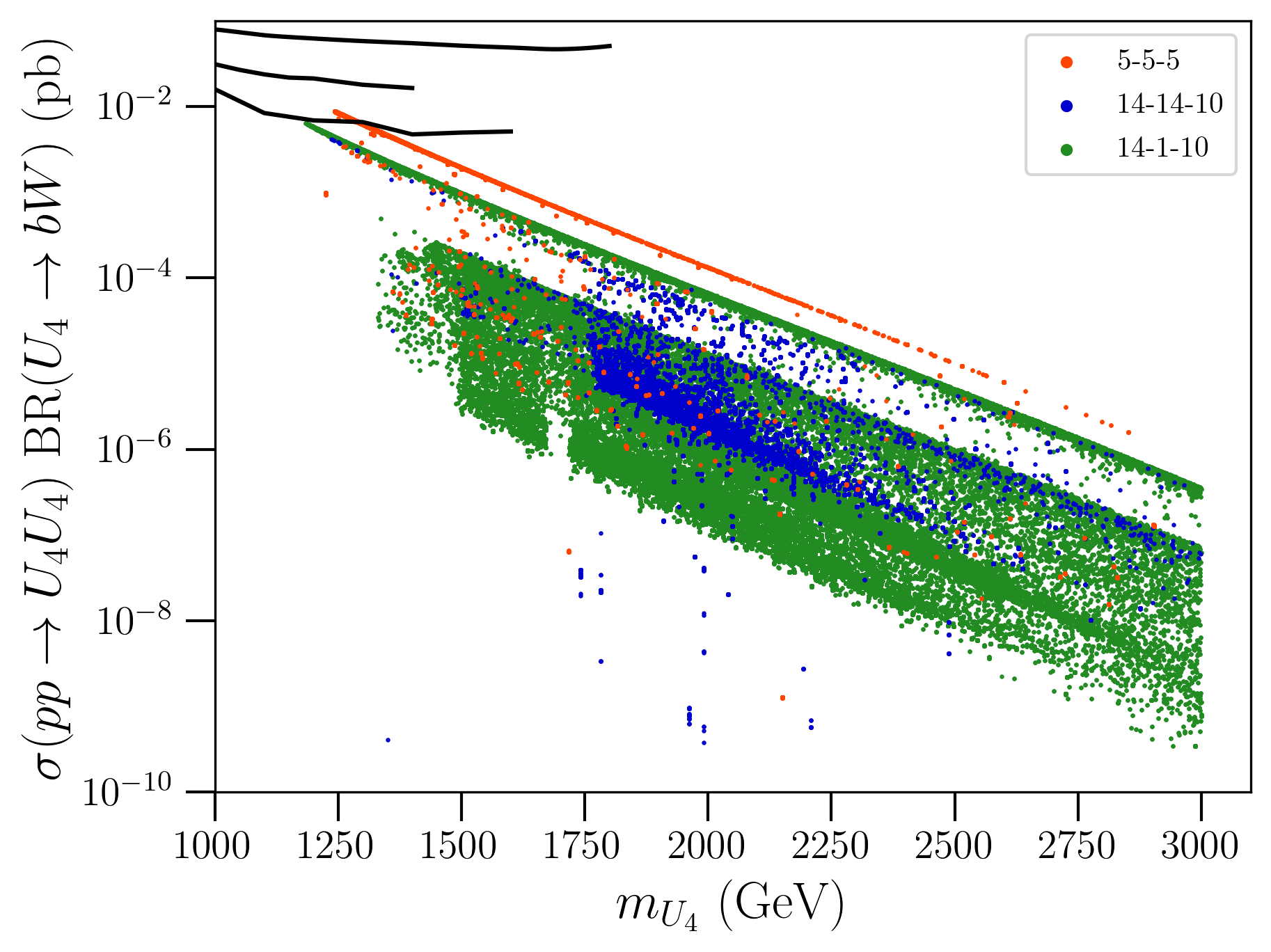}
\end{subfigure}
\begin{subfigure}{.49\linewidth}
  \centering
  \includegraphics[width=1\textwidth]{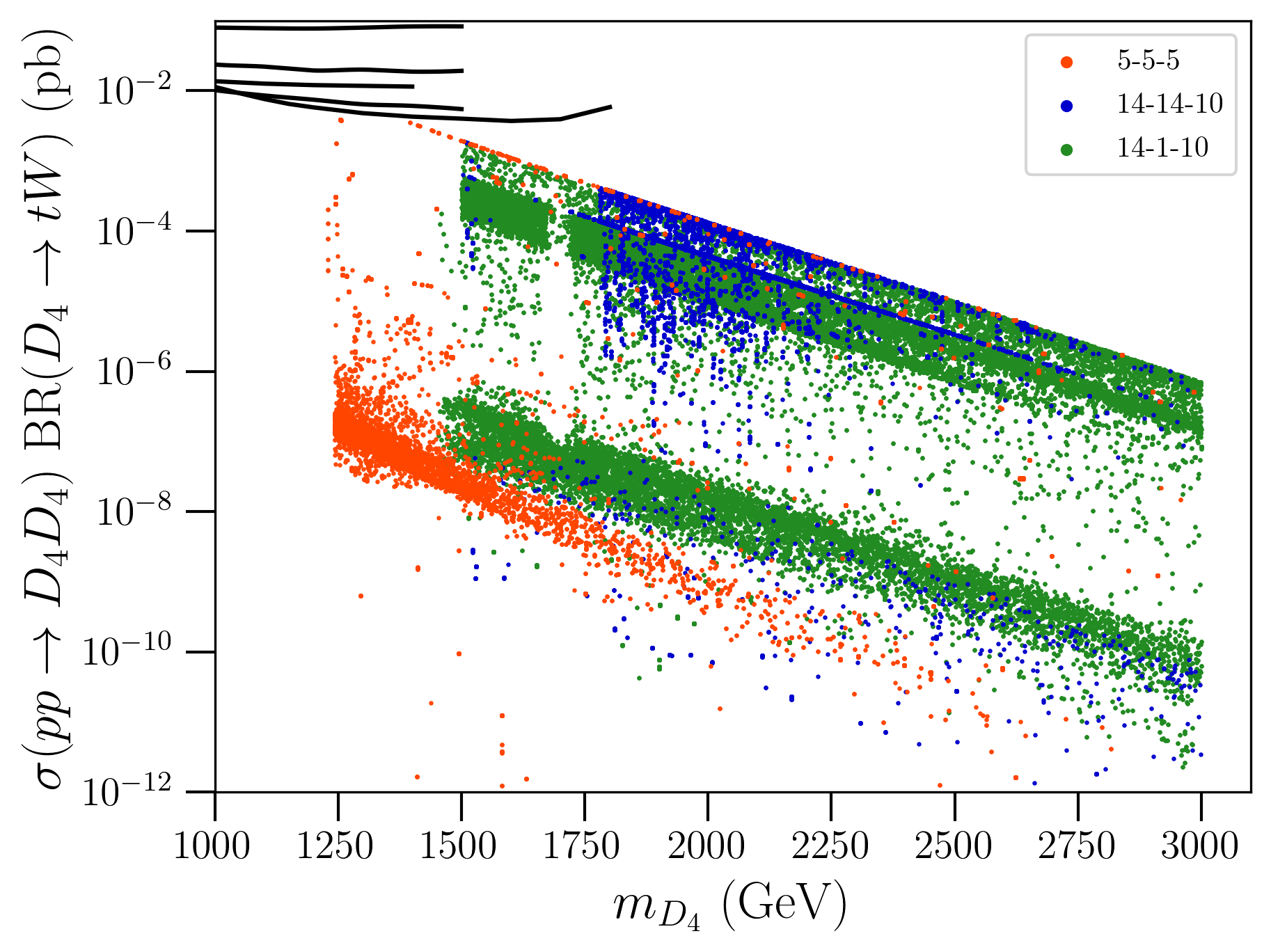}
\end{subfigure}
\begin{subfigure}{.49\linewidth}
  \centering
  \includegraphics[width=1\textwidth]{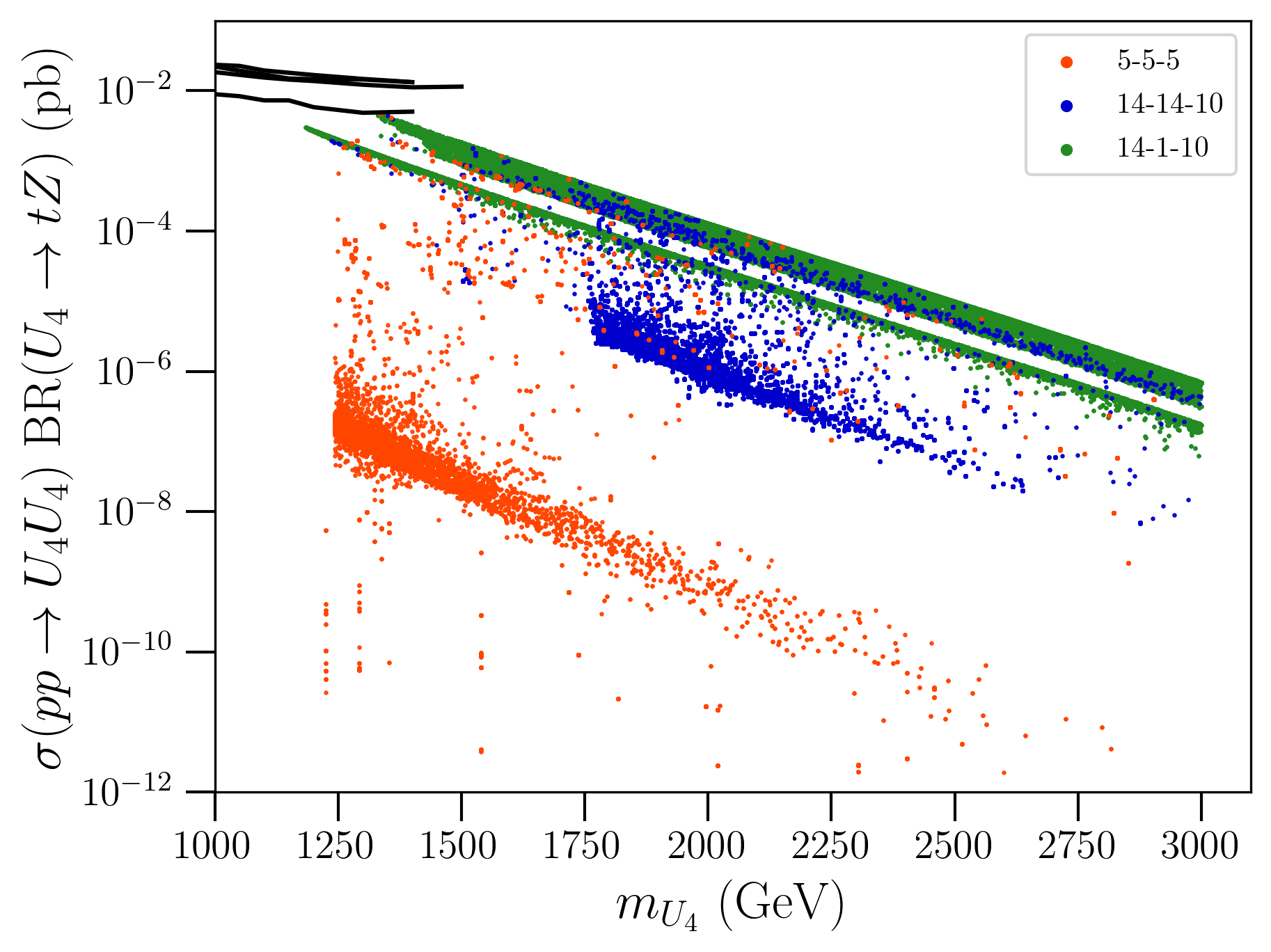}
\end{subfigure}
\begin{subfigure}{.49\linewidth}
  \centering
  \includegraphics[width=1\textwidth]{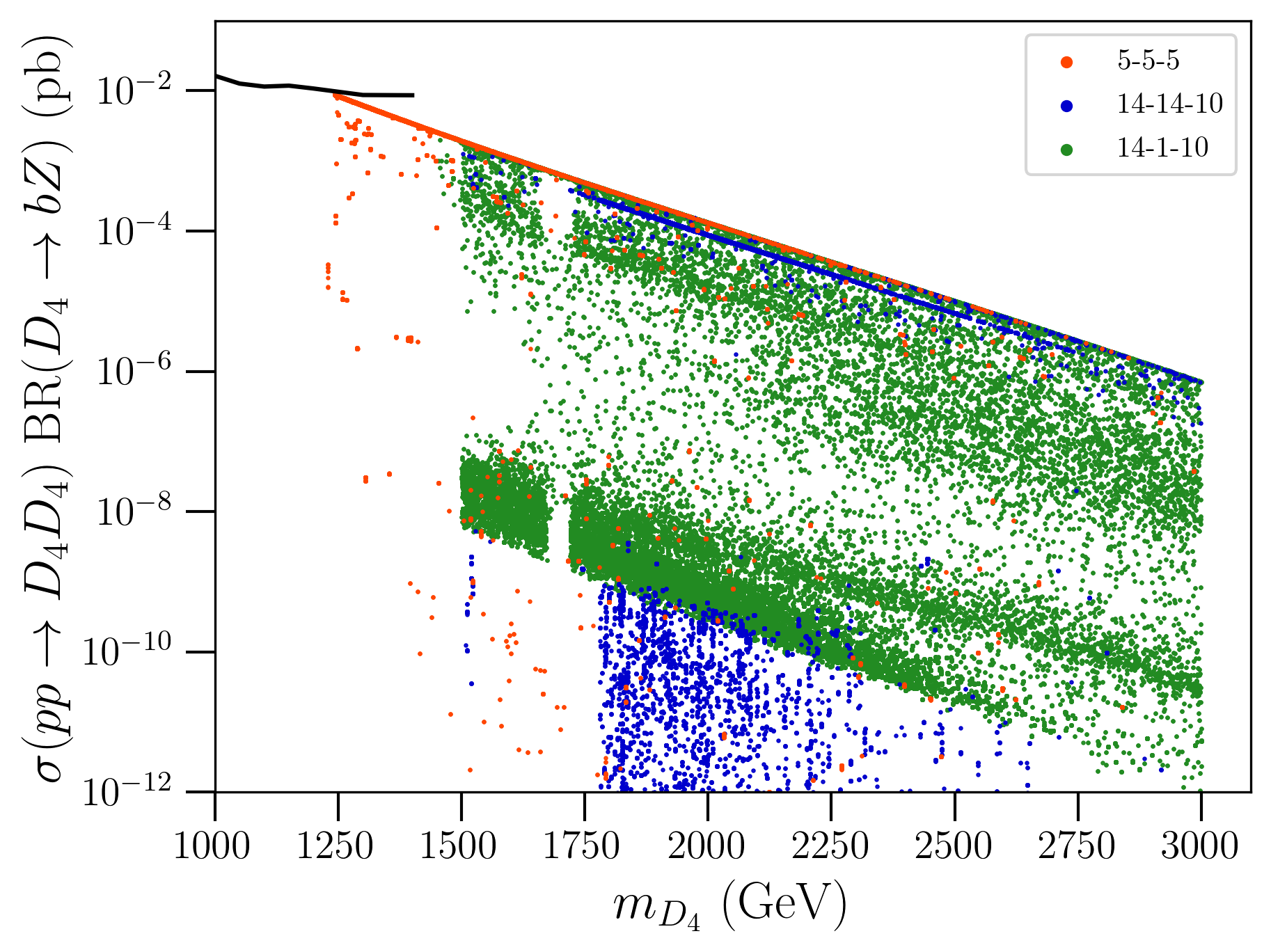}
\end{subfigure}
\begin{subfigure}{.49\linewidth}
  \centering
  \includegraphics[width=1\textwidth]{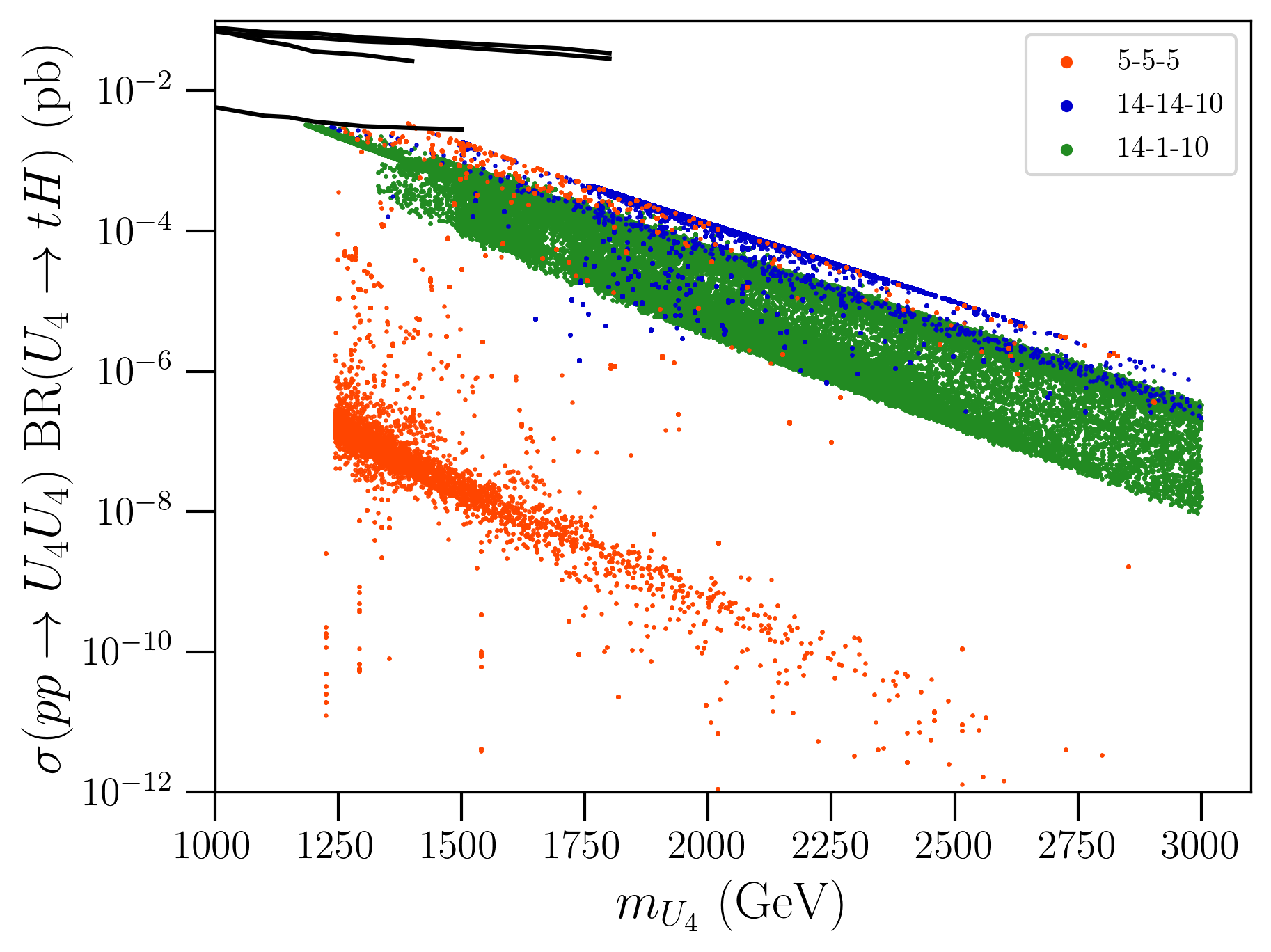}
\end{subfigure}
\begin{subfigure}{.49\linewidth}
  \centering
  \includegraphics[width=1\textwidth]{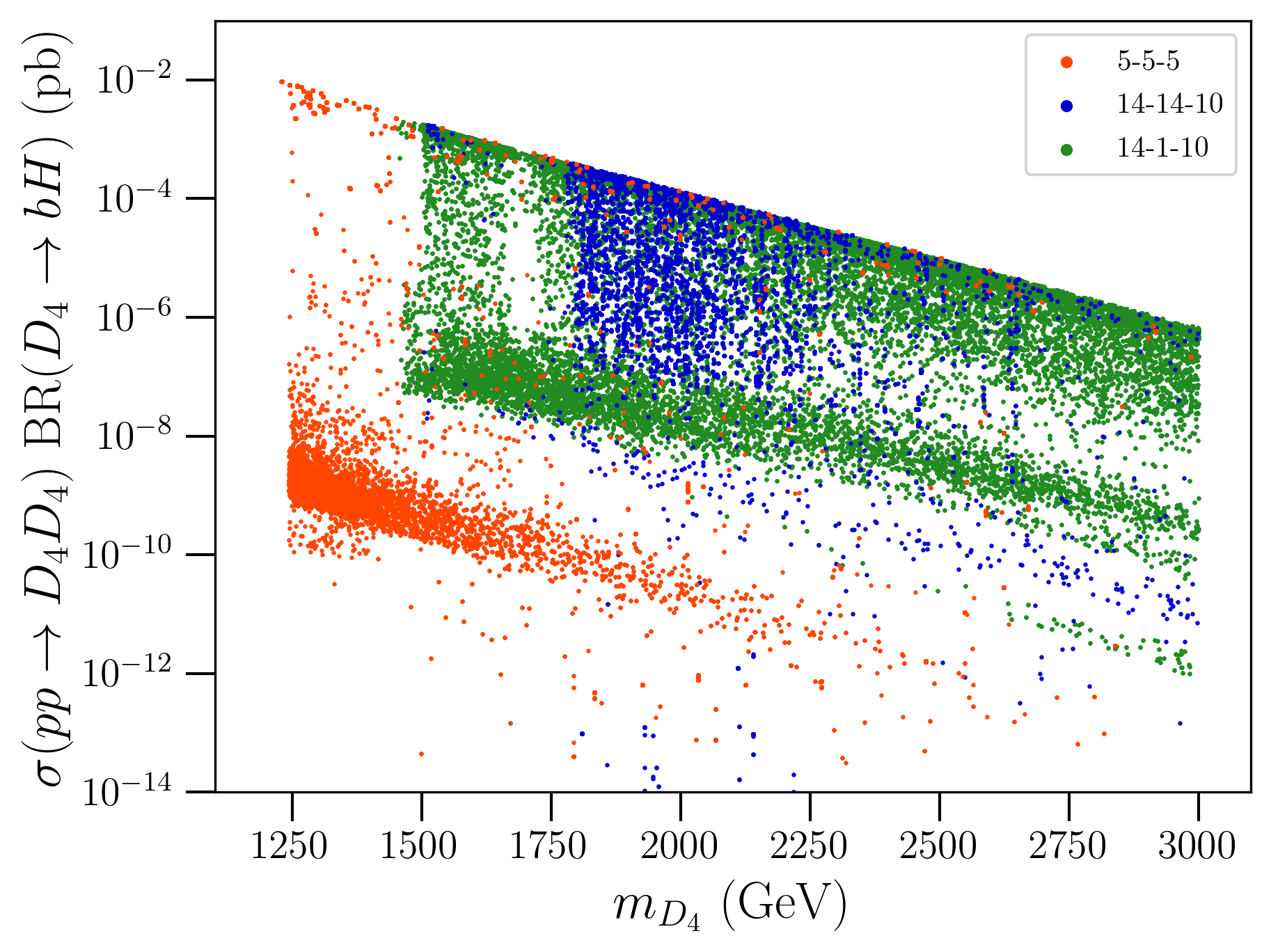}
\end{subfigure}
\caption{Cross sections for decays of the lightest up-type and down-type resonances U$_{4}$ and D$_{4}$ into SM final states at $\sqrt{s} = 13$~TeV for valid points in each model. Black lines mark the $95$\% CL upper bounds taken as constraints.}
\label{fig:all_model_UD_decays}
\end{figure}

We begin with the BSM particle spectra. Recall that these models contain several up-type (U) and down-type (D) composite fermions, exotic fermions Q$_{x}$ with electric charges $x \in \{4/3,\ 5/3,\ 8/3\}$, as well as composite gauge bosons with unit charge (W) and neutral charge\footnote{Not including the gluon sector.} (Z). Our convention for distinguishing the different members of each species is to label them with numerical subscripts in order from lightest to heaviest, including SM states. For example, the lightest neutral composite gauge boson will be called Z$_{3}$, being the lightest neutral boson after the photon (Z$_{1}$) and the electroweak Z boson (Z$_{2}$).

In the M4DCHM$^{5-5-5}$, the lightest composite U and D resonances share approximately the same mass range, from $1.2$~TeV to $5.3$~TeV, with U$_{4}$ typically being the lighter of the two. Similar lower bounds apply to the next three lightest of these species, as well as the lightest of each exotic species, all of which mostly lie below ${\sim}6$~TeV but can range upwards of $10$~TeV. The remaining fermions can be as light as $2$~TeV or $3$~TeV, but often lie well above $10$~TeV. There are no apparent correlations between the masses across fermion species, aside from the ones that follow from \Cref{fig:fermion_masses}. For all valid points, Z$_{3}$ is the lightest composite gauge boson, ranging from $1.4$~TeV upwards, while Z$_{4}$ is approximately degenerate with W$_{2}$ and W$_{3}$, which lie above $2.4$~TeV. All other gauge bosons tend to be heavier than $10$~TeV.

\begin{figure}[t]
\centering
\begin{subfigure}{.49\linewidth}
  \centering
  \includegraphics[width=1\textwidth]{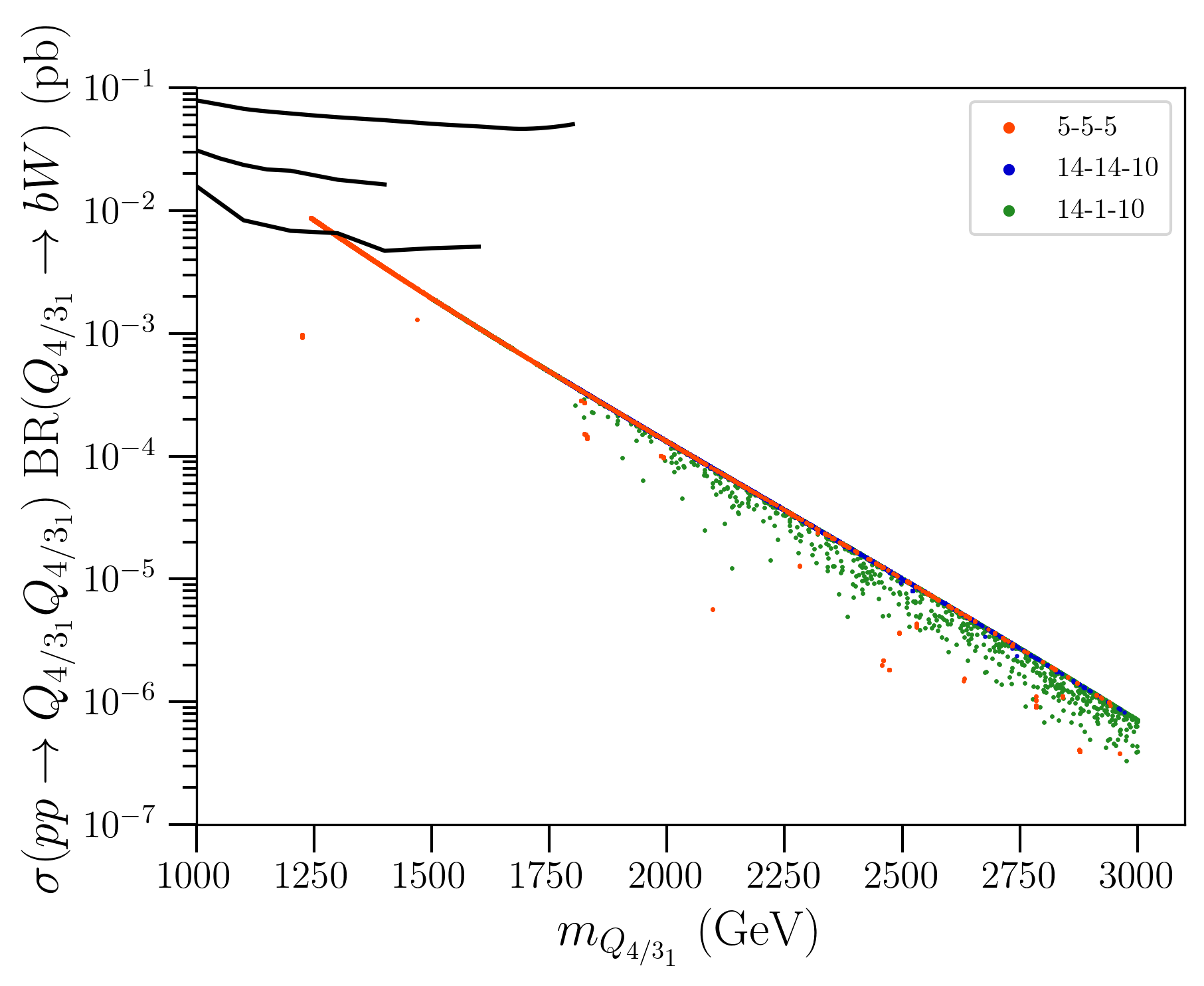}
\end{subfigure}
\begin{subfigure}{.49\linewidth}
  \centering
  \includegraphics[width=1\textwidth]{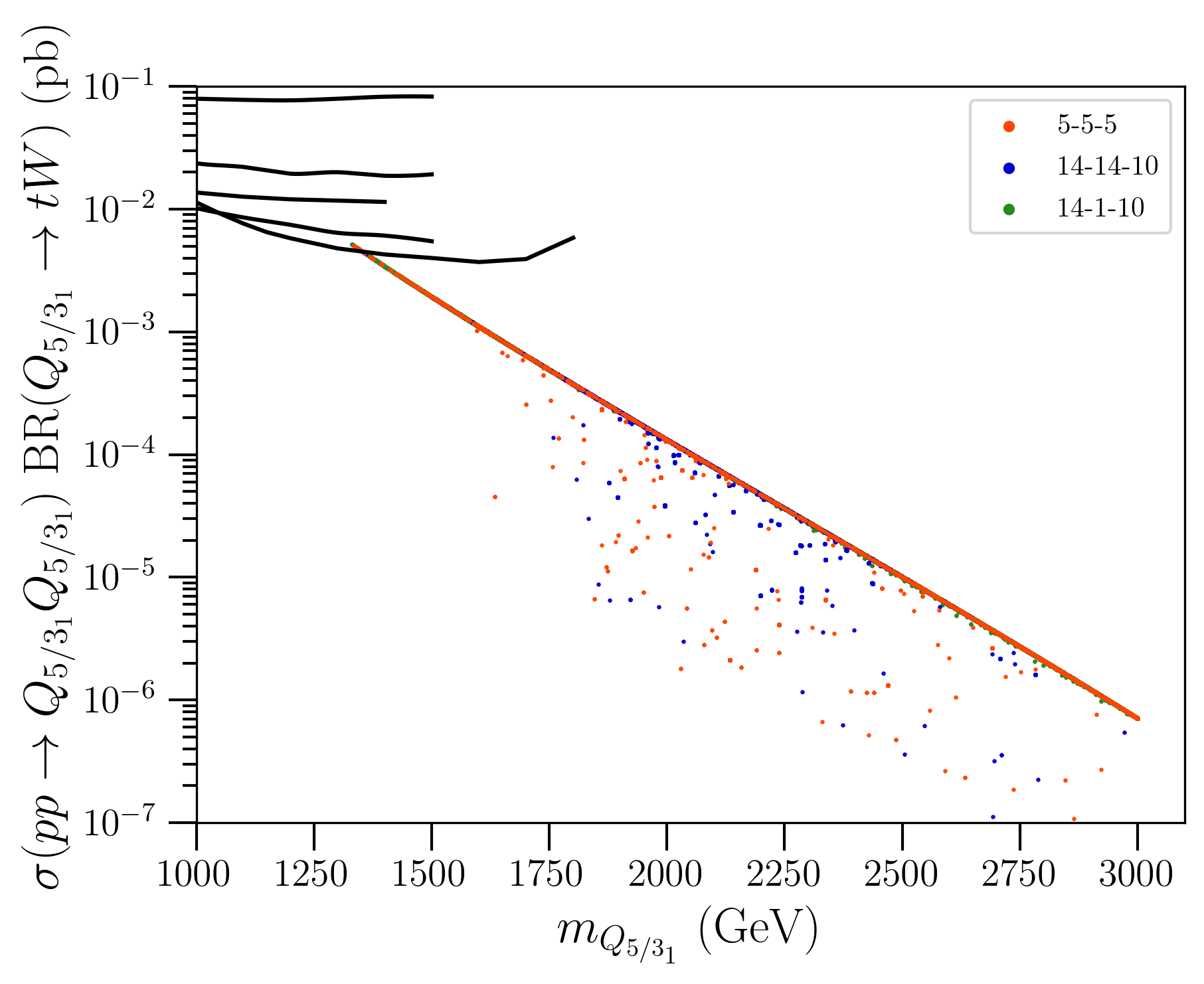}
\end{subfigure}
\caption{Cross sections for decays of the lightest resonances Q$_{x}$ with exotic electric charge $x$ into SM final states at $\sqrt{s} = 13$~TeV for valid points in each model. Most of the points in each model lie on the prominent lines in these plots. Black lines mark the $95$\% CL upper bounds taken as constraints.}
\label{fig:all_model_Q_decays}
\end{figure}

\begin{figure}[h]
\centering
\begin{subfigure}{.49\linewidth}
  \centering
  \includegraphics[width=1\textwidth]{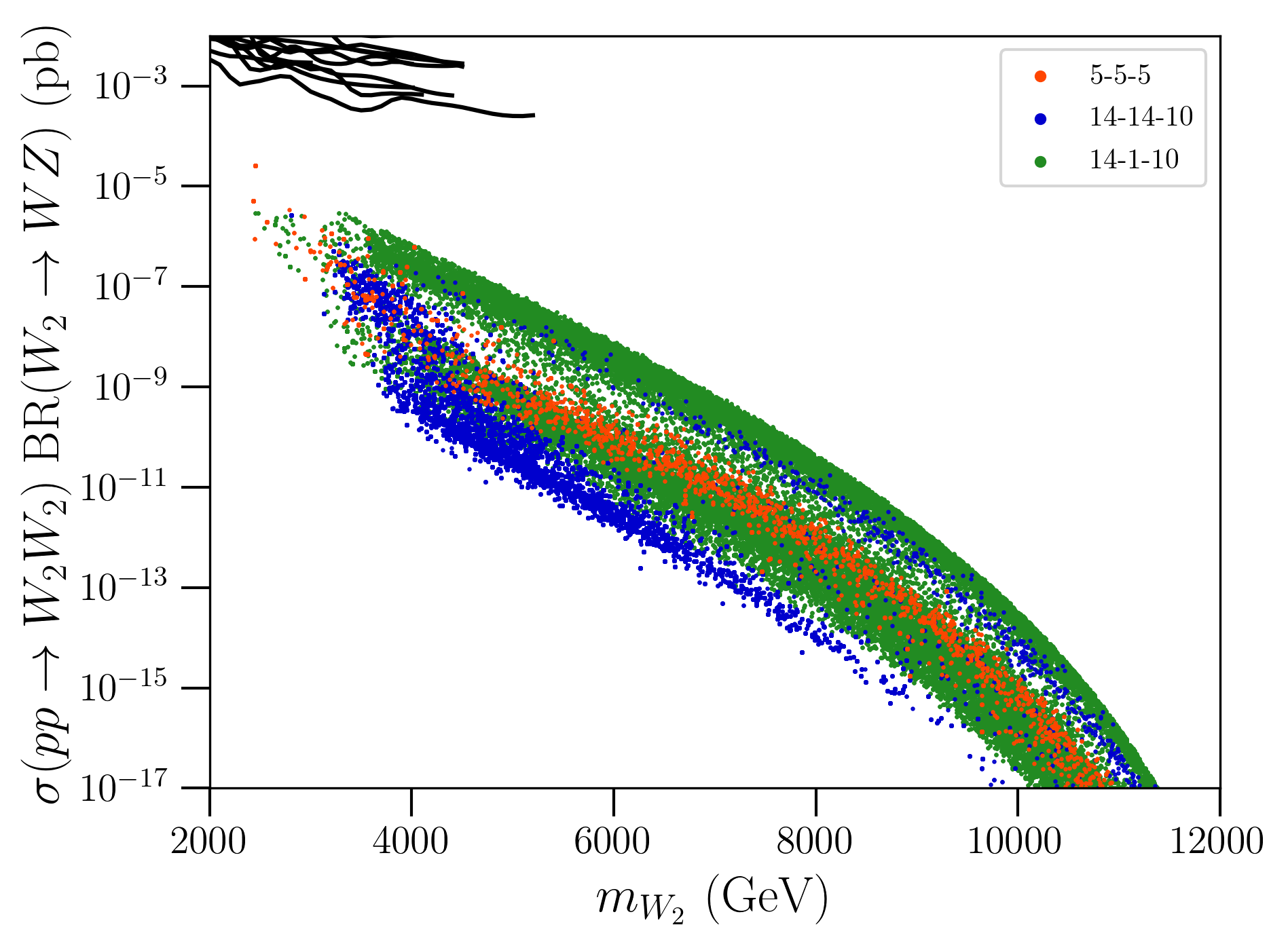}
\end{subfigure}
\begin{subfigure}{.49\linewidth}
  \centering
  \includegraphics[width=1\textwidth]{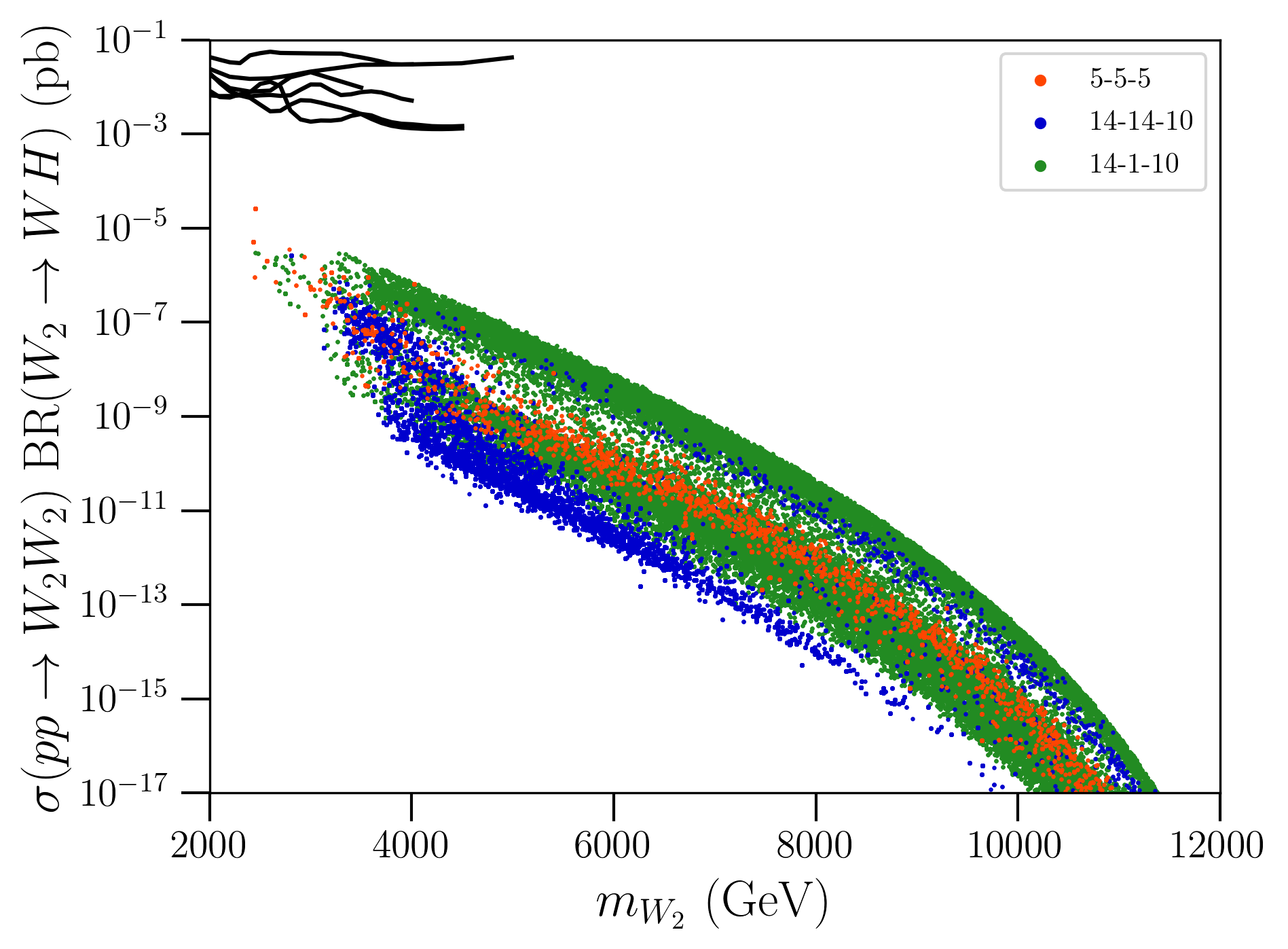}
\end{subfigure}
\caption{Cross sections for decays of the lightest charged composite vector boson W$_{2}$ into SM final states at $\sqrt{s} = 13$~TeV for valid points in each model. Black lines mark the $95$\% CL upper bounds taken as constraints.}
\label{fig:all_model_W_decays}
\end{figure}

For the M4DCHM$^{14-14-10}$, recall from \Cref{fig:fermion_masses} that two pairs each of U, D, and Q$_{5/3}$ particles, and one pair each of Q$_{4/3}$ and Q$_{8/3}$ particles, come with masses of approximately $M_{\pm}(m_{q},m_{t},m_{Y_{t}},0)$. These are found to typically be among the lightest composite resonances, with masses respectively lying above $1.4$~TeV and $1.8$~TeV, so there is much approximate degeneracy among the lightest states. This is excepting the U$_{4}$ resonance, which can be as light as $1.3$~TeV. The remaining fermions (save for the two heaviest U resonances) can be found below $3$~TeV, but are usually significantly heavier. Gauge boson masses have the same pattern in this model as in the M4DCHM$^{5-5-5}$ but at a heavier scale, with Z$_{3}$ only being found above $2.1$~TeV and W$_{2}$ above $3$~TeV.

\begin{figure}[t]
\centering
\begin{subfigure}{.49\linewidth}
  \centering
  \includegraphics[width=1\textwidth]{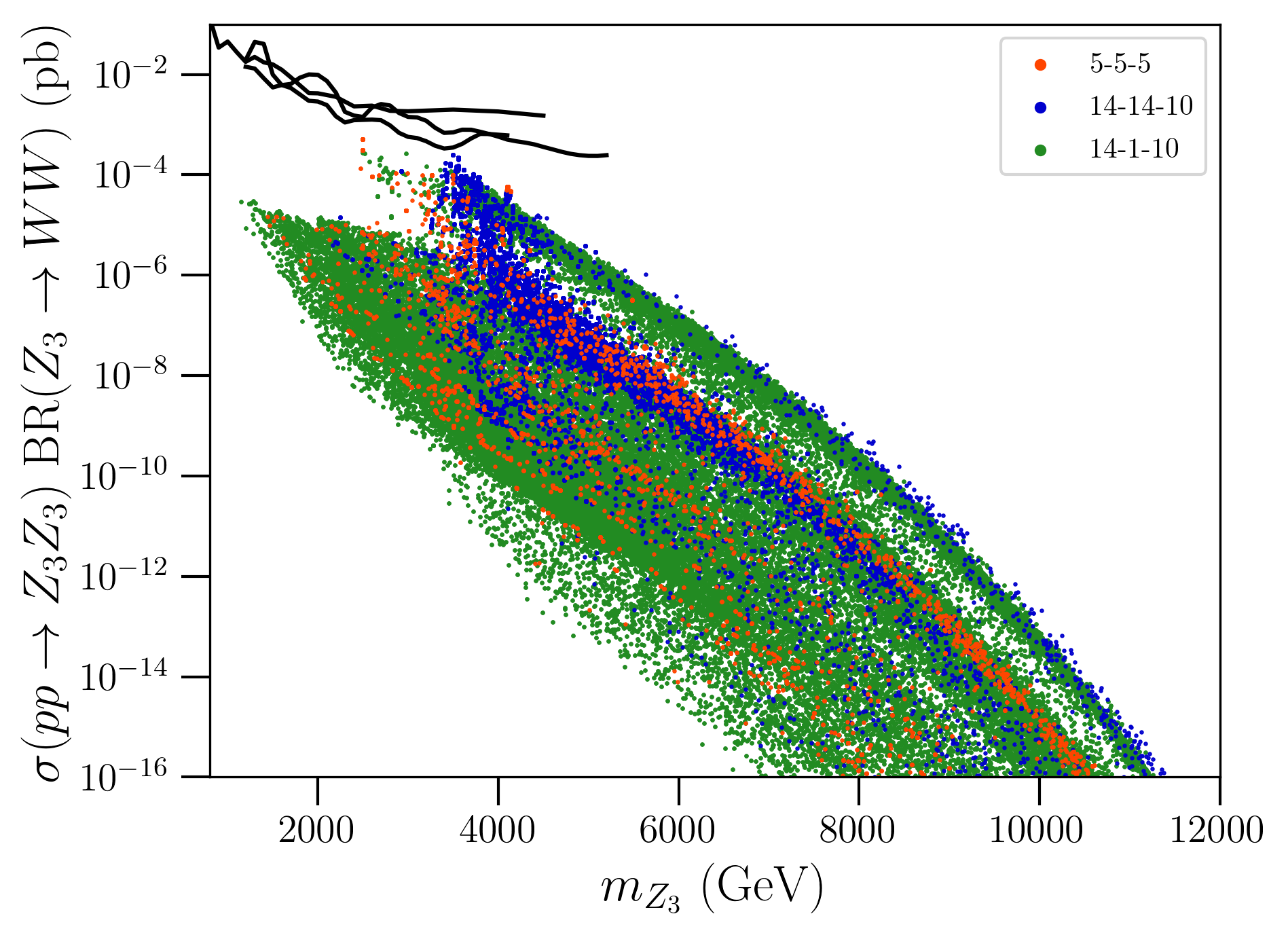}
\end{subfigure}
\begin{subfigure}{.49\linewidth}
  \centering
  \includegraphics[width=1\textwidth]{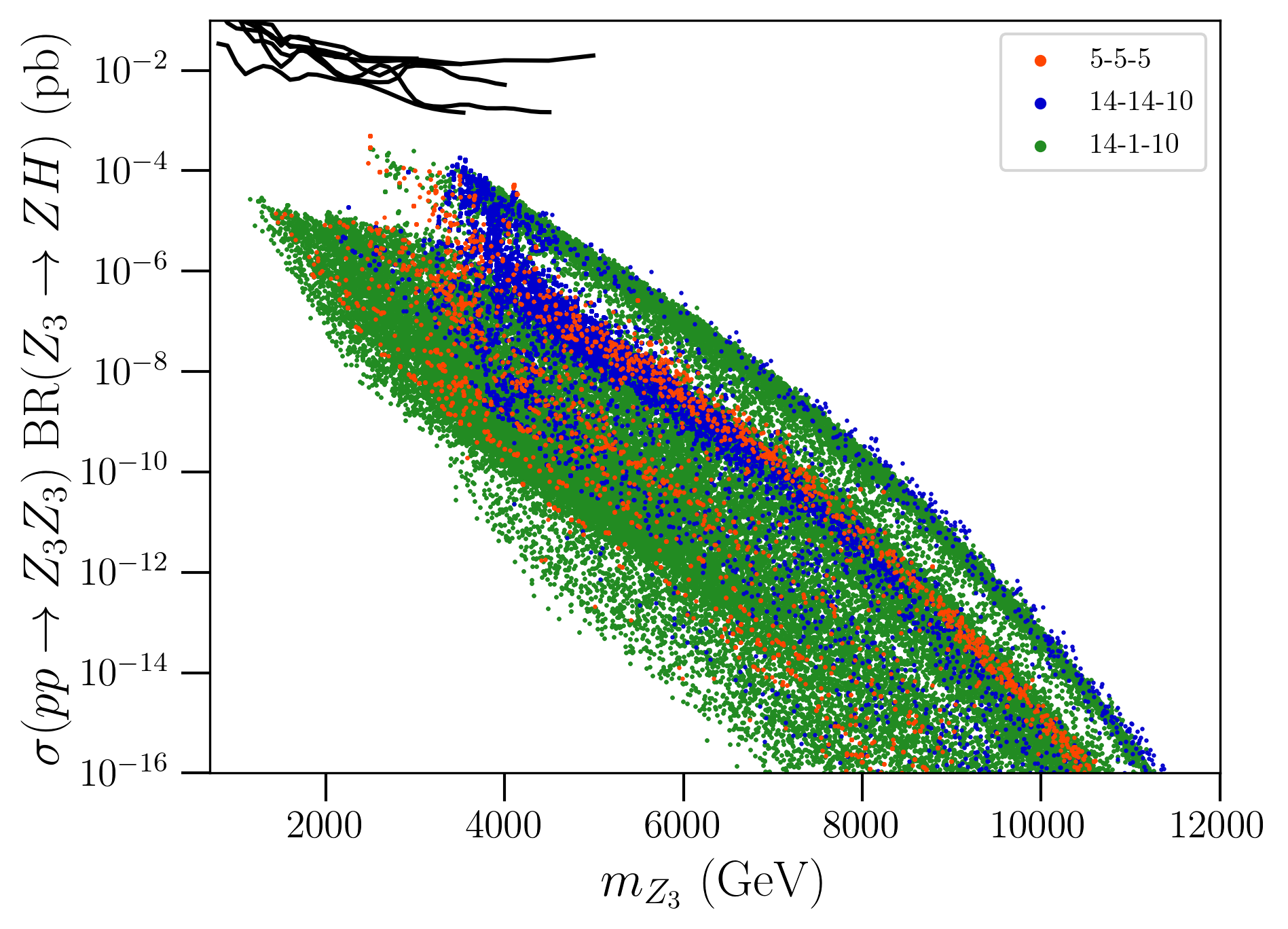}
\end{subfigure}
\begin{subfigure}{.49\linewidth}
  \centering
  \includegraphics[width=1\textwidth]{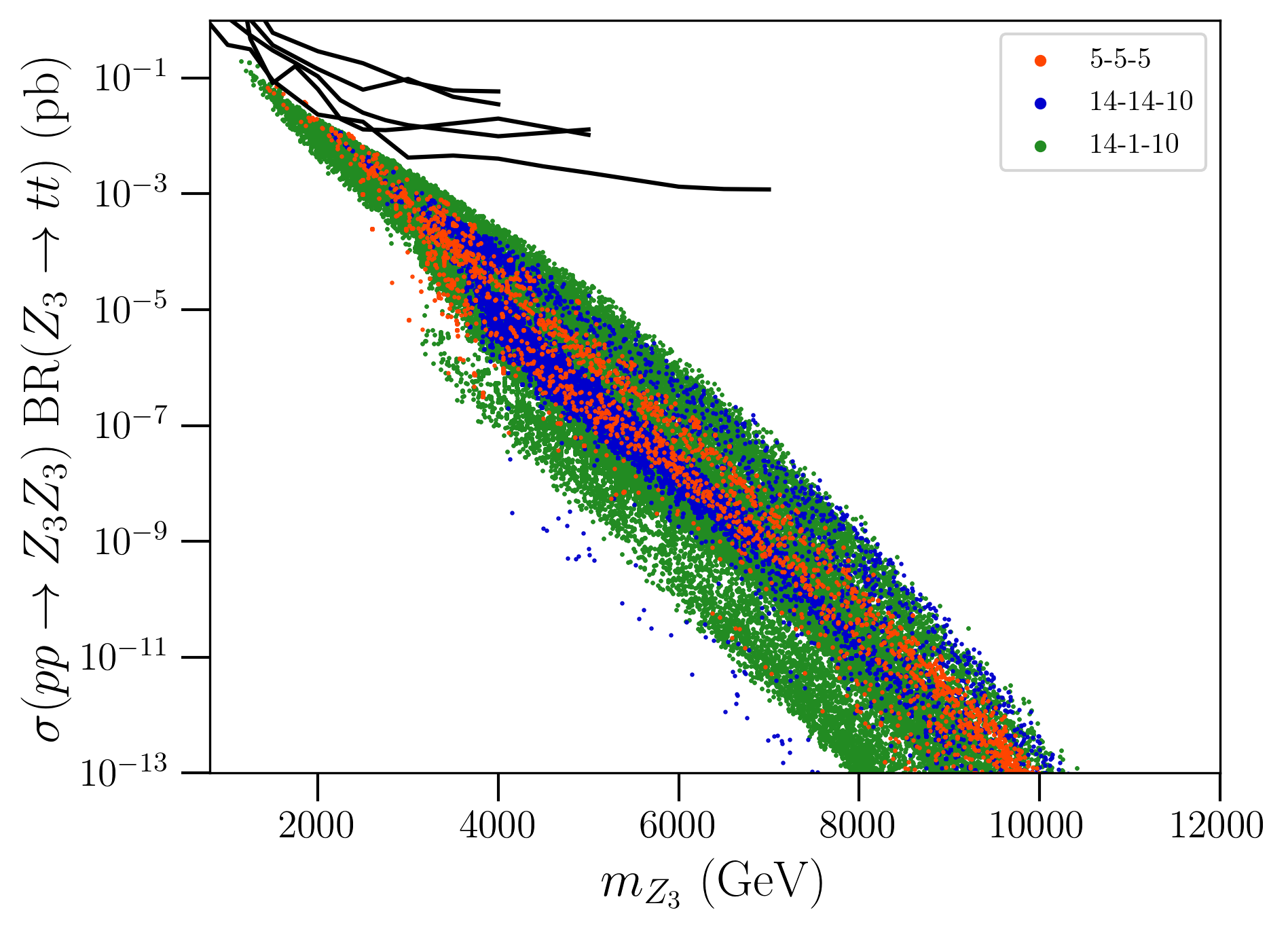}
\end{subfigure}
\caption{Cross sections for decays of the lightest neutral composite vector boson Z$_{3}$ into SM final states at $\sqrt{s} = 13$~TeV for valid points in each model. Black lines mark the $95$\% CL upper bounds taken as constraints.}
\label{fig:all_model_Z_decays}
\end{figure}

The analysis for the M4DCHM$^{14-1-10}$ is the most straightforward since the masses of all fermions are given by the simple formulae in \Cref{fig:fermion_masses}. The lightest fermion tends to be U$_{4}$, with a mass $M_{-}(m_{q},m_{t},2Y_{t}/\sqrt{5},\Delta_{t}) \gtrsim 1.1$~TeV. The next lightest states tend to be those with masses $M_{-}(m_{q},m_{b},Y_{b}/2,0)$, followed closely by those with masses $m_{b}$ and $m_{q}$, all reaching as low as ${\sim}1.4$~TeV. As in the M4DCHM$^{14-14-10}$, all fermions here can lie well above $10$~TeV. The pattern of gauge boson masses is the same here as before, with $1.2\tev \lesssim \text{Z}_{3} \lesssim 60\tev$, and $2.4\tev \lesssim \text{Z}_{4} \approx \text{W}_{2} \approx \text{W}_{3} \lesssim 60\tev$. The remaining bosons, including the heavy gluons, once again tend to lie above $10$~TeV.

Such resonances would be discovered in collider experiments through the observation of excesses in the invariant mass distributions of the resonance decay products. For the valid points we have found, we give the cross sections of such processes involving the lightest composite resonances that would be measured at the $13$~TeV LHC in \Cref{fig:all_model_UD_decays,fig:all_model_Q_decays,fig:all_model_W_decays,fig:all_model_Z_decays}.

It is seen from \Cref{fig:all_model_UD_decays} that the majority of valid points have predicted U$_{4}$ and D$_{4}$ cross sections that are far below current collider sensitivity. Exceptions exist in the $U_{4} \rightarrow bW^{+}$ decay channel for the M4DCHM$^{5-5-5}$ and the M4DCHM$^{14-1-10}$, and in the $D_{4} \rightarrow bZ$ channel for the M4DCHM$^{5-5-5}$, where even a slight increase in sensitivity would be sufficient to rule out resonances up to ${\sim}1.5$~TeV. The exotic decay channels, in \Cref{fig:all_model_Q_decays}, are more promising, with all three models making the same remarkably precise predictions for the $Q_{4/3_{1}} \rightarrow \bar{b}W^{+}$ and $Q_{5/3_{1}} \rightarrow tW^{+}$ decays, with cross sections comparable to the upper bounds towards the lighter end of the mass spectrum. Probing these channels further can easily confirm or rule out exotic resonances of certain masses and thereby constrain the models rather straightforwardly, although the cross sections fall off rapidly at higher masses, so much of the space remains out of reach of the LHC.

Based on \Cref{fig:all_model_W_decays}, we would not expect to detect any charged composite gauge boson decays in the analysed channels under any of the models. We also considered $W_{2} \rightarrow t \bar{b}$ decays, but the predicted cross sections for this process were all negligible and not worth presenting. There is better hope for the neutral gauge boson decays in \Cref{fig:all_model_Z_decays}, with all models predicting similar cross sections. Promising channels here are $Z_{3} \rightarrow t \bar{t}$, and to a lesser extent $Z_{3} \rightarrow W^{+}W^{-}$. However, many points predict resonance masses too large to be probed in the near future, so failure to detect any resonances at the LHC, for example, would not be sufficient to rule out any of these models. See Refs.~\cite{Barducci:2012kk,Barducci:2012sk} for more in-depth analyses of the detection prospects for vector bosons in the M4DCHM$^{5-5-5}$.

\subsection{Higgs signal strengths}

Other important theoretical predictions of the M4DCHMs that we analyse are the (gluon-fusion produced) Higgs signal strengths. Modifications to the Higgs couplings in a composite Higgs framework have been subject to considerable analysis. To summarise the points that concern us here, in CHMs there are three sources of new physics that can modify the gluon-fusion production of a Higgs boson compared to the SM: non-linearities of the pNGB Higgs (dependent only on $v/f$), modified Yukawa couplings of SM particles, and loop contributions from composite resonances \cite{giudice2007}. If the contributions from the Higgs non-linearities were to dominate, as might be expected in our models \cite{Low:2010mr}, the signal strengths would be of the form
\begin{align}
    \mu^{gg}_{XX} = 1 - C_{{}_{X}}  \frac{v^{2}}{f^{2}}+ \mathcal{O}\brackets{\frac{v^{4}}{f^{4}}},
\label{eqn:signal_strength_modification}
\end{align}
for some channel-dependent (and model-dependent) constants $C_{{}_{X}}$ \cite{giudice2007}. Further discussion on this analytic approximation and the subleading contributions to the signal strengths can be found in Ref.~\cite{Carena}.

\begin{figure}[t]
\centering
\begin{subfigure}{.32\linewidth}
  \centering
  \includegraphics[width=1.0\textwidth]{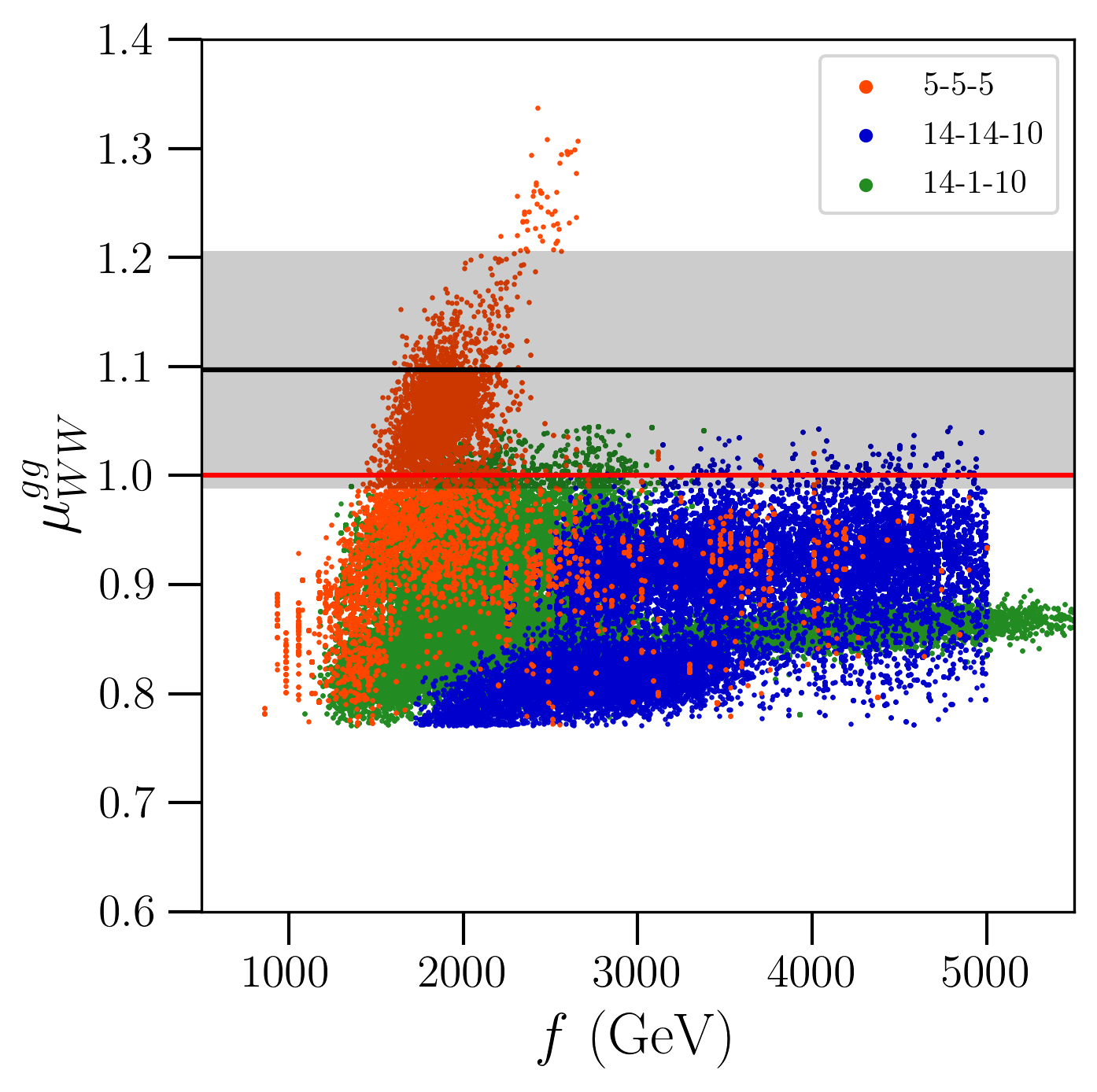}
\end{subfigure}
\begin{subfigure}{.32\linewidth}
  \centering
  \includegraphics[width=1.0\textwidth]{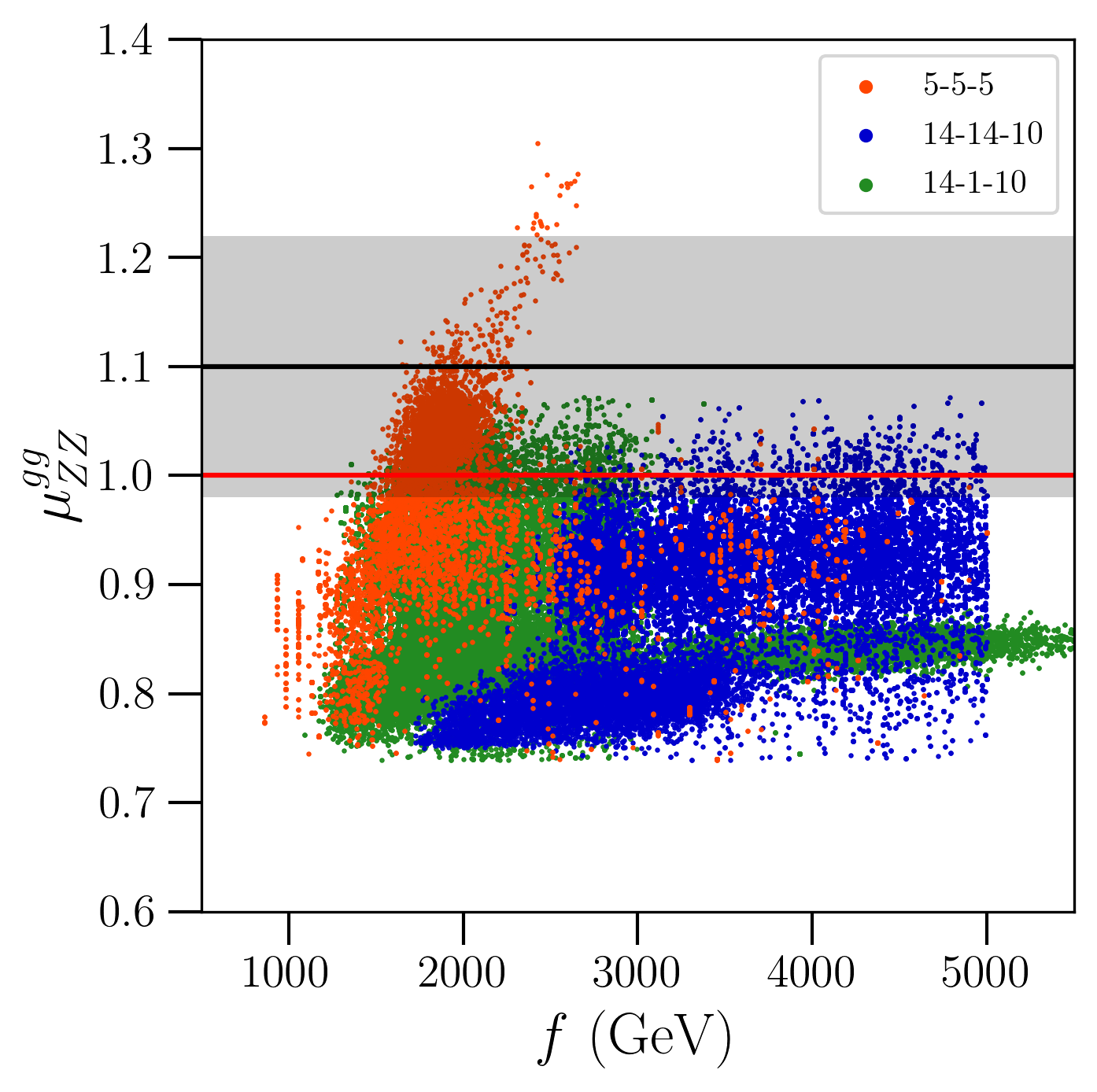}
\end{subfigure}
\begin{subfigure}{.32\linewidth}
  \centering
  \includegraphics[width=1.0\textwidth]{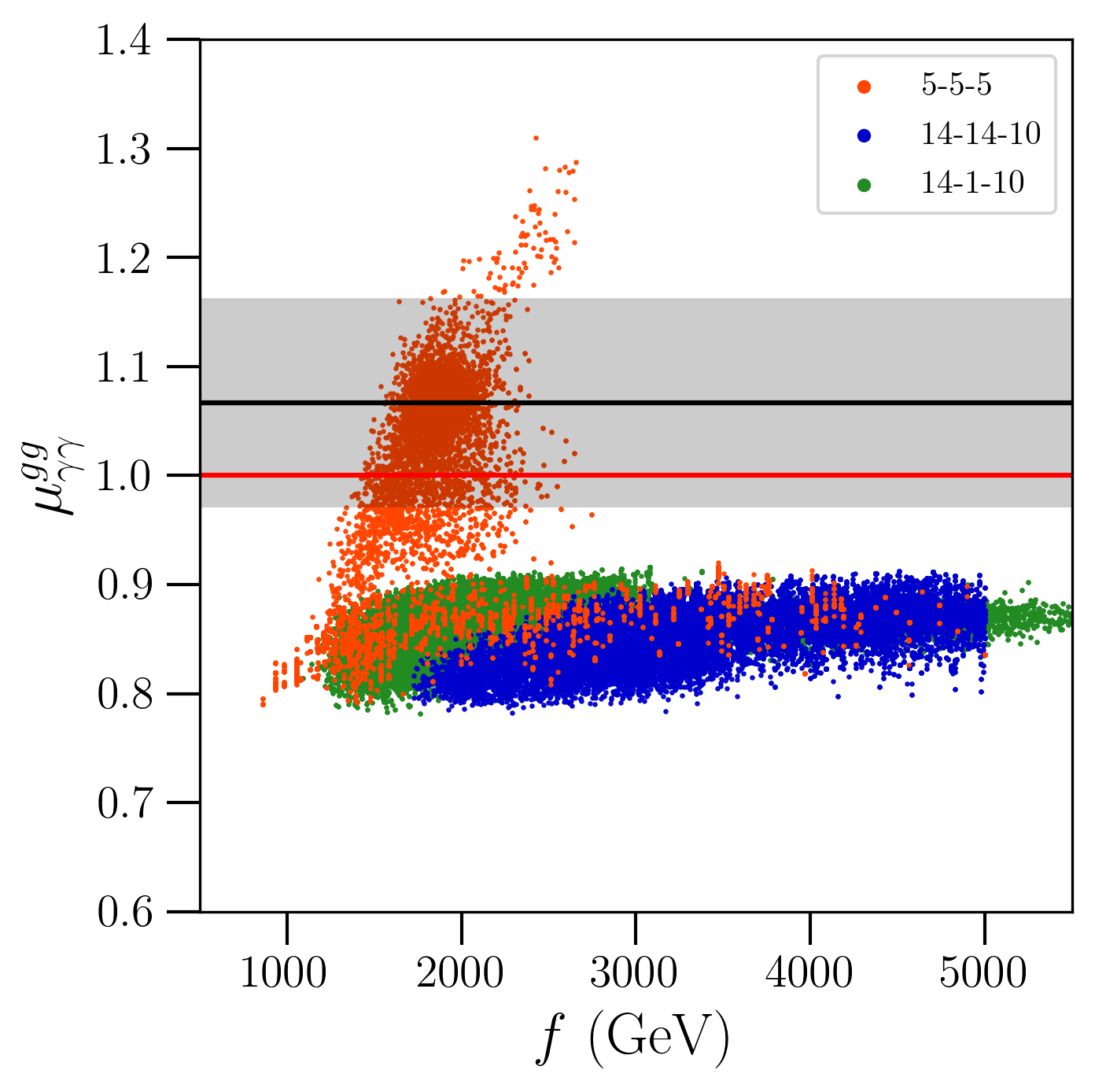}
\end{subfigure}
\caption{Higgs signal strengths for valid points in the various M4DCHMs. Red lines show the SM predictions $\mu^{gg}_{XX} = 1$, while black lines mark the best-fit values, and the shaded regions the $1\sigma$ uncertainties, of the combined measurements of these processes.}
\label{fig:signal_strengths}
\end{figure}

We give the signal strengths\footnote{The signal strengths for the $h \rightarrow \tau \tau$ decay channel are not presented because this channel is subject to large uncertainties, but the predicted values tend to be similar to those for the $h \rightarrow \gamma \gamma$ channel.} of the valid points in each of our models in \Cref{fig:signal_strengths}. The signal strengths $\mu^{gg}_{ZZ}$ and $\mu^{gg}_{WW}$ are approximately equal due to the custodial symmetry of the M4DCHMs. It is immediately apparent in \Cref{fig:signal_strengths} that the approximation \Cref{eqn:signal_strength_modification} does not hold well, especially in the M4DCHM$^{5-5-5}$, but there do seem to be some structures of points that loosely follow the $-C_{{}_{X}} v^{2}/f^{2}$ deviation from unity, which is most clearly seen for $\mu^{gg}_{\gamma\gamma}$ in the M4DCHM$^{14-14-10}$ and M4DCHM$^{14-1-10}$. This is in contrast to Ref.~\cite{Niehoff:2015iaa}, where clear curves of points were found. It is possible that our points do not follow this distribution cleanly because we consider larger values of $f$, so the curves are less pronounced and deviations from \Cref{eqn:signal_strength_modification} are more apparent.

Among the valid points in the M4DCHM$^{5-5-5}$, the predicted signal strengths in \Cref{fig:signal_strengths} span a wide range of values from $0.8$ to $1.3$. This agrees with the results of Ref.~\cite{Barducci:2013wjc}, where it was found that the enhancement in this channel is primarily due to the reduction of the Higgs total width from down-type quark mixing effects. It is therefore unlikely that further probing the signal strengths could rule out or provide evidence for the M4DCHM$^{5-5-5}$. Contrastingly, all the valid points in the M4DCHM$^{14-14-10}$ and M4DCHM$^{14-1-10}$ give signal strengths that are less than the SM prediction of unity, excepting a small portion of points for $\mu^{gg}_{WW}$ and $\mu^{gg}_{ZZ}$. Furthermore, there is a remarkable prediction in these two models: that the signal strength $\mu^{gg}_{\gamma\gamma}$ lies in a narrow range from ${\sim}0.8$ to ${\sim}0.9$. These clear predictions may serve as tests of the M4DCHM$^{14-14-10}$ and M4DCHM$^{14-1-10}$ upon more precise measurements of Higgs decays. It is projected that $\mu^{gg}_{\gamma\gamma}$ will be measured to an uncertainty of around $5$\% by each ATLAS and CMS once they achieve integrated luminosities of $3000$~fb$^{-1}$ in the future high-luminosity run of the LHC \cite{CMS-DP-2016-064,Savin:2015gda}. At the same time, $\mu^{gg}_{ZZ}$ will see a similar improvement in its measurement, while $\mu^{gg}_{WW}$ will have only modest improvements. So the diphoton decay channel will serve as the best test of the M4DCHMs, and it is highly likely, based on current measurements, that this channel will disfavour both of these models in the coming years.

It should be stressed that these are the predictions we have found by scanning over the most \textit{natural} regions of parameter space, under the definition of naturalness encoded by our priors, where low values of $f$ are favoured. It could be the case that more realistic signal strength predictions occur at higher values of $f$. If the above predictions were to be found false, then, that would only rule out the most natural regions of these models, and not necessarily the models outright. But again, there is a trade-off between accuracy and naturalness that guides the search of CHMs. It would be interesting to investigate the signal strength predictions at larger $f$ in future work.

\begin{figure}[t]
\begin{subfigure}{.5\linewidth}
  \centering
  \includegraphics[width=1.0\textwidth]{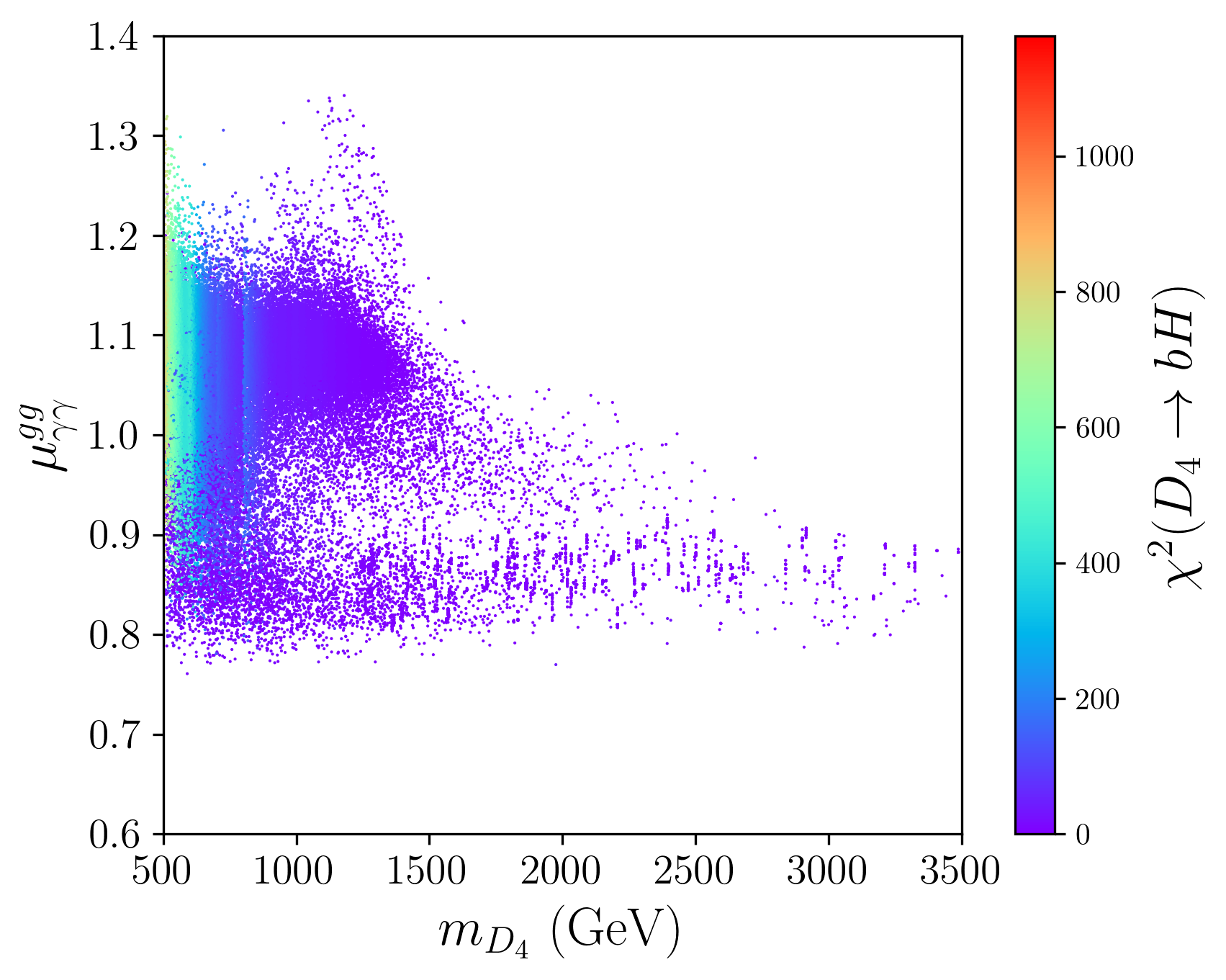}
\end{subfigure}
\begin{subfigure}{.5\linewidth}
  \centering
  \includegraphics[width=1.0\textwidth]{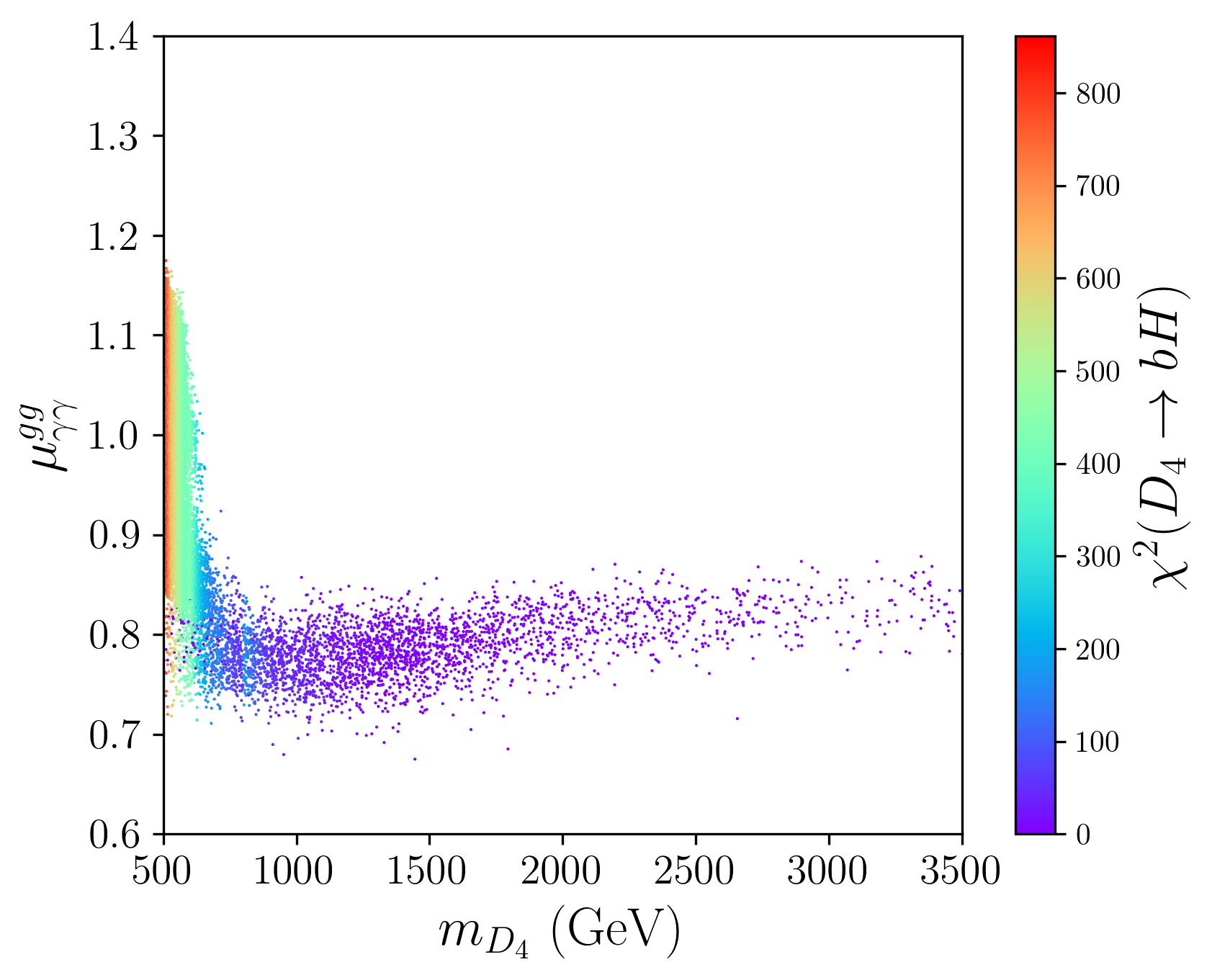}
\end{subfigure}
\caption{Relationship between the Higgs signal strength $\mu^{gg}_{\gamma\gamma}$ and the mass of the lightest down-type resonance in (\textit{left}) the M4DCHM$^{5-5-5}$, and (\textit{right}) the M4DCHM$^{14-14-10}$. Points included here satisfy all constraints, excluding the collider constraints and Higgs signal strength constraints, at the $3 \sigma$ level. The colour indicates the $\chi^{2}$ contribution from the $D \rightarrow bH$ collider bounds.}
\label{fig:mu_gg_gaga_vs_mD4}
\end{figure}

Although the M4DCHM$^{14-14-10}$ and the M4DCHM$^{14-1-10}$ give the same predictions for $\mu^{gg}_{\gamma \gamma}$, they arrive at these predictions for different reasons. In the case of the M4DCHM$^{14-14-10}$, it is the direct collider search constraints, in combination with the other implicit and explicit constraints, that seemingly require $\mu^{gg}_{\gamma \gamma}$ to be significantly less than unity: more realistic signal strengths $\mu^{gg}_{\gamma \gamma} \gtrsim 0.9$ only occur for otherwise valid points when there are low-mass resonances that are excluded by experiment, as shown in \Cref{fig:mu_gg_gaga_vs_mD4}. Such a drastic fall-off of $\mu^{gg}_{\gamma \gamma}$ for larger resonance masses is not seen in the M4DCHM$^{5-5-5}$, allowing it to have more realistic signal strengths. For the M4DCHM$^{14-1-10}$, on the other hand, it is the SM mass constraints that favour the lower values of $\mu^{gg}_{\gamma \gamma}$ - particularly the bottom mass, and the top quark and Higgs masses to a lesser degree. Satisfying all three mass constraints simultaneously, or fixing the correct quark Yukawa couplings and Higgs mass, results in all valid points having $\mu^{gg}_{\gamma \gamma} \lesssim 0.9$. The fact that the SM mass constraints were included in the scans, while the direct search constraints were not, might explain why the M4DCHM$^{14-1-10}$ had more difficulty satisfying the signal strength constraints than the M4DCHM$^{14-14-10}$. But note that if this unfair advantage were to be fixed, the disparity between the evidences of the M4DCHM$^{14-14-10}$ and the M4DCHM$^{14-1-10}$ would only be exacerbated.

\section{Conclusions}
\label{sec:conclusions}

\sloppy We have performed extensive numerical explorations of the M4DCHM$^{5-5-5}$, M4DCHM$^{14-14-10}$, and M4DCHM$^{14-1-10}$ that constitute the first convergent global fits of realistic CHMs. Our fits constrain the parameter spaces into those regions that best reproduce, collectively, the SM masses, the electroweak scale, electroweak precision observables, and various Z boson decay ratios and Higgs signal strengths. Software developed for Refs.~\cite{Niehoff:2015iaa,niehoff2017electroweak} was employed to calculate these observables, which we modified with updated experimental values. The fits are carried out under a Bayesian statistical framework, aided by the advanced nested sampling program \texttt{PolyChord}, which uses slice sampling to explore the spaces in search of the most natural viable regions.

We determined the Bayesian evidence for each model and took a novel approach in measuring the fine-tuning within a composite Higgs framework, quantifying it with the Kullback-Leibler divergence between the posterior and the prior.
With the imposed priors, the M4DCHM$^{5-5-5}$ was found to have a Bayesian evidence many orders of magnitude greater than those of the other models, and both the EWSB scale and the top quark mass are incorporated more naturally.
Tension between the Higgs signal strength and SM mass constraints was the main obstruction to a higher evidence for the M4DCHM$^{14-1-10}$, while a considerable fine-tuning resulted in the M4DCHM$^{14-14-10}$ having the lowest evidence.
We found that these results are strongly prior dependent and we stress that a future conclusive model comparison requires a detailed analysis of the impact of the prior distributions and bounds.
Interestingly, it was found that the posterior occupies surfaces of only ${\sim}7-12$ effective dimensions in the ($18+$ dimensional) parameter spaces of the models.

Each model was found to be capable of satisfying all constraints individually to within $3\sigma$, and their collider phenomenology was analysed in these viable regions. Our results suggest that the fermionic resonances in these models are excluded below ${\sim}1.1$~TeV. We expect the models to have different signals for the various decay modes of the up-type and down-type resonances at the $\sqrt{s} = 13$~TeV LHC, but remarkably all models give the same precise predictions for the exotic fermion decays $Q_{5/3} \rightarrow tW^{+}$ and $Q_{4/3} \rightarrow \bar{b}W^{+}$. These exotic decays constitute the most promising channels for probing these models in future collider searches. The lightest composite vector bosons range in mass from ${\sim}1$~TeV to over $10$~TeV, and neutral bosons on the lighter end of the spectrum can be probed effectively in the $Z_{3} \rightarrow t \bar{t}$ decay channel. Finally, the Higgs signal strengths predicted by the M4DCHMs were analysed, and the models M4DCHM$^{14-14-10}$ and M4DCHM$^{14-1-10}$ were found to have a clear signature: they predict the $gg \rightarrow H \rightarrow \gamma \gamma$ cross section to be between ${\sim}80$\% and ${\sim}90$\% of the value predicted by the SM. This is already in tension with experiment at around the $2\sigma$ level, and has the potential of being strongly ruled out in the future high-luminosity run of the LHC.

\section{Acknowledgements}
% \Wei{update your grant.}
MJW, AGW and WS are funded by the ARC Discovery Project DP180102209, the ARC Centre of Excellence for Dark Matter Particle Physics CE200100008 and the Centre for the Subatomic Structure of Matter (CSSM).
%\acknowledgement

\appendix

\section{$SO(5)$ generators and representations}
\label{SO5_appendix}

Here we give the conventions for $SO(5)$ generators and embeddings we use in this work. This section makes extensive use of local isomorphism $SO(4) \cong SU(2)_L \times SU(2)_R$, since this allows representations of $SO(5) \cong SO(4) \times SO(5)/SO(4)$ to be decomposed into those of $SU(2)_L \times SU(2)_R$, making the embedded fields' quantum numbers easily identifiable.

\subsection*{Generators}
To exploit the local isomorphism, the generators are conveniently split into $(T^a_{L,R})^3_{a=1}$, which generate the subgroup locally isomorphic to $SU(2)_{L,R}$, and the broken generators $(\hat{T}^b)^4_{b=1}$ associated with the $SO(5)/SO(4)$ coset. In the fundamental representation they are given by
\begin{align}
    (T^a_L)_{ij} &= -\frac{i}{2} \left( \epsilon^{abc} \delta^b_i \delta^c_j  + \left( \delta^a_i \delta^4_j - \delta^4_i \delta^a_j \right) \right), \\
    (T^a_R)_{ij} &= -\frac{i}{2} \left( \epsilon^{abc} \delta^b_i \delta^c_j  - \left( \delta^a_i \delta^4_j - \delta^4_i \delta^a_j \right) \right), \\
    (\hat{T}^b)_{ij} &= -\frac{i}{\sqrt{2}} \left( \delta^b_i \delta^5_j - \delta^b_j \delta^5_i \right).
\end{align}
Note this is consistent with our choice of the $SO(4)$-invariant vector $\Phi_0 = (0,0,0,0,1)^\intercal$ that implicitly embeds $SO(4)$ into the top left $4{\times}4$ block of $SO(5)$ matrices.

\subsection*{Field Embeddings}
The representations of $SO(5)$ that we are interested in decompose under $SO(4)$ and subsequently under $SU(2)_L \times SU(2)_R$ as
\begin{align}
    \begin{array}{rll}
    \mathbf{5}   & \rightarrow \mathbf{4} \oplus \mathbf{1} & \rightarrow (\mathbf{2},\mathbf{2}) \oplus (\mathbf{1},\mathbf{1}) ,\\
    \mathbf{10} & \rightarrow \mathbf{6} \oplus \mathbf{4} & \rightarrow (\mathbf{3},\mathbf{1}) \oplus (\mathbf{1},\mathbf{3}) \oplus (\mathbf{2},\mathbf{2}), \\
    \mathbf{14} & \rightarrow \mathbf{9} \oplus \mathbf{4} \oplus \mathbf{1} & \rightarrow (\mathbf{3},\mathbf{3}) \oplus (\mathbf{2},\mathbf{2}) \oplus (\mathbf{1},\mathbf{1}).
    \end{array}
\end{align}

Multiplets in the representation $\mathbf{r}$ will be denoted $\Psi_{\mathbf{r}}$. Representations of $SO(4)$ are embedded inside those of $SO(5)$ as \cite{Montull:2013mla}
\begin{align}
    \Psi_{\mathbf{5}} = \left( \begin{matrix}
        {} \\
        \Psi_{\mathbf{4}}\\
        {} \\
        \hline
        \Psi_{\mathbf{1}}
    \end{matrix} \right), \quad \Psi_{\mathbf{10}}  = \left( \begin{array}{c|c}
        {} \\
        \Psi_{\mathbf{6}} & \frac{1}{\sqrt{2}} \Psi_{\mathbf{4}} \\
        {} \\
        \hline
        -\frac{1}{\sqrt{2}} \Psi_{\mathbf{4}}^\intercal & 0 \\
    \end{array} \right), \quad \Psi_{\mathbf{14}}  = \left( \begin{array}{c|c}
        {} \\
        \Psi_{\mathbf{9}} - \frac{1}{2 \sqrt{5}} \mathbb{1} \Psi_{\mathbf{1}} & \frac{1}{\sqrt{2}} \Psi_{\mathbf{4}} \\
        {} \\
        \hline
        \frac{1}{\sqrt{2}} \Psi_{\mathbf{4}}^\intercal & \frac{2}{\sqrt{5}} \Psi_{\mathbf{1}} \\
    \end{array} \right).
\end{align}
Here we are treating $\Psi_{\mathbf{10,14}}$ as matrices, which are acted upon by the generators through the commutator:
\begin{align}
    T^A \Psi_{\mathbf{10,14}} \equiv [T^A, \Psi_{\mathbf{10,14}}].
\end{align}
The $SO(4)$ multiplets are populated with fields $\Psi^{n_L, n_R}$ having $SU(2)_L \times SU(2)_R$ quantum numbers\footnote{Essentially the $SO(5)$ multiplets are being expressed as linear combinations of eigenvectors of $T^3_{L,R}$ with eigenvalues $n_{L,R}$, and $\Psi^{n_L, n_R}$ are the coefficients.} $(n_L, n_R)$ according to
\begin{align}
    \Psi_{\mathbf{1}} &= \Psi^{0,0}, \nonumber \\[2pt]
    \Psi_{\mathbf{4}} &= \frac{1}{\sqrt{2}} \left( \begin{matrix}
        i \Psi^{-,-} - i \Psi^{+,+} \\
          \Psi^{-,-} +  \Psi^{+,+}\\
        i \Psi^{+,-} -  \Psi^{-,+} \\
        i \Psi^{-,+} -  \Psi^{+,-}
    \end{matrix} \right), \nonumber \\[3pt]
    \Psi_{\mathbf{6}} &=\ \frac{1}{2}\ \left( \begin{matrix}
     0 &\  \Psi^{0,0}_{+}\ &\ -i(\Psi^{\pm,0}_{-} + \Psi^{0,\pm}_{-})\ &\ \Psi^{\pm,0}_{+} - \Psi^{0,\pm}_{+}  \\[2pt]
    {} & 0 &\ \Psi^{\pm,0}_{+} + \Psi^{0,\pm}_{+}\ &\ i(\Psi^{\pm,0}_{-} - \Psi^{0,\pm}_{-}) \\[2pt]
    {} & {} & 0 &\ -i \Psi^{0,0}_{-} \\[2pt]
    {} & {} & {} & 0
    \end{matrix} \right), \\[3pt]
    \Psi_{\mathbf{9}} &=\ \frac{1}{2}\ \left( \begin{matrix}
     \Psi^{\pm,\pm}_{+} - \Psi^{0,0}\ &\  i \Psi^{\pm,\pm}_{+} & \Psi^{\pm,0}_{+}  + i \Psi^{0,\pm}_{+}\ &\ i \Psi^{\pm,0}_{-} - \Psi^{0,\pm}_{+} \\[3pt]
    {} &\ -(\Psi^{\pm,\pm}_{+} + \Psi^{0,0})\ &\ i \Psi^{\pm,0}_{-} - \Psi^{0,\pm}_{-}\ &\ i \Psi^{0,\pm}_{+} - \Psi^{\pm,0}_{+} \\[3pt]
    {} & {} &\ \Psi^{0,0} -  \Psi^{\mp,\pm}_{-}\ &\ i \Psi^{\mp,\pm}_{+} \\[3pt]
    {} & {} & {} &\ \Psi^{0,0} +  \Psi^{\mp,\pm}_{-}
    \end{matrix} \right), \nonumber
\label{eq:SO4_embeddings}
\end{align}
where $\Psi_{\mathbf{6}}$ is antisymmetric and $\Psi_{\mathbf{9}}$ symmetric. The superscript $\pm$ denotes $\pm 1/2$ in $\Psi_{\mathbf{4}}$ and $\pm 1$ in the other multiplets, and we have defined
\begin{align}
\begin{array}{lr}
    \Psi^{0,0}_{\textcolor{blue}{\pm}} = \Psi^{0,0}_{1} \textcolor{blue}{\pm} \Psi^{0,0}_{2}, \qquad & \Psi^{\textcolor{orange}{\pm},0}_{\textcolor{blue}{\pm}} = (\Psi^{\textcolor{orange}{+},0} \textcolor{blue}{\pm} \Psi^{\textcolor{orange}{-},0}) / \sqrt{2}, \\[2pt]
    \Psi^{\textcolor{orange}{\mp},\textcolor{red}{\pm}}_{\textcolor{blue}{\pm}} = \Psi^{\textcolor{orange}{-},\textcolor{red}{+}} \textcolor{blue}{\pm} \Psi^{\textcolor{orange}{+},\textcolor{red}{-}},        \qquad & \Psi^{0,\textcolor{red}{\pm}}_{\textcolor{blue}{\pm}} = (\Psi^{0,\textcolor{red}{+}} \textcolor{blue}{\pm} \Psi^{0,\textcolor{red}{-}}) / \sqrt{2}, \\[2pt]
    \Psi^{\textcolor{orange}{\pm},\textcolor{red}{\pm}}_{\textcolor{blue}{\pm}} = \Psi^{\textcolor{orange}{+},\textcolor{red}{+}} \textcolor{blue}{\pm} \Psi^{\textcolor{orange}{-},\textcolor{red}{-}}. 
\end{array}
\end{align}
Two different fields have $(n_L,n_R) = (0,0)$ in $\Psi_{\mathbf{6}}$.

\section{Correlators}
\label{appendix_correlators}

Here we provide expressions for the correlators $\Pi_{{t,b}_{L,R}}$ and $M_{t,b}$ for each model of the M4DCHM, originally given in Ref.~\cite{Carena}. See \Cref{eqn:form_factor_Lagrangian} for the definitions of these functions. In all models, the form factors are given in terms of the functions
\begin{align}
    & A_{R}(m_{1},m_{2},m_{3},m_{4},\Delta) := \Delta^{2} (m_{1}^{2} m_{2}^{2} + m_{2}^{2} m_{3}^{2} - p^{2} (m_{1}^{2} + m_{2}^{2} + m_{3}^{2} + m_{4}^{2}) + p^{4}), \nonumber\\
    & A_{L}(m_{1},m_{2},m_{3},m_{4},\Delta) := \Delta^{2} m_{1}^{2} m_{4}^{2} + A_{R}(m_{1},m_{2},m_{3},m_{4},\Delta), \nonumber\\
    & A_{M}(m_{1},m_{2},m_{3},m_{4},\Delta_{1},\Delta_{2}) := \Delta_{1} \Delta_{2} m_{1} m_{2} m_{4} (m_{3}^{2} - p^{2}), \nonumber\\
    & B(m_{1},m_{2},m_{3},m_{4},m_{5}) := m_{1}^{2} m_{2}^{2} m_{3}^{2} - p^{2} (m_{1}^{2} m_{2}^{2} + m_{1}^{2} m_{3}^{2} + m_{2}^{2} m_{3}^{2} + m_{2}^{2} m_{5}^{2} + m_{3}^{2} m_{4}^{2}) \nonumber\\
    &\hphantom{B(m_{1},m_{2},m_{3},m_{4},m_{5}) :=} + p^{4} (m_{1}^{2} + m_{2}^{2} + m_{3}^{2} + m_{4}^{2} + m_{5}^{2}) - p^{6},
\end{align}
in Minkowski space.

\subsection{M4DCHM$^{5-5-5}$}

In this model,
\begin{align}
    \Pi_{t_{L}} &= \Pi^{(4)}_{q_{t}} + \Pi^{(4)}_{q_{b}} + \frac{1}{2} \brackets{\Pi^{(1)}_{q_{t}} - \Pi^{(4)}_{q_{t}}} s^{2}_{h}, && M_{t} = \brackets{M^{(1)}_{t} - M^{(4)}_{t}} \sqrt{\frac{1-s^{2}_{h}}{2}} s_{h}, \nonumber\\
    \Pi_{t_{R}} &= \Pi^{(1)}_{t} - \brackets{\Pi^{(1)}_{t} - \Pi^{(4)}_{t}} s^{2}_{h},
\end{align}
where
\begin{align}
    \Pi^{(1)}_{q_{t}} &= \frac{A_{L}(m_{\tilde{t}},0,m_{Y_{t}} + Y_{t},0,\Delta_{tL})}{B(m_{t},m_{\tilde{t}},0,m_{Y_{t}} + Y_{t},0)},  && \Pi^{(4)}_{q_{t}} = \frac{A_{L}(m_{\tilde{t}},0,m_{Y_{t}},0,\Delta_{tL})}{B(m_{t},m_{\tilde{t}},0,m_{Y_{t}},0)}, \nonumber\\
    \Pi^{(1)}_{t} &= \frac{A_{L}(m_{t},0,m_{Y_{t}} + Y_{t},0,\Delta_{tR})}{B(m_{t},m_{\tilde{t}},0,m_{Y_{t}} + Y_{t},0)},  && \Pi^{(4)}_{t} = \frac{A_{L}(m_{t},0,m_{Y_{t}},0,\Delta_{tR})}{B(m_{t},m_{\tilde{t}},0,m_{Y_{t}},0)}, \nonumber\\
    M^{(1)}_{t} &= \frac{A_{M}(m_{t},m_{\tilde{t}},0,m_{Y_{t}} + Y_{t},\Delta_{tL},\Delta_{tR})}{B(m_{t},m_{\tilde{t}},0,m_{Y_{t}} + Y_{t},0)},  && M^{(4)}_{t} = \frac{A_{M}(m_{t},m_{\tilde{t}},0,m_{Y_{t}},\Delta_{tL},\Delta_{tR})}{B(m_{t},m_{\tilde{t}},0,m_{Y_{t}},0)}. \nonumber\\
\end{align}
The form factors for the bottom sector are obtained by interchanging all $t$ and $\tilde{t}$ subscripts with $b$ and $\tilde{b}$.

\subsection{M4DCHM$^{14-14-10}$}

In this model,
\begin{align}
    \Pi_{t_{L}} &= \Pi^{(4)}_{q} - \brackets{\Pi^{(4)}_{q} - \Pi^{(9)}_{q}} \frac{s^{2}_{h}}{2} + \brackets{5\Pi^{(1)}_{q} - 8\Pi^{(4)}_{q} + 3\Pi^{(9)}_{q}} \frac{s^{2}_{h} (1 - s^{2}_{h})}{4}, \nonumber\\
    \Pi_{t_{R}} &= \frac{1}{5} \brackets{8\Pi^{(4)}_{t} - 3 \Pi^{(9)}_{t}} - \frac{3}{2} \brackets{\Pi^{(4)}_{t} - \Pi^{(9)}_{t}} s^{2}_{h} + \brackets{5\Pi^{(1)}_{t} - 8\Pi^{(4)}_{t} + 3\Pi^{(9)}_{t}} \frac{(4 - 5s^{2}_{h})^{2}}{5}, \nonumber\\
    M_{t} &= i \frac{\sqrt{5}}{2} \brackets{ \brackets{M^{(1)}_{t} - M^{(4)}_{t}} - \brackets{5M^{(1)}_{t} - 8M^{(4)}_{t} + 3M^{(9)}_{t}}\frac{s^{2}_{h}}{4} } s_{h} \sqrt{1-s^{2}_{h}},
\end{align}
and
\begin{align}
    \Pi_{b_{L}} &= \Pi^{(4)}_{q} - \brackets{\Pi^{(4)}_{q} - \Pi^{(9)}_{q}} s^{2}_{h}, && M_{b} = - M^{(4)}_{b} s_{h} \sqrt{\frac{1-s^{2}_{h}}{2}}, \nonumber\\
    \Pi_{b_{R}} &= \Pi^{(6)}_{b} + \brackets{\Pi^{(4)}_{b} - \Pi^{(6)}_{b}} \frac{s^{2}_{h}}{2},
\end{align}
where
\begin{align}
    \Pi^{(1)}_{q} &= \frac{A_{L}(m_{t},0,m^{(1)}_{Y_{t}},0,\Delta_{q})}{B(m_{q},m_{t},0,m^{(1)}_{Y_{t}},0)}, && \Pi^{(1)}_{t} = \frac{A_{R}(m_{q},0,m^{(1)}_{Y_{t}},0,\Delta_{t})}{B(m_{q},m_{t},0,m^{(1)}_{Y_{t}},0)}, \nonumber\\
    \Pi^{(4)}_{q} &= \frac{A_{L}(m_{t},m_{b},m^{(4)}_{Y_{t}},Y_{b}/2,\Delta_{q})}{B(m_{q},m_{t},m_{b},m^{(4)}_{Y_{t}},Y_{b}/2)}, && \Pi^{(4)}_{t} = \frac{A_{R}(m_{q},m_{b},m^{(4)}_{Y_{t}},Y_{b}/2,\Delta_{t})}{B(m_{q},m_{t},m_{b},m^{(4)}_{Y_{t}},Y_{b}/2)}, \nonumber\\
    \Pi^{(9)}_{q} &= \frac{A_{L}(m_{t},0,m_{Y_{t}},0,\Delta_{q})}{B(m_{q},m_{t},0,m_{Y_{t}},0)}, && \Pi^{(9)}_{t} = \frac{A_{R}(m_{q},0,m_{Y_{t}},0,\Delta_{t})}{B(m_{q},m_{t},0,m_{Y_{t}},0)}, \nonumber\\
    \Pi^{(4)}_{b} &= \frac{A_{R}(m_{q},m_{t},Y_{b}/2,m^{(4)}_{Y_{t}},\Delta_{b})}{B(m_{q},m_{t},m_{b},m^{(4)}_{Y_{t}},Y_{b}/2)}, && M^{(1)}_{t} = \frac{A_{M}(m_{q},m_{t},0,m^{(1)}_{Y_{t}},\Delta_{q},\Delta_{t})}{B(m_{q},m_{t},0,m^{(1)}_{Y_{t}},0)}, \nonumber\\
    \Pi^{(6)}_{b} &= \frac{A_{R}(m_{q},0,0,0,\Delta_{b})}{B(m_{q},m_{b},0,0,0)}, && M^{(4)}_{t} = \frac{A_{M}(m_{q},m_{t},m_{b},m^{(4)}_{Y_{t}},\Delta_{q},\Delta_{t})}{B(m_{q},m_{t},m_{b},m^{(4)}_{Y_{t}},Y_{b}/2)}, \nonumber\\
    M^{(4)}_{b} &= - i \frac{A_{M}(m_{q},m_{b},m_{t},Y_{b}/2,\Delta_{q},\Delta_{b})}{B(m_{q},m_{t},m_{b},m^{(4)}_{Y_{t}},Y_{b}/2)}, && M^{(9)}_{t} = \frac{A_{M}(m_{q},m_{t},0,m_{Y_{t}},\Delta_{q},\Delta_{t})}{B(m_{q},m_{t},0,m_{Y_{t}},0)},
\end{align}
having defined
\begin{align}
    m^{(1)}_{Y_{t}} &:= m_{Y_{t}} + 4Y_{t}/5 + 4\tilde{Y}_{t}/5, \nonumber\\
    m^{(4)}_{Y_{t}} &:= m_{Y_{t}} + \frac{1}{2} Y_{t},
\end{align}
for convenience.

\subsection{M4DCHM$^{14-1-10}$}

In this model,
\begin{align}
    \Pi_{t_{L}} &= \Pi^{(4)}_{q} - \brackets{\Pi^{(4)}_{q} - \Pi^{(9)}_{q}} \frac{s^{2}_{h}}{2} + \brackets{5\Pi^{(1)}_{q} - 8\Pi^{(4)}_{q} + 3\Pi^{(9)}_{q}} \frac{s^{2}_{h} (1 - s^{2}_{h})}{4}, \nonumber\\
    \Pi_{t_{R}} &= \Pi^{(1)}_{t}, \nonumber\\
    M_{t} &= - \frac{\sqrt{5}}{4} M^{(1)}_{t} s_{h},
\end{align}
and
\begin{align}
    \Pi_{b_{L}} &= \Pi^{(4)}_{q} - \brackets{\Pi^{(4)}_{q} - \Pi^{(9)}_{q}} s^{2}_{h}, && M_{b} = - i M^{(4)}_{b} s_{h} \sqrt{\frac{1-s^{2}_{h}}{2}}, \nonumber\\
    \Pi_{b_{R}} &= \Pi^{(6)}_{b} + \brackets{\Pi^{(4)}_{b} - \Pi^{(6)}_{b}} \frac{s^{2}_{h}}{2},
\end{align}
where
\begin{align}
    \Pi^{(1)}_{q} &= \frac{A_{L}(m_{t},0,2Y_{t}/\sqrt{5},0,\Delta_{q})}{B(m_{q},m_{t},0,2Y_{t}/\sqrt{5},0)},  && \Pi^{(1)}_{t} = \frac{A_{R}(m_{q},0,2Y_{t}/\sqrt{5},0,\Delta_{t})}{B(m_{q},m_{t},0,2Y_{t}/\sqrt{5},0)}, \nonumber\\
    \Pi^{(4)}_{q} &= \frac{A_{L}(0,m_{b},0,Y_{b}/2,\Delta_{q})}{B(m_{q},0,m_{b},0,Y_{b}/2)}, && \Pi^{(4)}_{b} = \frac{A_{R}(m_{q},0,Y_{b}/2,0,\Delta_{b})}{B(m_{q},0,m_{b},0,Y_{b}/2)}, \nonumber\\
    \Pi^{(9)}_{q} &= \frac{A_{L}(0,0,0,0,\Delta_{q})}{B(m_{q},0,0,0,0)}, && \Pi^{(6)}_{b} = \frac{A_{R}(0,0,0,0,\Delta_{b})}{B(0,m_{b},0,0,0)}, \nonumber\\
    M^{(1)}_{t} &= - \frac{A_{M}(m_{q},m_{t},0,2Y_{t}/\sqrt{5},\Delta_{q},\Delta_{t})}{B(m_{q},m_{t},0,2Y_{t}/\sqrt{5},0)}, && M^{(4)}_{b} = - i \frac{A_{M}(m_{q},m_{b},0,Y_{b}/2,\Delta_{q},\Delta_{b})}{B(m_{q},0,m_{b},0,Y_{b}/2)}.
\end{align}

\section{Mass matrices}
\label{appendix_mass_matrices}

Here we give the explicit mass matrices for each model of the M4DCHM. They are given in the Site 0 holographic gauge, in bases consisting of the fields with definite $SU(2)_L \times SU(2)_R$ quantum numbers presented in \Cref{SO5_appendix}. Only particles of the same electric charge can mix, so the mass matrices are separated according to the particles' charges.

\subsection{Boson sector}
The mixing terms between the gauge bosons are independent of the fermion representations, so they are the same across all models considered in this work. Their mixings are represented by symmetric matrices, whose singular values are the squared masses of the resonances.

Expansion of the Lagrangian in \Cref{eq:boson_lagrangian} reveals the $SU(3)$ resonances only mix among themselves, through the matrix
\begin{align}
 M^{2}_{\text{gluon}} = \left(\begin{array}{c|cc}
    {} & G_{\mu}^{0} & \rho_{G_{\mu}}\\[2pt]
    \hline\\\\[-4.5\medskipamount]
    G_{\mu}^{0} & \frac{1}{2} (g^0_{s})^2 f^2_G & -\frac{1}{2} g^{0}_{s} g_G f^2_G \\[2pt]
    \rho_{G_{\mu}} &  {} & \frac{1}{2} g^2_{G} f^2_G \end{array}\right).
\end{align}
Note one singular value of this matrix is 0, as expected for the massless SM gluons.

Of the remaining bosons, four are charged and seven are uncharged. The charged bosons $\{W^{0^{\pm}}_{\mu},  \rho_{L_{\mu}}^{\pm},  \rho_{R_{\mu}}^{\pm}, \mathfrak{a}^{\pm}_{\mu}\}$ are linear combinations of the gauge fields that come paired with the generators (using our choice of generators) given by the form
\begin{align}
    X^{\pm}_{\mu} := \frac{X^{1}_{\mu} \mp i X^{2}_{\mu}}{\sqrt{2}}.
\end{align}
Their mass matrix is given by \Cref{eq:charged_bosons}, and the uncharged bosons have the mass matrix \Cref{eq:neutral_bosons}.\\

\begin{sideways}
\parbox{\textheight}{
\begin{align}
 M^2_{\text{charged}}& =  \nonumber \\
 &\left(\begin{array}{c|cccc}
    {} & W_{\mu}^{0^{+}} & \rho_{L_{\mu}}^{+} & \rho_{R_{\mu}}^{+} & \mathfrak{a}^{+}_{\mu} \\
    \hline\\\\[-4.5\medskipamount]
    W^{0^{-}\mu} & {\frac{1}{2} g_{0}^{2} f_{1}^{2}} & {-\frac{1}{2} g_{0} g_{\rho} f_{1}^{2} \cos ^{2}\left(\frac{h}{2 f}\right)} & {-\frac{1}{2} g_{0} g_{\rho} f_{1}^{2} \sin ^{2}\left(\frac{h}{2 f}\right)} & {-\frac{1}{2 \sqrt{2}} g_{0} g_{\rho} f_{1}^{2} \sin \left(\frac{h}{f}\right)} \\[2pt]
    {\rho_{L}^{- \mu}} & {} & {\frac{1}{2} g_{\rho}^{2} f_{1}^{2}} & {0} & {0} \\[2pt]
    {\rho_{R}^{- \mu}} & {} & {} & {\frac{1}{2} g_{\rho}^{2} f_{1}^{2}} & {0} \\[2pt]
    \mathfrak{a}^{- \mu} & {} & {} & {} & {\frac{1}{2} g_{\rho}^{2} \frac{f_{1}^{4}}{f_{1}^{2}-f^{2}}}\end{array}\right),
\label{eq:charged_bosons}\\[36pt]
M^2_{\text{neutral}}& =  \nonumber \\
 &\left( \begin{array}{c|ccccccc}
    {} & W_{\mu}^{0^{3}} & B^0_{\mu} & \rho^3_{L_{\mu}} & \rho^3_{R_{\mu}} & \mathfrak{a}^{3}_{\mu} & \rho_{X_{\mu}} & \mathfrak{a}^{4}_{\mu} \\
    \hline\\\\[-4.5\medskipamount]
    W^{0^{3}\mu} & {\frac{1}{2} g_{0}^{2} f_{1}^{2}} & 0 & {-\frac{1}{2} g_{0} g_{\rho} f_{1}^{2} \cos ^{2}\left(\frac{h}{2 f}\right)} & {-\frac{1}{2} g_{0} g_{\rho} f_{1}^{2} \sin ^{2}\left(\frac{h}{2 f}\right)} & {-\frac{1}{2 \sqrt{2}} g_{0} g_{\rho} f_{1}^{2} \sin \left(\frac{h}{f}\right)} & 0 & 0 \\[2pt]
    B^{0^{\mu}} & {} & {\frac{1}{2} g^{\prime 2}_{0}(f_{1}^{2}+f_{x}^{2})} & {-\frac{1}{2} g'_{0} g_{\rho} f_{1}^{2} \sin ^{2}\left(\frac{h}{2 f}\right)} & {-\frac{1}{2} g'_{0} g_{\rho} f_{1}^{2} \cos ^{2}\left(\frac{h}{2 f}\right)} & {\frac{1}{2 \sqrt{2}} g'_{0} g_{\rho} f_{1}^{2} \sin \left(\frac{h}{f}\right)} & -\frac{1}{2} g'_0 g_X f^2_X & {0} \\[2pt]
    \rho^{3^{\mu}}_{L} & {} & {} & {\frac{1}{2} g_{\rho}^{2} f_{1}^{2}} & {0} & {0} & {0} & {0} \\[2pt]
    \rho^{3^{\mu}}_{R} & {} & {} & {} & {\frac{1}{2} g_{\rho}^{2} f_{1}^{2}} & {0} & {0} & {0} \\[2pt]
    \mathfrak{a}^{3^{\mu}} & {} & {} & {} & {} & \frac{1}{2} g_{\rho}^2 \frac{f^4_1}{f^2_1 - f^2} & {0} & {0} \\[2pt]
    \rho_{X}^{\mu}  & {} & {} & {} & {} & {} & \frac{1}{2} g_{X}^2 f_{X}^2 & {0} \\[2pt]
    \mathfrak{a}^{4^{\mu}} & {} & {} & {} & {} & {} & {} & \frac{1}{2} g_{\rho}^2 \frac{f^4_1}{f^2_1 - f^2}\end{array}\right).
\label{eq:neutral_bosons}
\end{align}}
\end{sideways}

\subsection{Fermion sector}

The Lagrangians \Cref{eq:fundamental_fermion_lagrangian_5-5-5,eq:fundamental_fermion_lagrangian_14-14-10,eq:fundamental_fermion_lagrangian_14-1-10} lead to the quark partner mass matrices for each model below. Subscripts denote the matrices for the up-type (U) and down-type (D) fields, as well as the fields of exotic charges $4/3$, $5/3$, and $8/3$. The subscripts/tildes of the fields match those of the multiplets in which they are contained. Where necessary to avoid ambiguity, subscripts denote the representations of $SO(5)$ or $SO(4)$ in which the fields reside. Fields may share a label with those in other matrices due to the different definitions of $\pm$ in superscripts for different $SO(4)$ representations (see discussion after \Cref{eq:SO4_embeddings}), but the electric charges distinguish them and remove ambiguity about the representation each comes from.

\noindent To save space in writing the matrices, we define the quantities
\begin{align}
    s_{xh} = \sin \brackets{x\frac{h}{f}},\quad c_{xh} = \cos \brackets{x\frac{h}{f}} \quad \text{for } x \in \mathbb{R},
\end{align}
and
\begin{align}
    c_{\pm} = \frac{c_{h} \pm c_{2h}}{2},\ \tilde{c} = \frac{3 + 5 c_{2h}}{8},  \text{ and } \tilde{s} = \frac{\sqrt{5}}{4} s_{2h}.
\end{align}

\subsubsection{M4DCHM$^{5-5-5}$}

A caveat: in this model a different convention was used for the field embeddings than was presented in \Cref{SO5_appendix}. The fields in $\Psi_{\mathbf{4}}$ in \Cref{SO5_appendix} have been redefined here as $\Psi^{-,-} \rightarrow -i \Psi^{-,-}$, $\Psi^{+,+} \rightarrow i \Psi^{+,+}$, and $\Psi^{+,-} \rightarrow -i \Psi^{+,-}$.

\begin{align}
M^{5-5-5}_{4/3}& = \left(\begin{array}{c|cc}
    {} & \Psi^{-,-}_{bR} & \tilde{\Psi}^{-,-}_{bR}  \\[2pt]
    \hline\\\\[-4.5\medskipamount]
    \bar{\Psi}^{-,-}_{bL} & m_b & m_{Y_b}  \\
    \bar{\tilde{\Psi}}^{-,-}_{bL} &  0 &  m_{\tilde{b}} \end{array}\right),\\[20pt]
\quad M^{5-5-5}_{5/3}& = \left(\begin{array}{c|cc}
    {} & \Psi^{+,+}_{tR} & \tilde{\Psi}^{+,+}_{tR}  \\[2pt]
    \hline\\\\[-4.5\medskipamount]
    \bar{\Psi}^{+,+}_{tL} & m_t & m_{Y_t}  \\
    \bar{\tilde{\Psi}}^{+,+}_{tL} &  0 &  m_{\tilde{t}} \end{array}\right),
\end{align}

\vfill

\begin{sideways}
\parbox{\textheight}{
\begin{align}
M^{5-5-5}_{U}& = \left( \begin{array}{c|ccccccccc}
    {} & t^0_R & \Psi^{+,-}_{tR} & \tilde{\Psi}^{+,-}_{tR} & \Psi^{-,+}_{tR} & \tilde{\Psi}^{-,+}_{tR}  & \Psi^{+,+}_{bR} & \tilde{\Psi}^{+,+}_{bR} & \Psi^{0,0}_{tR} & \tilde{\Psi}^{0,0}_{tR} \\[2pt]
    \hline\\\\[-4.5\medskipamount]
    \bar{t}^0_L & 0 & - c^2_{h/2} \Delta_{tL} & 0 & s^2_{h/2} \Delta_{tL} & 0 &  - \Delta_{bL} & 0 & \frac{i}{\sqrt{2}} s_{h} \Delta_{tL} & 0 \\
    \bar{\Psi}^{+,-}_{tL} & 0 & m_t & m_{Y_t} & 0 & 0 & 0 & 0 & 0 & 0 \\
    \bar{\tilde{\Psi}}^{+,-}_{tL} & - \frac{i}{\sqrt{2}} s_{h} \Delta^{\dagger}_{tR} & 0 & m_{\tilde{t}} & 0 & 0 & 0 & 0 & 0 & 0 \\
    \bar{\Psi}^{-,+}_{tL} & 0 & 0 & 0 & m_t & m_{Y_t} & 0 & 0 & 0 & 0 \\
    \bar{\tilde{\Psi}}^{-,+}_{tL} &  - \frac{i}{\sqrt{2}} s_{h} \Delta^{\dagger}_{tR} & 0 & 0 & 0 & m_{\tilde{t}} & 0 & 0 & 0 & 0 \\
    \bar{\Psi}^{+,+}_{bL} & 0 & 0 & 0 & 0 & 0 & m_b & m_{Y_b} & 0 & 0 \\
    \bar{\tilde{\Psi}}^{+,+}_{bL} & 0 & 0 & 0 & 0 & 0 & 0 & m_{\tilde{b}} & 0 & 0 \\
    \bar{\Psi}^{0,0}_{tL} & 0 & 0 & 0 & 0 & 0 & 0 & 0 & m_t & m_{Y_t} + Y_t \\
    \bar{\tilde{\Psi}}^{0,0}_{tL} & - c_{h} \Delta^{\dagger}_{tR} & 0 & 0 & 0 & 0 & 0 & 0 & 0 & m_{\tilde{t}} \\
    \end{array}\right), \\[20pt]
M^{5-5-5}_{D}& = \left( \begin{array}{c|ccccccccc}
    {} & b^0_R & \Psi^{+,-}_{bR} & \tilde{\Psi}^{+,-}_{bR} & \Psi^{-,+}_{bR} & \tilde{\Psi}^{-,+}_{bR}  & \Psi^{-,-}_{tR} & \tilde{\Psi}^{-,-}_{tR} & \Psi^{0,0}_{bR} & \tilde{\Psi}^{0,0}_{bR} \\[2pt]
    \hline\\\\[-4.5\medskipamount]
    \bar{b}^0_L & 0 & s^{2}_{h/2} \Delta_{bL} & 0 & - c^{2}_{h/2} \Delta_{bL} & 0 &  - \Delta_{tL} & 0 & \frac{i}{\sqrt{2}} s_{h} \Delta_{bL} & 0 \\
    \bar{\Psi}^{+,-}_{bL} & 0 & m_b & m_{Y_b} & 0 & 0 & 0 & 0 & 0 & 0 \\
    \bar{\tilde{\Psi}}^{+,-}_{bL} & - \frac{i}{\sqrt{2}} s_{h} \Delta^{\dagger}_{bR} & 0 & m_{\tilde{b}} & 0 & 0 & 0 & 0 & 0 & 0 \\
    \bar{\Psi}^{-,+}_{bL} & 0 & 0 & 0 & m_b & m_{Y_b} & 0 & 0 & 0 & 0 \\
    \bar{\tilde{\Psi}}^{-,+}_{dL} &  - \frac{i}{\sqrt{2}} s_{h} \Delta^{\dagger}_{bR} & 0 & 0 & 0 & m_{\tilde{b}} & 0 & 0 & 0 & 0 \\
    \bar{\Psi}^{-,-}_{tL} & 0 & 0 & 0 & 0 & 0 & m_t & m_{Y_t} & 0 & 0 \\
    \bar{\tilde{\Psi}}^{-,-}_{tL} & 0 & 0 & 0 & 0 & 0 & 0 & m_{\tilde{t}} & 0 & 0 \\
    \bar{\Psi}^{0,0}_{bL} & 0 & 0 & 0 & 0 & 0 & 0 & 0 & m_b & m_{Y_b} + Y_b \\
    \bar{\tilde{\Psi}}^{0,0}_{bL} & - c_{h} \Delta^{\dagger}_{bR} & 0 & 0 & 0 & 0 & 0 & 0 & 0 & m_{\tilde{b}}
    \end{array}\right).
\end{align}}
\end{sideways}

\subsubsection{M4DCHM$^{14-14-10}$}

\setlength{\arraycolsep}{1.3pt}
\begin{align}
&M^{14-14-10}_{4/3} = \left(\begin{array}{c|cc}
    {} & \Psi^{-,-}_L & \tilde{\Psi}^{-,-}_{\mathbf{14}_L}  \\[2pt]
    \hline\\\\[-4.5\medskipamount]
    \bar{\Psi}^{-,-}_R & m_q & m_{Y_t} \\[2pt]
    \bar{\tilde{\Psi}}^{-,-}_{\mathbf{14}_R} & 0 & m_t \end{array}\right), \quad M^{14-14-10}_{8/3} = \left(\begin{array}{c|cc}
    {} & \Psi^{+,+}_L & \tilde{\Psi}^{+,+}_{\mathbf{14}_L}  \\[2pt]
    \hline\\\\[-4.5\medskipamount]
    \bar{\Psi}^{+,+}_R & m_q & m_{Y_t} \\[2pt]
    \bar{\tilde{\Psi}}^{+,+}_{\mathbf{14}_R} & 0 & m_t \end{array}\right),\\[20pt]
&M^{14-14-10}_{5/3} = \left(\begin{array}{c|ccccccccc}
    {} & \tilde{\Psi}^{+,+}_{\mathbf{10}_L} & \Psi^{+,+}_L & \tilde{\Psi}^{+,+}_{\mathbf{14}_L} & \Psi^{+,0}_L & \tilde{\Psi}^{+,0}_{\mathbf{14}_L} & \Psi^{0,+}_L & \tilde{\Psi}^{0,+}_{\mathbf{14}_L} & \tilde{\Psi}^{+,0}_{\mathbf{10}_L} & \tilde{\Psi}^{0,+}_{\mathbf{10}_L}  \\[2pt]
    \hline\\\\[-4.5\medskipamount]
    \bar{\tilde{\Psi}}^{+,+}_{\mathbf{10}_R} & m_b & 0 & 0 & 0 & 0 & 0 & 0 & 0 & 0 \\[3pt]
    \bar{\Psi}^{+,+}_R & \frac{1}{2} Y_b & m_q & m_{Y_t} + \frac{1}{2} Y_t & 0 & 0 & 0 & 0 & 0 & 0 \\[3pt]
    \bar{\tilde{\Psi}}^{+,+}_{\mathbf{14}_R} & 0 & 0 & m_t & 0 & 0 & 0 & 0 & 0 & 0 \\[3pt]
    \bar{\Psi}^{+,0}_R & 0 & 0 & 0 & m_q & m_{Y_t} & 0 & 0 & 0 & 0 \\[3pt]
    \bar{\tilde{\Psi}}^{+,0}_{\mathbf{14}_R} & 0 & 0 & 0 & 0 & m_t & 0 & 0 & 0 & 0 \\[3pt]
    \bar{\Psi}^{0,+}_R & 0 & 0 & 0 & 0 & 0 & m_q & m_{Y_t} & 0 & 0 \\[3pt]
    \bar{\tilde{\Psi}}^{0,+}_{\mathbf{14}_R} & 0 & 0 & 0 & 0 & 0 & 0 & m_t & 0 & 0 \\[3pt]
    \bar{\tilde{\Psi}}^{+,0}_{\mathbf{10}_R} & 0 & 0 & 0 & 0 & 0 & 0 & 0 & m_b & 0 \\[3pt]
    \bar{\tilde{\Psi}}^{0,+}_{\mathbf{10}_R} & 0 & 0 & 0 & 0 & 0 & 0 & 0 & 0 & m_b \end{array}\right),\\[20pt]
&M^{14-14-10}_{D} = \nonumber\\
&\quad \left(\begin{array}{c|cccccccccc}
    {} & b^0_L & \tilde{\Psi}^{-,-}_{\mathbf{10}_L} & \Psi^{-,-}_L & \tilde{\Psi}^{-,-}_{\mathbf{14}_L} & \Psi^{-,0}_L & \tilde{\Psi}^{-,0}_{\mathbf{14}_L} & \Psi^{0,-}_L & \tilde{\Psi}^{0,-}_{\mathbf{14}_L} & \tilde{\Psi}^{-,0}_{\mathbf{10}_L} & \tilde{\Psi}^{0,-}_{\mathbf{10}_L} \\[3pt]
    \hline\\\\[-4.5\medskipamount]
     \bar{b}^0_R & 0 & 0 & - c_{h} \Delta_q & 0 & -\frac{1}{\sqrt{2}} s_{h} \Delta_q & 0 & \frac{i}{\sqrt{2}} s_{h} \Delta_q & 0 & 0 & 0 \\[3pt]
     \bar{\tilde{\Psi}}^{-,-}_{\mathbf{10}_R} & -\frac{i}{\sqrt{2}} s_{h} \Delta^{\dagger}_b & m_b & 0 & 0 & 0 & 0 & 0 & 0 & 0 & 0 \\[3pt]
     \bar{\Psi}^{-,-}_R & 0 & \frac{1}{2} Y_b & m_q & m_{Y_t} + \frac{1}{2} Y_t & 0 & 0 & 0 & 0 & 0 & 0 \\[2pt]
     \bar{\tilde{\Psi}}^{-,-}_{\mathbf{14}_R} & 0 & 0 & 0 & m_t & 0 & 0 & 0 & 0 & 0 & 0 \\[3pt]
     \bar{\Psi}^{-,0}_R & 0 & 0 & 0 & 0 & m_q & m_{Y_t} & 0 & 0 & 0 & 0 \\[3pt]
     \bar{\tilde{\Psi}}^{-,0}_{\mathbf{14}_R} & 0 & 0 & 0 & 0 & 0 & m_t & 0 & 0 & 0 & 0 \\[3pt]
     \bar{\Psi}^{0,-}_R & 0 & 0 & 0 & 0 & 0 & 0 & m_q & m_{Y_t} & 0 & 0 \\[3pt]
     \bar{\tilde{\Psi}}^{0,-}_{\mathbf{14}_R} & 0 & 0 & 0 & 0 & 0 & 0 & 0 & m_t & 0 & 0 \\[3pt]
     \bar{\tilde{\Psi}}^{-,0}_{\mathbf{10}_R} & -s^{2}_{h/2}\Delta^{\dagger}_b & 0 & 0 & 0 & 0 & 0 & 0 & 0 & m_b & 0 \\[3pt]
     \bar{\tilde{\Psi}}^{0,-}_{\mathbf{10}_R} & -c^{2}_{h/2}\Delta^{\dagger}_b & 0 & 0 & 0 & 0 & 0 & 0 & 0 & 0 & m_b \end{array}\right),
\end{align}

\raggedbottom

\begin{sideways}
\parbox{\textheight}{
\setlength{\arraycolsep}{0.2pt}
\begin{align}
&M^{14-14-10}_{U} = \nonumber \\[6pt]
&\left(\begin{array}{c|ccccccccccccccccc}
    {} & t^0_L & \Psi^{0,0}_{\mathbf{1}_L} & \tilde{\Psi}^{0,0}_{\mathbf{1}_L} & \tilde{\Psi}^{+,-}_{\mathbf{10}_L} & \Psi^{+,-}_{\mathbf{4}_L} & \tilde{\Psi}^{+,-}_{\mathbf{14}_L} & \tilde{\Psi}^{-,+}_{\mathbf{10}_L} & \Psi^{-,+}_{\mathbf{4}_L} & \tilde{\Psi}^{-,+}_{\mathbf{14}_L} & \Psi^{-,+}_{\mathbf{9}_L} & \tilde{\Psi}^{-,+}_{\mathbf{9}_L} & \Psi^{0,0}_{\mathbf{9}_L} & \tilde{\Psi}^{0,0}_{\mathbf{9}_L} & \Psi^{+,-}_{\mathbf{9}_L} & \tilde{\Psi}^{+,-}_{\mathbf{9}_L} & \tilde{\Psi}^{0,0}_{1_{\mathbf{10}_L}} & \tilde{\Psi}^{0,0}_{2_{\mathbf{10}_L}} \\[3pt]
    \hline\\\\[-4.5\medskipamount]
     \bar{t}^0_R & 0 & \tilde{s} \Delta_q & 0 & 0 & - c_{+} \Delta_q & 0 & 0 & -i c_{-} \Delta_q & 0 & s_{h} s^{2}_{h/2} \Delta_q & 0 & - \frac{s_{2h}}{4} \Delta_q & 0 & s_{h} c^{2}_{h/2} \Delta_q & 0 & 0 & 0 \\[3pt]
     \bar{\Psi}^{0,0}_{\mathbf{1}_R} & 0 & m_q & m_{Y_t} + \frac{4}{5} (Y_t + \tilde{Y}_t) & 0 & 0 & 0 & 0 & 0 & 0 & 0 & 0 & 0 & 0 & 0 & 0 & 0 & 0 \\[3pt]
     \bar{\tilde{\Psi}}^{0,0}_{\mathbf{1}_R} & - \tilde{c} \Delta^{\dagger}_t & 0 & m_t & 0 & 0 & 0 & 0 & 0 & 0 & 0 & 0 & 0 & 0 & 0 & 0 & 0 & 0 \\[3pt]
     \bar{\tilde{\Psi}}^{+,-}_{\mathbf{10}_R} & 0 & 0 & 0 & m_b & 0 & 0 & 0 & 0 & 0 & 0 & 0 & 0 & 0 & 0 & 0 & 0 & 0 \\[3pt]
     \bar{\Psi}^{+,-}_{\mathbf{4}_R} & 0 & 0 & 0 & \frac{1}{2} Y_b & m_q & m_{Y_t} + \frac{1}{2} Y_t & 0 & 0 & 0 & 0 & 0 & 0 & 0 & 0 & 0 & 0 & 0 \\[3pt]
     \bar{\tilde{\Psi}}^{+,-}_{\mathbf{14}_R} & - \tilde{s} \Delta^{\dagger}_t & 0 & 0 & 0 & 0 & m_t & 0 & 0 & 0 & 0 & 0 & 0 & 0 & 0 & 0 & 0 & 0 \\[3pt]
     \bar{\tilde{\Psi}}^{-,+}_{\mathbf{10}_R} & 0 & 0 & 0 & 0 & 0 & 0 & m_b & 0 & 0 & 0 & 0 & 0 & 0 & 0 & 0 & 0 & 0 \\[3pt]
     \bar{\Psi}^{-,+}_{\mathbf{4}_R} & 0 & 0 & 0 & 0 & 0 & 0 & \frac{1}{2} Y_b & m_q & m_{Y_t} + \frac{1}{2} Y_t & 0 & 0 & 0 & 0 & 0 & 0 & 0 & 0 \\[3pt]
     \bar{\tilde{\Psi}}^{-,+}_{\mathbf{14}_R} & -i \tilde{s} \Delta^{\dagger}_t & 0 & 0 & 0 & 0 & 0 & 0 & 0 & m_t & 0 & 0 & 0 & 0 & 0 & 0 & 0 & 0 \\[3pt]
     \bar{\Psi}^{-,+}_{\mathbf{9}_R} & 0 & 0 & 0 & 0 & 0 & 0 & 0 & 0 & 0 & m_q & m_{Y_t} & 0 & 0 & 0 & 0 & 0 & 0 \\[3pt]
     \bar{\tilde{\Psi}}^{-,+}_{\mathbf{9}_R} & - \frac{\sqrt{5}}{4} s^{2}_{h} \Delta^{\dagger}_t & 0 & 0 & 0 & 0 & 0 & 0 & 0 & 0 & 0 & m_t & 0 & 0 & 0 & 0 & 0 & 0 \\[3pt]
     \bar{\Psi}^{0,0}_{\mathbf{9}_R} & 0 & 0 & 0 & 0 & 0 & 0 & 0 & 0 & 0 & 0 & 0 & m_q & m_{Y_t} & 0 & 0 & 0 & 0 \\[3pt]
     \bar{\tilde{\Psi}}^{0,0}_{\mathbf{9}_R} & - \frac{\sqrt{5}}{4} s^{2}_{h} \Delta^{\dagger}_t & 0 & 0 & 0 & 0 & 0 & 0 & 0 & 0 & 0 & 0 & 0 & m_t & 0 & 0 & 0 & 0 \\[3pt]
     \bar{\Psi}^{+,-}_{\mathbf{9}_R} & 0 & 0 & 0 & 0 & 0 & 0 & 0 & 0 & 0 & 0 & 0 & 0 & 0 & m_q & m_{Y_t} & 0 & 0 \\[3pt]
     \bar{\tilde{\Psi}}^{+,-}_{\mathbf{9}_R} & \frac{\sqrt{5}}{4} s^{2}_{h} \Delta^{\dagger}_t & 0 & 0 & 0 & 0 & 0 & 0 & 0 & 0 & 0 & 0 & 0 & 0 & 0 & m_t & 0 & 0 \\[3pt]
     \bar{\tilde{\Psi}}^{0,0}_{1_{\mathbf{10}_R}} & 0 & 0 & 0 & 0 & 0 & 0 & 0 & 0 & 0 & 0 & 0 & 0 & 0 & 0 & 0 & m_b & 0 \\[3pt]
     \bar{\tilde{\Psi}}^{0,0}_{2_{\mathbf{10}_R}} & 0 & 0 & 0 & 0 & 0 & 0 & 0 & 0 & 0 & 0 & 0 & 0 & 0 & 0 & 0 & 0 & m_b \end{array}\right).
\end{align}}
\end{sideways}

\subsubsection{M4DCHM$^{14-1-10}$}

\setlength{\arraycolsep}{1.5pt}
\begin{align}
&M^{14-1-10}_{4/3} = \left(\begin{array}{c|c}
    {} & \Psi^{-,-}_L  \\[1.5pt]
    \hline\\\\[-4.5\medskipamount]
    \bar{\Psi}^{-,-}_R &  m_q \end{array}\right), \quad M^{14-1-10}_{8/3} = \left(\begin{array}{c|c}
    {} & \Psi^{+,+}_L  \\[1.5pt]
    \hline\\\\[-4.5\medskipamount]
    \bar{\Psi}^{+,+}_R &  m_q \end{array}\right),\\[20pt]
&M^{14-1-10}_{5/3} = \left(\begin{array}{c|cccccc}
    {} & \tilde{\Psi}^{+,+}_L & \Psi^{+,+}_L & \Psi^{+,0}_L & \Psi^{0,+}_L & \tilde{\Psi}^{+,0}_L & \tilde{\Psi}^{0,+}_L  \\[2pt]
    \hline\\\\[-4.5\medskipamount]
    \bar{\tilde{\Psi}}^{+,+}_R & m_b & 0 & 0 & 0 & 0 & 0 \\[1pt]
    \bar{\Psi}^{+,+}_R & \frac{1}{2} Y_b & m_q & 0 & 0 & 0 & 0 \\[1pt]
    \bar{\Psi}^{+,0}_R & 0 & 0 & m_q & 0 & 0 & 0 \\[1pt]
    \bar{\Psi}^{0,+}_R & 0 & 0 & 0 & m_q & 0 & 0 \\[1pt]
    \bar{\tilde{\Psi}}^{+,0}_R & 0 & 0 & 0 & 0 & m_b & 0 \\[1pt]
    \bar{\tilde{\Psi}}^{0,+}_R & 0 & 0 & 0 & 0 & 0 & m_b \end{array}\right),\\[20pt]
&M^{14-1-10}_{D} = \left(\begin{array}{c|ccccccc}
    {} & b^0_L & \tilde{\Psi}^{-,-}_L & \Psi^{-,-}_L & \Psi^{-,0}_L & \Psi^{0,-}_L & \tilde{\Psi}^{-,0}_L & \tilde{\Psi}^{0,-}_L \\[2pt]
    \hline\\\\[-4.5\medskipamount]
     \bar{b}^0_R & 0 & 0 & -c_{h} \Delta_q & -\frac{1}{\sqrt{2}} s_{h} \Delta_q & \frac{i}{\sqrt{2}} s_{h} \Delta_q & 0 & 0 \\[1pt]
     \bar{\tilde{\Psi}}^{-,-}_R & -\frac{i}{\sqrt{2}} s_{h} \Delta^{\dagger}_b & m_b & 0 & 0 & 0 & 0 & 0 \\[1pt]
     \bar{\Psi}^{-,-}_R & 0 & \frac{1}{2} Y_b & m_q & 0 & 0 & 0 & 0 \\[1pt]
     \bar{\Psi}^{-,0}_R & 0 & 0 & 0 & m_q & 0 & 0 & 0 \\[1pt]
     \bar{\Psi}^{0,-}_R & 0 & 0 & 0 & 0 & m_q & 0 & 0 \\[1pt]
     \bar{\tilde{\Psi}}^{-,0}_R & - s^{2}_{h/2} \Delta^{\dagger}_b & 0 & 0 & 0 & 0 & m_b & 0 \\[1pt]
     \bar{\tilde{\Psi}}^{0,-}_R & - c^{2}_{h/2} \Delta^{\dagger}_b & 0 & 0 & 0 & 0 & 0 & m_b \end{array}\right),\\[16pt]
&M^{14-1-10}_{U} =\nonumber\\
&\quad \left(\begin{array}{c|cccccccccccc}
    {} & t^0_L & \Psi^{0,0}_{\mathbf{1}_L} & \tilde{\Psi}^{0,0}_L & \tilde{\Psi}^{+,-}_L & \Psi^{+,-}_{\mathbf{4}_L} & \tilde{\Psi}^{-,+}_L & \Psi^{-,+}_{\mathbf{4}_L} & \Psi^{-,+}_{\mathbf{9}_L} & \Psi^{0,0}_{\mathbf{9}_L} & \Psi^{+,-}_{\mathbf{9}_L} & \tilde{\Psi}^{0,0}_{1_L} & \tilde{\Psi}^{0,0}_{2_L} \\[2pt]
    \hline\\\\[-4.5\medskipamount]
     \bar{t}^0_R & 0 & \tilde{s} \Delta_q & 0 & 0 & - c_{+} \Delta_q & 0 & -i c_{-} \Delta_q & s_{h} s^{2}_{h/2} \Delta_q & -\frac{s_{2h}}{4} \Delta_q & s_{h} c^{2}_{h/2} \Delta_q & 0 & 0 \\[1pt]
     \bar{\Psi}^{0,0}_{\mathbf{1}_R} & 0 & m_q & \frac{2}{\sqrt{5}} Y_t & 0 & 0 & 0 & 0 & 0 & 0 & 0 & 0 & 0 \\[1pt]
     \bar{\tilde{\Psi}}^{0,0}_R & -\Delta^{\dagger}_t & 0 & m_t & 0 & 0 & 0 & 0 & 0 & 0 & 0 & 0 & 0 \\[1pt]
     \bar{\tilde{\Psi}}^{+,-}_R & 0 & 0 & 0 & m_b & 0 & 0 & 0 & 0 & 0 & 0 & 0 & 0 \\[1pt]
     \bar{\Psi}^{+,-}_{\mathbf{4}_R} & 0 & 0 & 0 & \frac{1}{2} Y_b & m_q & 0 & 0 & 0 & 0 & 0 & 0 & 0 \\[1pt]
     \bar{\tilde{\Psi}}^{-,+}_R & 0 & 0 & 0 & 0 & 0 & m_b & 0 & 0 & 0 & 0 & 0 & 0 \\[1pt]
     \bar{\Psi}^{-,+}_{\mathbf{4}_R} & 0 & 0 & 0 & 0 & 0 & \frac{1}{2} Y_b & m_q & 0 & 0 & 0 & 0 & 0 \\[1pt]
     \bar{\Psi}^{-,+}_{\mathbf{9}_R} & 0 & 0 & 0 & 0 & 0 & 0 & 0 & m_q & 0 & 0 & 0 & 0 \\[1pt]
     \bar{\Psi}^{0,0}_{\mathbf{9}_R} & 0 & 0 & 0 & 0 & 0 & 0 & 0 & 0 & m_q & 0 & 0 & 0 \\[1pt]
     \bar{\Psi}^{+,-}_{\mathbf{9}_R} & 0 & 0 & 0 & 0 & 0 & 0 & 0 & 0 & 0 & m_q & 0 & 0 \\[1pt]
     \bar{\tilde{\Psi}}^{0,0}_{1_R} & 0 & 0 & 0 & 0 & 0 & 0 & 0 & 0 & 0 & 0 & m_b & 0 \\[1pt]
     \bar{\tilde{\Psi}}^{0,0}_{2_R} & 0 & 0 & 0 & 0 & 0 & 0 & 0 & 0 & 0 & 0 & 0 & m_b \end{array}\right).
\end{align}

\section{Collider search constraints}
\label{collider_search_appendix}

\begin{table}[H]
\begin{center}
\begin{tabular}{ @{}lllll @{} }
\toprule
Decay & Experiment & $\sqrt{s}$ (TeV) & Analysis &  Ref.\\
\midrule
\multirow{1}{*}{$V\rightarrow \ell\nu$} & ATLAS & 7 & EXOT-2012-02 & \cite{Aad:2012dm}\\
\midrule
\multirow{4}{*}{$V\rightarrow e\nu$} & ATLAS & 7 & EXOT-2012-02 & \cite{Aad:2012dm}\\
 & CMS & 13 & PAS-EXO-15-006 & \cite{CMS:2015kjy}\\
 & ATLAS & 13 & CONF-2016-061 & \cite{ATLAS:2016ecs}\\
 & ATLAS & 13 & CONF-2018-017 & \cite{ATLAS:2018lcz}\\
\midrule
\multirow{4}{*}{$V\rightarrow \mu\nu$} & ATLAS & 7 & EXOT-2012-02 & \cite{Aad:2012dm}\\
 & CMS & 13 & PAS-EXO-15-006 & \cite{CMS:2015kjy}\\
 & ATLAS & 13 & CONF-2016-061 & \cite{ATLAS:2016ecs}\\
 & ATLAS & 13 & CONF-2018-017 & \cite{ATLAS:2018lcz}\\
\midrule
\multirow{3}{*}{$V\rightarrow \tau\nu$} & CMS & 8 & EXO-12-011 & \cite{Khachatryan:2015pua}\\
 & CMS & 13 & PAS-EXO-16-006 & \cite{CMS:2016ppa}\\
 & CMS & 13 & PAS-EXO-16-006 & \cite{CMS:2016ppa}\\
\midrule
\multirow{8}{*}{$V\rightarrow ee$} & ATLAS & 8 & EXOT-2012-23 & \cite{Aad:2014cka}\\
 & CMS & 8 & EXO-12-061 & \cite{Khachatryan:2014fba}\\
 & CMS & 13 & EXO-18-006 & \cite{CMS:2018wsn}\\
 & CMS & 13 & PAS-EXO-16-031 & \cite{CMS:2016abv}\\
 & ATLAS & 13 & CONF-2016-045 & \cite{ATLAS:2016cyf}\\
 & ATLAS & 13 & EXOT-2016-05 & \cite{Aaboud:2017buh}\\
 & ATLAS & 13 & EXOT-2018-08 & \cite{Aad:2019fac}\\
 & ATLAS & 13 & EXOT-2018-08 & \cite{Aad:2019fac}\\
\midrule
\multirow{6}{*}{$V\rightarrow \mu\mu$} & ATLAS & 8 & EXOT-2012-23 & \cite{Aad:2014cka}\\
 & CMS & 8 & EXO-12-061 & \cite{Khachatryan:2014fba}\\
 & CMS & 13 & EXO-16-047 & \cite{Sirunyan:2018exx}\\
 & CMS & 13 & PAS-EXO-16-031 & \cite{CMS:2016abv}\\
 & ATLAS & 13 & EXOT-2016-05 & \cite{Aaboud:2017buh}\\
 & ATLAS & 13 & CONF-2016-045 & \cite{ATLAS:2016cyf}\\
\midrule
\multirow{3}{*}{$V\rightarrow \tau\tau$} & ATLAS & 8 & EXOT-2014-05 & \cite{Aad:2015osa}\\
 & CMS & 8 & EXO-12-046 & \cite{CMS:2015ufa}\\
 & CMS & 13 & PAS-EXO-16-008 & \cite{CMS:2016zxk}\\
\bottomrule
\end{tabular}
\caption{Experimental analyses of leptonic decay channels of vector resonances we use to judge the fitness of our points.}
\label{tab:V_l}
\end{center}
\end{table}

\begin{table}[H]
\begin{center}
\begin{tabular}{ @{}lllll @{} }
\toprule
Decay & Experiment &  $\sqrt{s}$ (TeV)  & Analysis & Ref. \\
\midrule
\multirow{2}{*}{$V\rightarrow qq$} & CMS & 13 & PAS-EXO-16-032 & \cite{CMS:2016wpz}\\
 & CMS & 13 & PAS-EXO-16-032 & \cite{CMS:2016wpz}\\
\midrule
\multirow{5}{*}{$V\rightarrow jj$} & CMS & 13 & EXO-15-001 & \cite{Khachatryan:2015dcf}\\
 & CMS & 13 & EXO-16-056 & \cite{Sirunyan:2018xlo}\\
 & ATLAS & 13 & EXOT-2015-02 & \cite{ATLAS:2015nsi}\\
 & ATLAS & 13 & EXOT-2018-05 & \cite{Aaboud:2019zxd}\\
\midrule
\multirow{5}{*}{$V\rightarrow tb$} & CMS & 8 & B2G-12-010 & \cite{Chatrchyan:2014koa}\\
 & CMS & 8 & B2G-12-009 & \cite{Khachatryan:2015edz}\\
 & CMS & 13 & PAS-B2G-16-009 & \cite{CMS:2016ude}\\
 & CMS & 13 & PAS-B2G-16-017 & \cite{CMS:2016wqa}\\
 & ATLAS & 13 & EXOT-2017-02 & \cite{Aaboud:2018juj}\\
\midrule
\multirow{7}{*}{$V\rightarrow tt$} & ATLAS & 8 & CONF-2015-009 & \cite{ATLAS:2015aka}\\
 & CMS & 8 & B2G-13-008 & \cite{Khachatryan:2015sma}\\
 & CMS & 13 & PAS-B2G-15-003 & \cite{CMS:2016ehh}\\
 & CMS & 13 & PAS-B2G-15-002 & \cite{CMS:2016zte}\\
 & CMS & 13 & B2G-17-017 & \cite{Sirunyan:2018ryr}\\
 & ATLAS & 13 & EXOT-2015-04 & \cite{Aaboud:2018mjh}\\
 & ATLAS & 13 & EXOT-2016-24 & \cite{Aaboud:2019roo}\\
\bottomrule
\end{tabular}
\caption{Experimental analyses of hadronic decay channels of vector resonances we use to judge the fitness of our points. Here, $j$ refers to a light quark or $b$ jet, and $q$ refers to a light quark jet.}
\label{tab:V_h}
\end{center}
\end{table}

\begin{table}[H]
\begin{center}
\begin{tabular}{ @{}lllll @{} }
\toprule
Decay & Experiment &  $\sqrt{s}$ (TeV)  & Analysis & Ref. \\
\midrule
\multirow{9}{*}{$V\rightarrow ZH$} & ATLAS & 8 & EXOT-2013-23 & \cite{Aad:2015yza}\\
 & CMS & 8 & EXO-13-007 & \cite{Khachatryan:2015ywa}\\
 & CMS & 13 & PAS-B2G-16-003 & \cite{CMS:2016dzw}\\
 & CMS & 13 & B2G-17-006 & \cite{Sirunyan:2018fuh}\\
 & CMS & 13 & B2G-17-002 & \cite{Sirunyan:2017wto}\\
 & CMS & 13 & B2G-17-004 & \cite{Sirunyan:2018qob}\\
 & ATLAS & 13 & EXOT-2015-18 & \cite{Aaboud:2016lwx}\\
 & ATLAS & 13 & CONF-2015-074 & \cite{TheATLAScollaboration:2015ulg}\\
 & ATLAS & 13 & CONF-2016-083 & \cite{ATLAS:2016kxc}\\
\midrule
\multirow{14}{*}{$V\rightarrow WZ$} & ATLAS & 8 & EXOT-2013-08 & \cite{Aad:2015owa}\\
 & ATLAS & 8 & EXOT-2013-01 & \cite{Aad:2015ufa}\\
 & ATLAS & 8 & EXOT-2013-07 & \cite{Aad:2014pha}\\
 & CMS & 8 & EXO-12-024 & \cite{Khachatryan:2014hpa}\\
 & CMS & 13 & PAS-EXO-15-002 & \cite{CMS:2015nmz}\\
 & CMS & 13 & PAS-B2G-16-020 & \cite{CMS:2016pfl}\\
 & CMS & 13 & B2G-16-029 & \cite{Sirunyan:2018iff}\\
 & CMS & 13 & B2G-17-001 & \cite{Sirunyan:2017acf}\\
 & CMS & 13 & B2G-17-005 & \cite{Sirunyan:2018ivv}\\
 & CMS & 13 & B2G-17-013 & \cite{Sirunyan:2018hsl}\\
 & CMS & 13 & B2G-18-002 & \cite{Sirunyan:2019jbg}\\
 & ATLAS & 13 & CONF-2016-055 & \cite{ATLAS:2016yqq}\\
 & ATLAS & 13 & CONF-2016-062 & \cite{ATLAS:2016cwq}\\
 & ATLAS & 13 & CONF-2016-082 & \cite{ATLAS:2016npe}\\
\midrule
\multirow{8}{*}{$V\rightarrow WH$} & ATLAS & 8 & EXOT-2013-23 & \cite{Aad:2015yza}\\
 & CMS & 8 & EXO-14-010 & \cite{Khachatryan:2016yji}\\
 & CMS & 13 & PAS-B2G-16-003 & \cite{CMS:2016dzw}\\
 & CMS & 13 & B2G-17-006 & \cite{Sirunyan:2018fuh}\\
 & CMS & 13 & B2G-17-002 & \cite{Sirunyan:2017wto}\\
 & CMS & 13 & B2G-17-004 & \cite{Sirunyan:2018qob}\\
 & ATLAS & 13 & EXOT-2015-18 & \cite{Aaboud:2016lwx}\\
 & ATLAS & 13 & CONF-2016-083 & \cite{ATLAS:2016kxc}\\
\midrule
\multirow{5}{*}{$V\rightarrow WW$} & ATLAS & 8 & EXOT-2013-01 & \cite{Aad:2015ufa}\\
 & CMS & 8 & EXO-13-009 & \cite{Khachatryan:2014gha}\\
 & CMS & 13 & B2G-17-001 & \cite{Sirunyan:2017acf}\\
 & CMS & 13 & B2G-18-002 & \cite{Sirunyan:2019jbg}\\
 & ATLAS & 13 & CONF-2016-062 & \cite{ATLAS:2016cwq}\\
\midrule
\multirow{1}{*}{$V\rightarrow WW+ZH$} & CMS & 13 & PAS-B2G-16-007 & \cite{CMS:2016wev}\\
\bottomrule
\end{tabular}
\caption{Experimental analyses of bosonic decay channels of vector resonances we use to judge the fitness of our points.}
\label{tab:V_b}
\end{center}
\end{table}

\begin{table}[H]
\begin{center}
\begin{tabular}{ @{}lllll @{} }
\toprule
Decay & Experiment &  $\sqrt{s}$ (TeV)  & Analysis & Ref. \\
\midrule
\multirow{2}{*}{$F\rightarrow jW$} & CDF & 1.96 & 10110 & \cite{CDF:10110}\\
 & ATLAS & 7 & EXOT-2011-28 & \cite{Aad:2012bt}\\
\midrule
\multirow{2}{*}{$F\rightarrow qW$} & ATLAS & 8 & EXOT-2014-10 & \cite{Aad:2015tba}\\
 & CMS & 8 & B2G-12-017 & \cite{CMS:2014dka}\\
\midrule
\multirow{9}{*}{$F\rightarrow bW$} & ATLAS & 7 & EXOT-12-07 & \cite{ATLAS:2012qe}\\
 & CMS & 7 & EXO-11-050 & \cite{CMS:2012ab}\\
 & CMS & 7 & EXO-11-099 & \cite{Chatrchyan:2012vu}\\
 & ATLAS & 8 & CONF-2015-012 & \cite{ATLAS:2015dka}\\
 & CMS & 8 & B2G-12-017 & \cite{CMS:2014dka}\\
 & CMS & 8 & B2G-13-005 & \cite{Khachatryan:2015oba}\\
 & ATLAS & 13 & CONF-2016-102 & \cite{ATLAS:2016cuv}\\
 & CMS & 13 & B2G-17-003 & \cite{Sirunyan:2017pks}\\
 & CMS & 13 & B2G-16-024 & \cite{Sirunyan:2017usq}\\
\midrule
\multirow{12}{*}{$F\rightarrow tW$} & CDF & 1.96 & 2009 & \cite{Aaltonen:2009nr}\\
 & CMS & 7 & B2G-12-004 & \cite{Chatrchyan:2012af}\\
 & CMS & 8 & B2G-12-012 & \cite{Chatrchyan:2013wfa}\\
 & CMS & 8 & B2G-13-003 & \cite{CMS:2013una}\\
 & CMS & 8 & B2G-13-006 & \cite{Khachatryan:2015gza}\\
 & ATLAS & 8 & EXOT-2013-16 & \cite{Aad:2015gdg}\\
 & ATLAS & 8 & EXOT-2014-17 & \cite{Aad:2015mba}\\
 & ATLAS & 13 & EXOT-2016-16 & \cite{Aaboud:2018xpj}\\
 & ATLAS & 13 & EXOT-2017-34 & \cite{Aaboud:2018uek}\\
 & CMS & 13 & PAS-B2G-15-006 & \cite{CMS:2015alb}\\
 & CMS & 13 & B2G-16-019 & \cite{CMS:2017jfv}\\
 & CMS & 13 & B2G-17-014 & \cite{Sirunyan:2018yun}\\
\bottomrule
\end{tabular}
\caption{Experimental analyses of heavy quark decays we use to judge the fitness of our points. Here, $j$ refers to a light quark or $b$ jet, and $q$ refers to a light quark jet.}
\label{tab:F_1}
\end{center}
\end{table}

\begin{table}[H]
\begin{center}
\begin{tabular}{ @{}lllll @{} }
\toprule
Decay & Experiment &  $\sqrt{s}$ (TeV)  & Analysis & Ref. \\
\midrule
\multirow{1}{*}{$F\rightarrow jZ$} & CDF & 1.96 & 2006 & \cite{Aaltonen:2007je}\\
\midrule
\multirow{4}{*}{$F\rightarrow bZ$} & CMS & 7 & EXO-11-066 & \cite{CMS:2012jwa}\\
 & CMS & 8 & B2G-13-003 & \cite{CMS:2013una}\\
 & CMS & 8 & B2G-13-006 & \cite{Khachatryan:2015gza}\\
 & ATLAS & 13 & EXOT-2016-35 & \cite{Aaboud:2018saj}\\
\midrule
\multirow{7}{*}{$F\rightarrow tZ$} & CMS & 7 & B2G-12-004 & \cite{Chatrchyan:2012af}\\
 & CMS & 7 & EXO-11-005 & \cite{Chatrchyan:2011ay}\\
 & CMS & 8 & B2G-13-005 & \cite{Khachatryan:2015oba}\\
 & ATLAS & 13 & CONF-2016-101 & \cite{ATLAS:2016qlg}\\
 & ATLAS & 13 & EXOT-2016-15 & \cite{Aaboud:2017qpr}\\
 & ATLAS & 13 & EXOT-2016-13 & \cite{Aaboud:2018xuw}\\
 & ATLAS & 13 & EXOT-2016-35 & \cite{Aaboud:2018saj}\\
\midrule
\multirow{4}{*}{$F\rightarrow bH$} & ATLAS & 8 & CONF-2015-012 & \cite{ATLAS:2015dka}\\
 & CMS & 8 & B2G-12-019 & \cite{CMS:2012hfa}\\
 & CMS & 8 & B2G-13-006 & \cite{Khachatryan:2015gza}\\
 & CMS & 8 & B2G-14-001 & \cite{CMS:2014afa}\\
\midrule
\multirow{5}{*}{$F\rightarrow tH$} & CMS & 8 & B2G-13-005 & \cite{Khachatryan:2015oba}\\
 & CMS & 13 & PAS-B2G-16-011 & \cite{CMS:2016dmr}\\
 & CMS & 13 & B2G-16-024 & \cite{Sirunyan:2017usq}\\
 & ATLAS & 13 & CONF-2016-013 & \cite{TheATLAScollaboration:2016gxs}\\
 & ATLAS & 13 & EXOT-2016-13 & \cite{Aaboud:2018xuw}\\
\bottomrule
\end{tabular}
\caption{(\textit{cont.}) Experimental analyses of heavy quark decays we use to judge the fitness of our points. Here, $j$ refers to a light quark or $b$ jet, and $q$ refers to a light quark jet.}
\label{tab:F_2}
\end{center}
\end{table}

\clearpage
\section{Scan comparisons}
\label{scan_agreement_appendix}

\subsection{M4DCHM$^{5-5-5}$}

\begin{figure}[h]
\centering
  \includegraphics[width=1\linewidth]{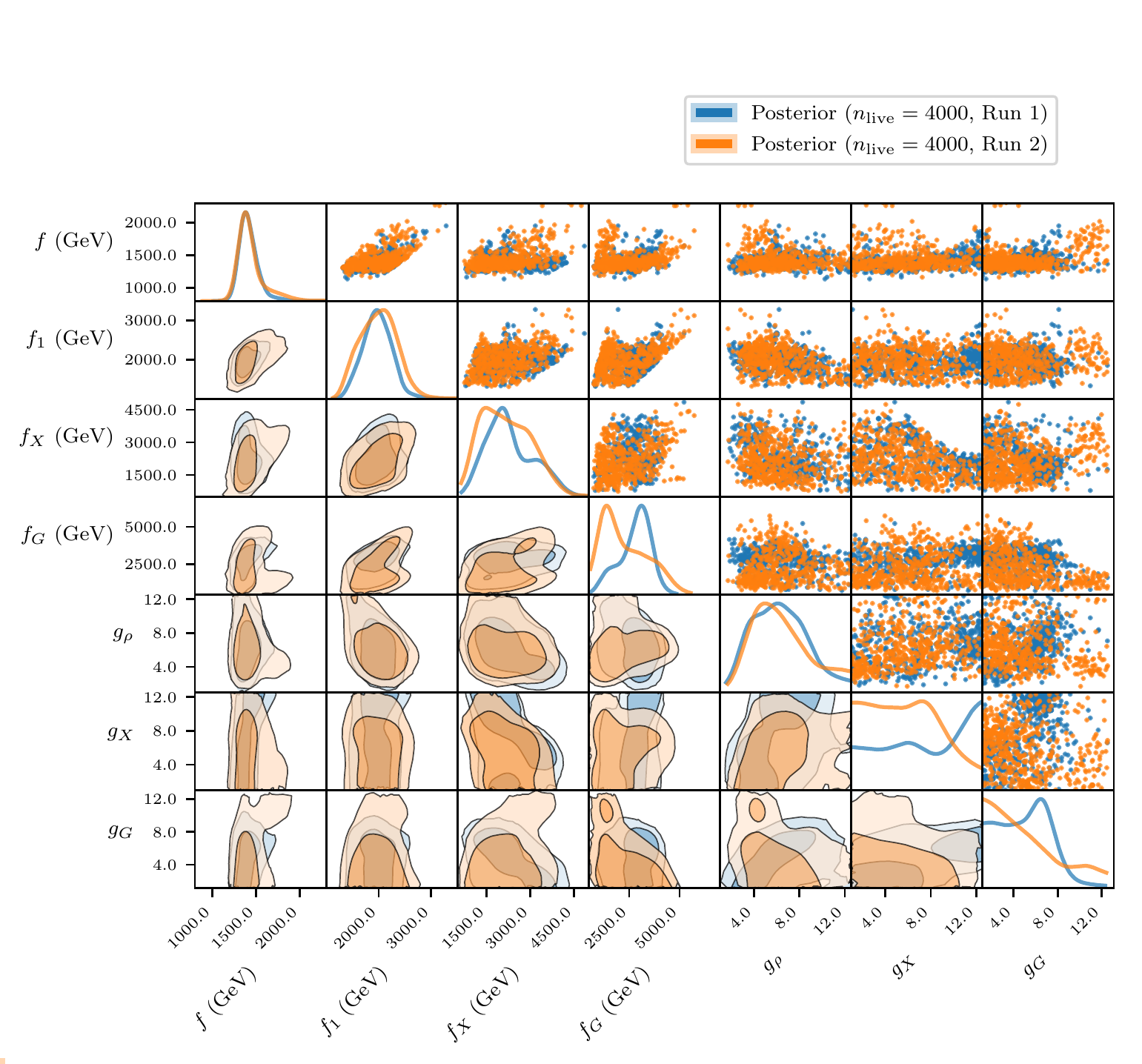}
\caption{1D and 2D marginalised posteriors for the gauge sector parameters in the M4DCHM$^{5-5-5}$ found in two different runs with $4000$ live points.}
\label{fig:5-5-5_gauge_posteriors}
\end{figure}

\begin{figure}[h]
\centering
  \includegraphics[width=1\linewidth]{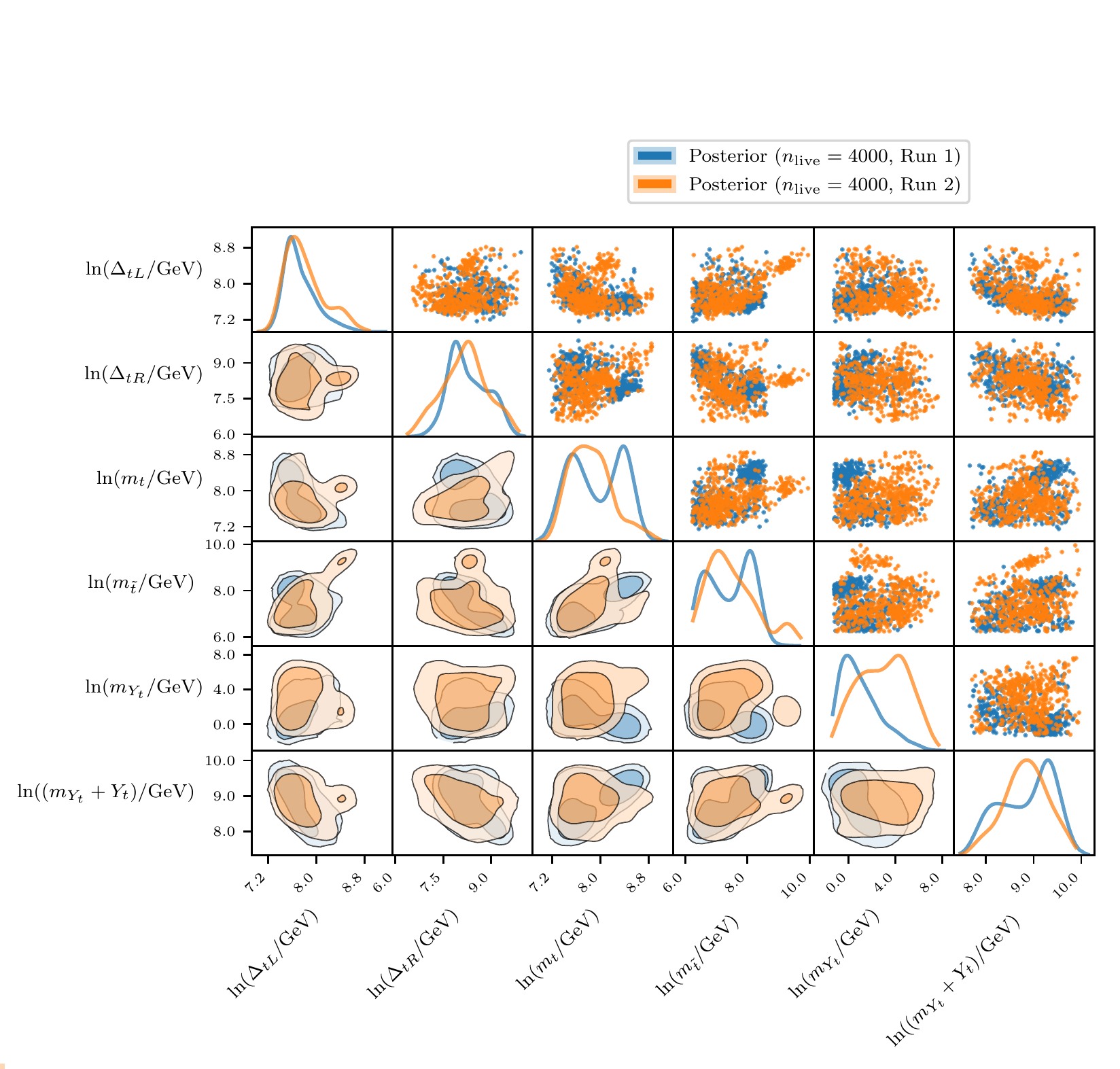}
\caption{1D and 2D marginalised posteriors for the top partner parameters in the M4DCHM$^{5-5-5}$ found in two different runs with $4000$ live points.}
\label{fig:5-5-5_T_posteriors}
\end{figure}

\begin{figure}[h]
\centering
  \includegraphics[width=1\linewidth]{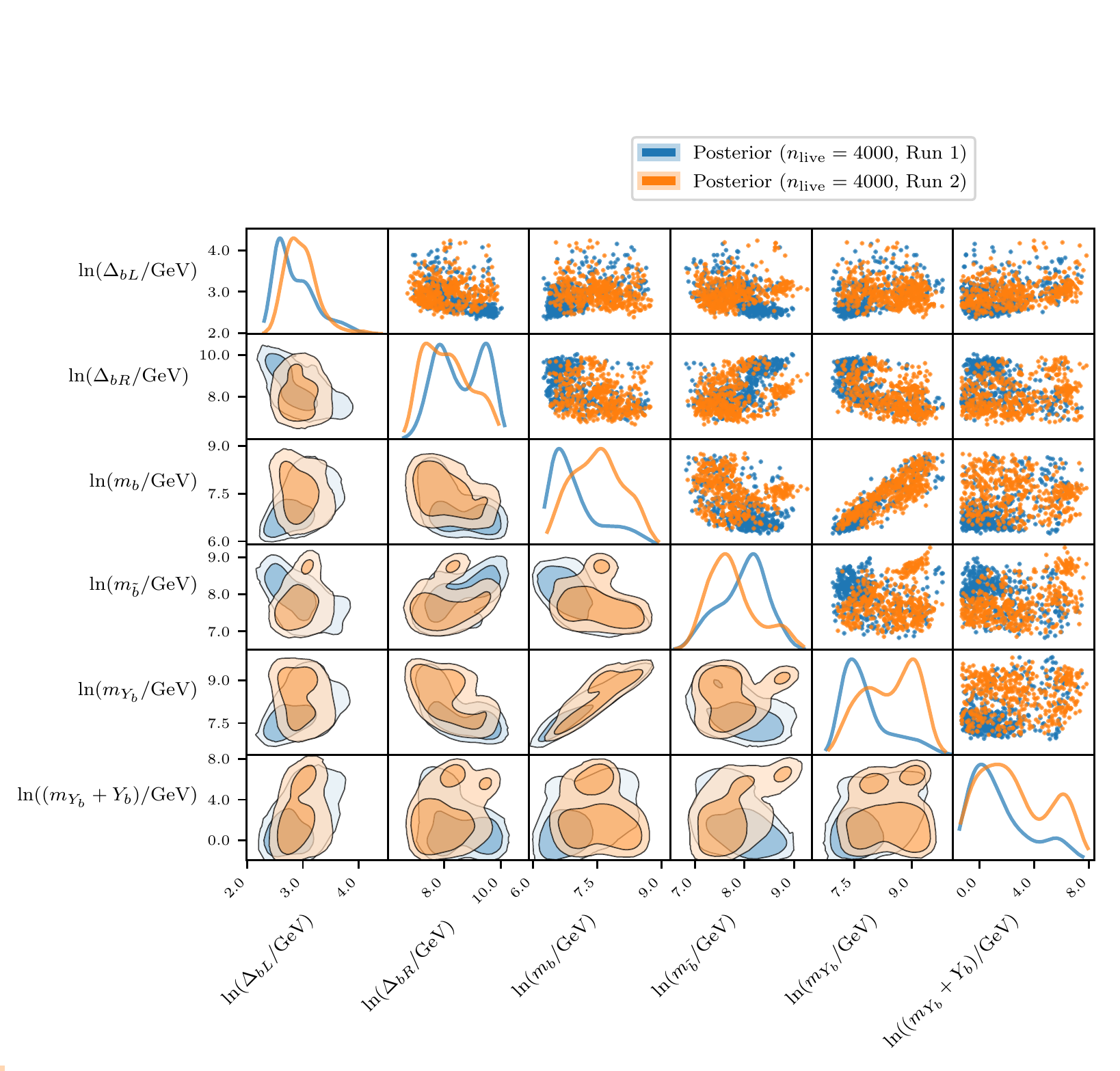}
\caption{1D and 2D marginalised posteriors for the bottom partner parameters in the M4DCHM$^{5-5-5}$ found in two different runs with $4000$ live points.}
\label{fig:5-5-5_B_posteriors}
\end{figure}

Our results for this model come from four different \texttt{PolyChord} scans. The first two scans, which each used $2000$ live points, did not display very good agreement with each other, but more agreement was seen after increasing the sampling density to $4000$ live points. The posteriors found by these $4000$ point scans are shown in \Cref{fig:5-5-5_gauge_posteriors,fig:5-5-5_T_posteriors,fig:5-5-5_B_posteriors}. There is not exact agreement in these posteriors - for example, the second run found a minor mode that the first did not, apparent in \Cref{fig:5-5-5_T_posteriors,fig:5-5-5_B_posteriors} - but there are no major differences in the general regions that each scan found, and specific features such as the correlation between $m_{b}$ and $m_{Y_{b}}$ are present in both scans. Parameters for which there is poor agreement, such as $g_{X}$ and $g_{G}$ in \Cref{fig:5-5-5_gauge_posteriors}, tend to be those that are are minimally affected by the constraints we impose and as such are not our primary interest. Given how difficult sampling the space has proven to be, we regard this as an acceptable level of agreement.

It is also encouraging that the Bayesian evidences found in the $4000$ point scans were in agreement:
\begin{align}
    \ln(\mathcal{Z})_{\text{Run 1}} &= -27.85 \pm 0.06, \nonumber\\
    \ln(\mathcal{Z})_{\text{Run 2}} &= -27.87 \pm 0.06,
\end{align}
which we take as further indication that the parameter space of this model has been reliably explored.

\clearpage
\subsection{M4DCHM$^{14-14-10}$}

\begin{figure}[h]
\centering
  \includegraphics[width=1\linewidth]{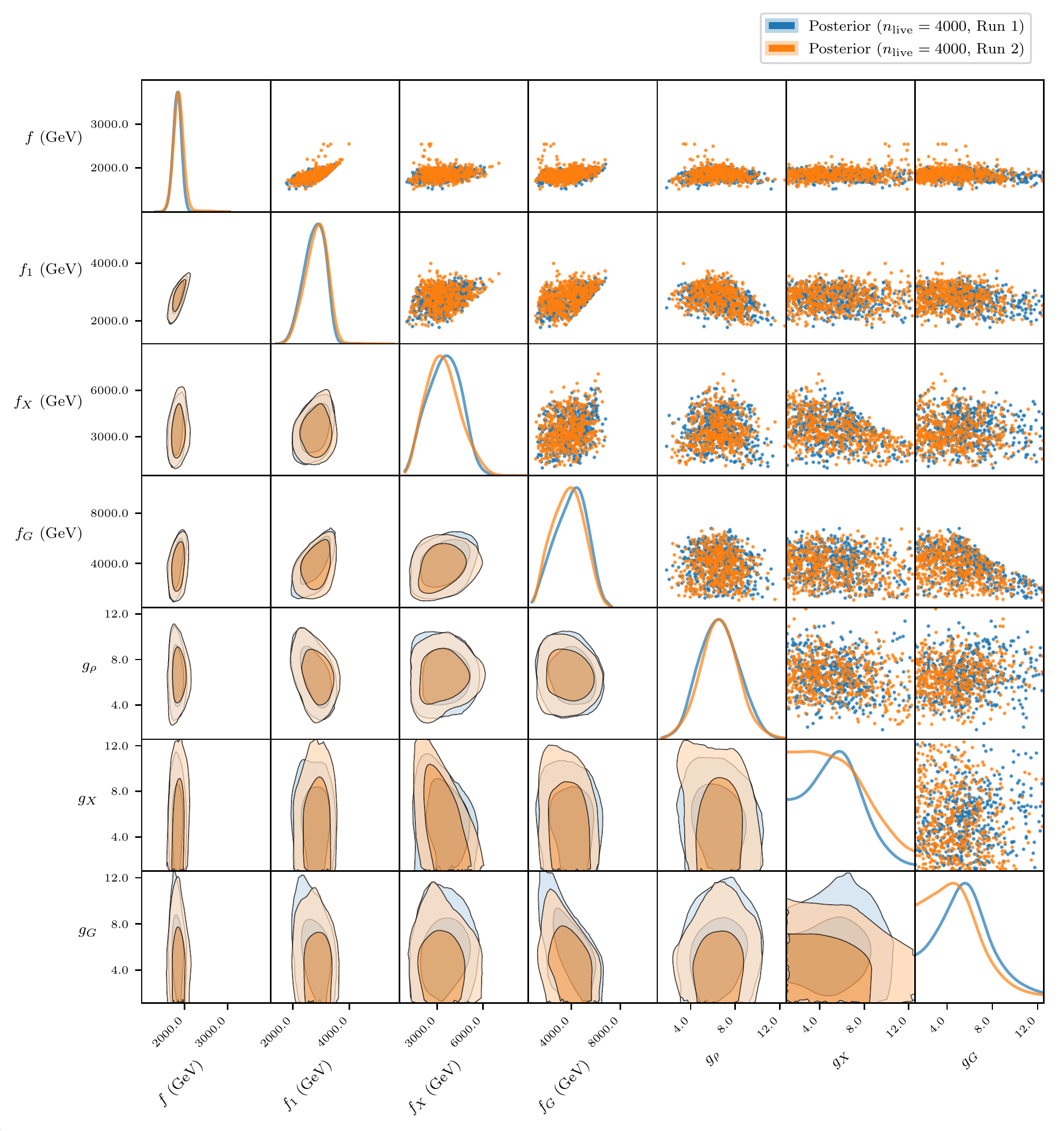}
\caption{1D and 2D marginalised posteriors for the gauge sector parameters in the M4DCHM$^{14-14-10}$ found in two different runs with $4000$ live points.}
\label{fig:14-14-10_gauge_posteriors}
\end{figure}

\begin{figure}[h]
\centering
  \includegraphics[width=1\linewidth]{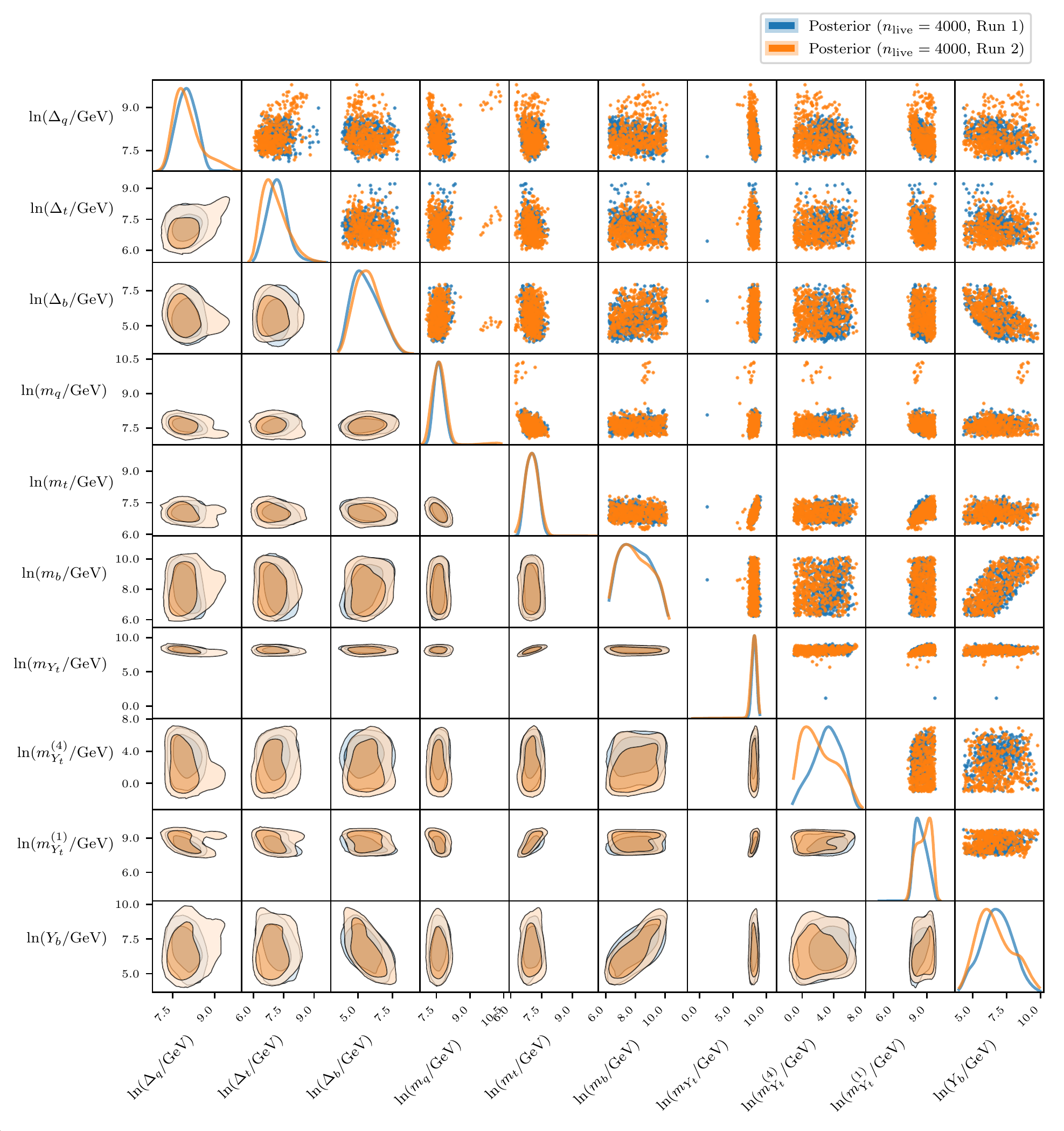}
\caption{1D and 2D marginalised posteriors for the fermion sector parameters in the M4DCHM$^{14-14-10}$ found in two different runs with $4000$ live points.}
\label{fig:14-14-10_fermion_posteriors}
\end{figure}

As with the M4DCHM$^{5-5-5}$, two $2000$ point scans and two $4000$ point scans were performed for the M4DCHM$^{14-14-10}$, with the latter two having significantly better results than the former two. In fact, posteriors from the two $4000$ point scans of this model display a significantly higher level of agreement than was found in the M4DCHM$^{5-5-5}$, as can be seen in \Cref{fig:14-14-10_gauge_posteriors,fig:14-14-10_fermion_posteriors}. Their evidences,
\begin{align}
    \ln(\mathcal{Z})_{\text{Run 1}} &= -37.56 \pm 0.08, \nonumber\\
    \ln(\mathcal{Z})_{\text{Run 2}} &= -37.87 \pm 0.08,
\end{align}
are acceptably consistent.

\clearpage
\subsection{M4DCHM$^{14-1-10}$}

\begin{figure}[h]
\centering
  \includegraphics[width=1\linewidth]{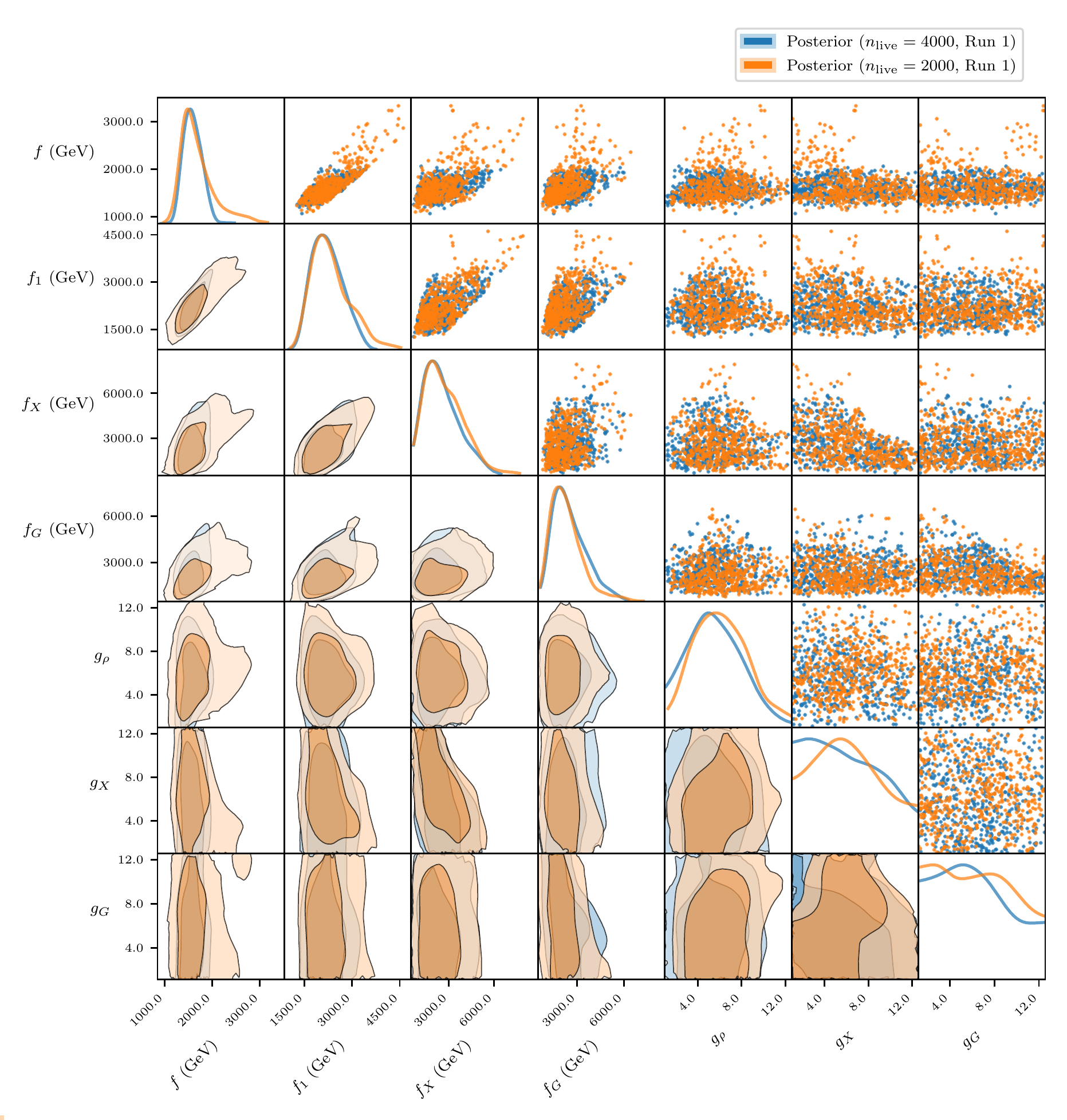}
\caption{1D and 2D marginalised posteriors for the gauge sector parameters in the M4DCHM$^{14-1-10}$ found in two different runs with $2000$ and $4000$ live points respectively.}
\label{fig:14-1-10_gauge_posteriors}
\end{figure}

\begin{figure}[h]
\centering
  \includegraphics[width=1\linewidth]{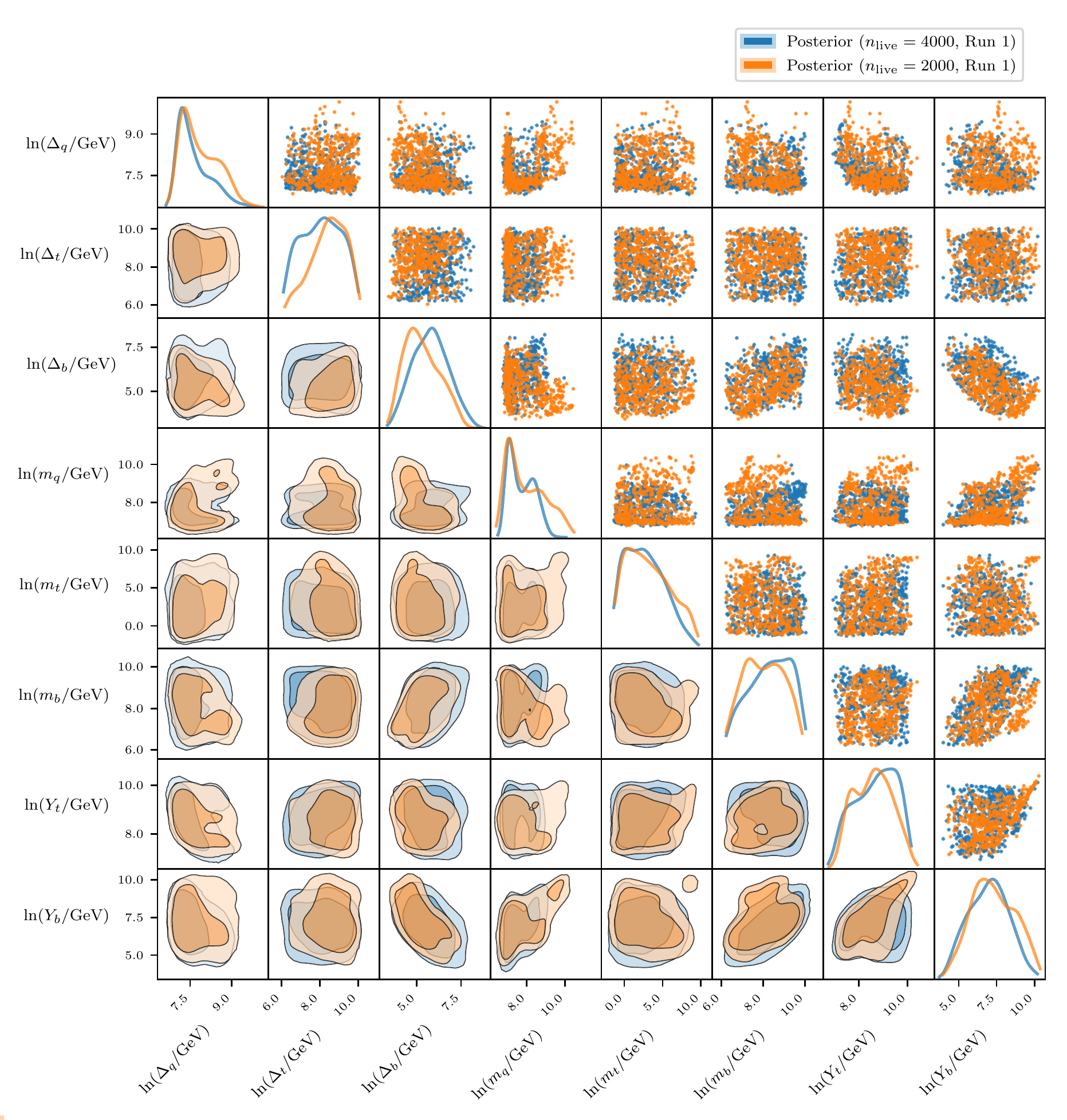}
\caption{1D and 2D marginalised posteriors for the fermion sector parameters in the M4DCHM$^{14-1-10}$ found in two different runs with $2000$ and $4000$ live points respectively.}
\label{fig:14-1-10_fermion_posteriors}
\end{figure}

For this model we performed four scans: the first three with $2000$ live points, and the last with $4000$ live points. Perhaps because of its smaller parameter space, this was by far the easiest of the three models to fit, with all four of these scans having good agreement in their results. Indeed, the posteriors of the $4000$ point run and the first $2000$ point run, for example, are displayed in \Cref{fig:14-1-10_gauge_posteriors,fig:14-1-10_fermion_posteriors} and are seen to match each other closely. The evidences also match quite well across scans:
\begin{align}
    \ln(\mathcal{Z})_{\text{Run 1}} &= -37.54 \pm 0.12, \nonumber\\
    \ln(\mathcal{Z})_{\text{Run 2}} &= -37.38 \pm 0.12, \nonumber\\
    \ln(\mathcal{Z})_{\text{Run 3}} &= -37.51 \pm 0.09, \nonumber\\
    \ln(\mathcal{Z})_{\text{Run 4}} &= -37.70 \pm 0.06.
\end{align}
As such, these results are quite reliable.

\clearpage

%\mciteErrorOnUnknownfalse
%\nocite{*}
\newpage
\bibliographystyle{JHEP}
\bibliography{tex/refs}

\providecommand{\href}[2]{#2}\begingroup\raggedright\begin{thebibliography}{100}

\bibitem{kaplan1984}
D.~B. Kaplan and H.~Georgi, \emph{{$SU(2) \times U(1)$ breaking by vacuum
  misalignment}},
  \href{https://doi.org/10.1016/0370-2693(84)91177-8}{\emph{Physics Letters B}
  {\bfseries 136} (1984) 183--186}.

\bibitem{kaplan1984b}
D.~B. Kaplan, H.~Georgi and S.~Dimopoulos, \emph{{Composite Higgs scalars}},
  \href{https://doi.org/10.1016/0370-2693(84)91178-X}{\emph{Physics Letters B}
  {\bfseries 136} (1984) 187--190}.

\bibitem{kaplan1985}
M.~J. Dugan, H.~Georgi and D.~B. Kaplan, \emph{{Anatomy of a composite Higgs
  model}}, \href{https://doi.org/10.1016/0550-3213(85)90221-4}{\emph{Nuclear
  Physics B} {\bfseries 254} (1985) 299--326}.

\bibitem{contino2006}
K.~Agashe, R.~Contino, L.~Da~Rold and A.~Pomarol, \emph{{A Custodial symmetry
  for $Zb \bar b$}},
  \href{https://doi.org/10.1016/j.physletb.2006.08.005}{\emph{Physics Letters}
  {\bfseries B641} (2006) 62--66},
  [\href{https://arxiv.org/abs/hep-ph/0605341}{{\ttfamily hep-ph/0605341}}].

\bibitem{contino2007b}
R.~Contino and A.~Pomarol, \emph{{The holographic composite Higgs}},
  \href{https://doi.org/10.1016/j.crhy.2006.12.007}{\emph{Comptes Rendus
  Physique} {\bfseries 8} (2007) 1058--1067}.

\bibitem{contino2003}
R.~Contino, Y.~Nomura and A.~Pomarol, \emph{{Higgs as a holographic
  pseudo-Goldstone boson}},
  \href{https://doi.org/10.1016/j.nuclphysb.2003.08.027}{\emph{Nuclear Physics}
  {\bfseries B671} (2003) 148--174},
  [\href{https://arxiv.org/abs/hep-ph/0306259}{{\ttfamily hep-ph/0306259}}].

\bibitem{agashe2005}
K.~Agashe, R.~Contino and A.~Pomarol, \emph{{The Minimal composite Higgs
  model}}, \href{https://doi.org/10.1016/j.nuclphysb.2005.04.035}{\emph{Nuclear
  Physics} {\bfseries B719} (2005) 165--187},
  [\href{https://arxiv.org/abs/hep-ph/0412089}{{\ttfamily hep-ph/0412089}}].

\bibitem{Contino:2006nn}
R.~Contino, T.~Kramer, M.~Son and R.~Sundrum, \emph{{Warped/composite
  phenomenology simplified}},
  \href{https://doi.org/10.1088/1126-6708/2007/05/074}{\emph{Journal of High
  Energy Physics} {\bfseries 05} (2007) 074},
  [\href{https://arxiv.org/abs/hep-ph/0612180}{{\ttfamily hep-ph/0612180}}].

\bibitem{giudice2007}
G.~F. Giudice, C.~Grojean, A.~Pomarol and R.~Rattazzi, \emph{{The
  Strongly-Interacting Light Higgs}},
  \href{https://doi.org/10.1088/1126-6708/2007/06/045}{\emph{Journal of High
  Energy Physics} {\bfseries 06} (2007) 045},
  [\href{https://arxiv.org/abs/hep-ph/0703164}{{\ttfamily hep-ph/0703164}}].

\bibitem{contino2007}
R.~Contino, L.~Da~Rold and A.~Pomarol, \emph{{Light custodians in natural
  composite Higgs models}},
  \href{https://doi.org/10.1103/PhysRevD.75.055014}{\emph{Physical Review D}
  {\bfseries 75} (2007) 055014}.

\bibitem{Panico:2007qd}
G.~Panico and A.~Wulzer, \emph{{Effective action and holography in 5D gauge
  theories}},
  \href{https://doi.org/10.1088/1126-6708/2007/05/060}{\emph{Journal of High
  Energy Physics} {\bfseries 05} (2007) 060},
  [\href{https://arxiv.org/abs/hep-th/0703287}{{\ttfamily hep-th/0703287}}].

\bibitem{ArkaniHamed:2001ca}
N.~Arkani-Hamed, A.~G. Cohen and H.~Georgi, \emph{{(De)constructing
  dimensions}},
  \href{https://doi.org/10.1103/PhysRevLett.86.4757}{\emph{Physical Review
  Letters} {\bfseries 86} (2001) 4757--4761},
  [\href{https://arxiv.org/abs/hep-th/0104005}{{\ttfamily hep-th/0104005}}].

\bibitem{Hill:2000mu}
C.~T. Hill, S.~Pokorski and J.~Wang, \emph{{Gauge Invariant Effective
  Lagrangian for Kaluza-Klein Modes}},
  \href{https://doi.org/10.1103/PhysRevD.64.105005}{\emph{Phys. Rev. D}
  {\bfseries 64} (2001) 105005},
  [\href{https://arxiv.org/abs/hep-th/0104035}{{\ttfamily hep-th/0104035}}].

\bibitem{kaplan1991}
D.~B. Kaplan, \emph{{Flavor at SSC energies: a new mechanism for dynamically
  generated fermion masses}},
  \href{https://doi.org/https://doi.org/10.1016/S0550-3213(05)80021-5}{\emph{Nuclear
  Physics B} {\bfseries 365} (1991) 259--278}.

\bibitem{de2012}
S.~De~Curtis, M.~Redi and A.~Tesi, \emph{{The 4D Composite Higgs}},
  \href{https://doi.org/10.1007/JHEP04(2012)042}{\emph{Journal of High Energy
  Physics} {\bfseries 04} (2012) 1--30}.

\bibitem{panico2011}
G.~Panico and A.~Wulzer, \emph{{The Discrete Composite Higgs Model}},
  \href{https://doi.org/10.1007/JHEP09(2011)135}{\emph{JHEP} {\bfseries 09}
  (2011) 135}, [\href{https://arxiv.org/abs/1106.2719}{{\ttfamily 1106.2719}}].

\bibitem{MarzoccaGeneralCHMs}
D.~Marzocca, M.~Serone and J.~Shu, \emph{{General Composite Higgs Models}},
  \href{https://doi.org/10.1007/JHEP08(2012)013}{\emph{Journal of High Energy
  Physics} {\bfseries 08} (2012) 013},
  [\href{https://arxiv.org/abs/1205.0770}{{\ttfamily 1205.0770}}].

\bibitem{Panico:2015jxa}
G.~Panico and A.~Wulzer, \emph{{The Composite Nambu-Goldstone Higgs}},
  \href{https://doi.org/10.1007/978-3-319-22617-0}{\emph{Lecture Notes in
  Physics} (2016) }.

\bibitem{Bellazzini:2014yua}
B.~Bellazzini, C.~Csáki and J.~Serra, \emph{{Composite Higgses}},
  \href{https://doi.org/10.1140/epjc/s10052-014-2766-x}{\emph{European Physical
  Journal C} {\bfseries C74} (2014) 2766},
  [\href{https://arxiv.org/abs/1401.2457}{{\ttfamily 1401.2457}}].

\bibitem{gripaios2009beyond}
B.~Gripaios, A.~Pomarol, F.~Riva and J.~Serra, \emph{{Beyond the Minimal
  Composite Higgs Model}},
  \href{https://doi.org/10.1088/1126-6708/2009/04/070}{\emph{Journal of High
  Energy Physics} {\bfseries 04} (2009) 070},
  [\href{https://arxiv.org/abs/0902.1483}{{\ttfamily 0902.1483}}].

\bibitem{Redi:2012ha}
M.~Redi and A.~Tesi, \emph{{Implications of a Light Higgs in Composite
  Models}}, \href{https://doi.org/10.1007/JHEP10(2012)166}{\emph{Journal of
  High Energy Physics} {\bfseries 10} (2012) 166},
  [\href{https://arxiv.org/abs/1205.0232}{{\ttfamily 1205.0232}}].

\bibitem{Banerjee:2017qod}
A.~Banerjee, G.~Bhattacharyya and T.~S. Ray, \emph{{Improving Fine-tuning in
  Composite Higgs Models}},
  \href{https://doi.org/10.1103/PhysRevD.96.035040}{\emph{Physical Review}
  {\bfseries D96} (2017) 035040},
  [\href{https://arxiv.org/abs/1703.08011}{{\ttfamily 1703.08011}}].

\bibitem{panico2012}
G.~Panico, M.~Redi, A.~Tesi and A.~Wulzer, \emph{{On the Tuning and the Mass of
  the Composite Higgs}},
  \href{https://doi.org/10.1007/JHEP03(2013)051}{\emph{Journal of High Energy
  Physics} {\bfseries 03} (2013) 051},
  [\href{https://arxiv.org/abs/1210.7114}{{\ttfamily 1210.7114}}].

\bibitem{carmona2015}
A.~Carmona and F.~Goertz, \emph{{A naturally light Higgs without light top
  partners}}, \href{https://doi.org/10.1007/JHEP05(2015)002}{\emph{Journal of
  High Energy Physics} {\bfseries 05} (2015) 1--63}.

\bibitem{BarnardFT}
J.~Barnard, D.~Murnane, M.~White and A.~G. Williams, \emph{{Constraining fine
  tuning in Composite Higgs Models with partially composite leptons}},
  \href{https://doi.org/10.1007/JHEP09(2017)049}{\emph{Journal of High Energy
  Physics} {\bfseries 09} (2017) 049},
  [\href{https://arxiv.org/abs/1703.07653}{{\ttfamily 1703.07653}}].

\bibitem{Carena}
M.~Carena, L.~Da~Rold and E.~Pont{\'o}n, \emph{{Minimal Composite Higgs Models
  at the LHC}}, \href{https://doi.org/10.1007/JHEP06(2014)159}{\emph{Journal of
  High Energy Physics} {\bfseries 06} (2014) 159},
  [\href{https://arxiv.org/abs/1402.2987}{{\ttfamily 1402.2987}}].

\bibitem{Niehoff:2015iaa}
C.~Niehoff, P.~Stangl and D.~M. Straub, \emph{{Direct and indirect signals of
  natural composite Higgs models}},
  \href{https://doi.org/10.1007/JHEP01(2016)119}{\emph{Journal of High Energy
  Physics} {\bfseries 01} (2016) 119},
  [\href{https://arxiv.org/abs/1508.00569}{{\ttfamily 1508.00569}}].

\bibitem{BarnardCC}
J.~Barnard and M.~White, \emph{{Collider constraints on tuning in composite
  Higgs models}}, \href{https://doi.org/10.1007/JHEP10(2015)072}{\emph{Journal
  of High Energy Physics} {\bfseries 10} (2015) 072},
  [\href{https://arxiv.org/abs/1507.02332}{{\ttfamily 1507.02332}}].

\bibitem{niehoff2017electroweak}
C.~Niehoff, P.~Stangl and D.~M. Straub, \emph{{Electroweak symmetry breaking
  and collider signatures in the next-to-minimal composite Higgs model}},
  \href{https://doi.org/10.1007/JHEP04(2017)117}{\emph{Journal of High Energy
  Physics} {\bfseries 04} (2017) 117},
  [\href{https://arxiv.org/abs/1611.09356}{{\ttfamily 1611.09356}}].

\bibitem{matsedonskyi2012}
O.~Matsedonskyi, G.~Panico and A.~Wulzer, \emph{{Light Top Partners for a Light
  Composite Higgs}},
  \href{https://doi.org/10.1007/JHEP01(2013)164}{\emph{Journal of High Energy
  Physics} {\bfseries 01} (2013) 164},
  [\href{https://arxiv.org/abs/1204.6333}{{\ttfamily 1204.6333}}].

\bibitem{Blasi:2019jqc}
S.~Blasi and F.~Goertz, \emph{{Softened Symmetry Breaking in Composite Higgs
  Models}},
  \href{https://doi.org/10.1103/PhysRevLett.123.221801}{\emph{Physical Review
  Letters} {\bfseries 123} (2019) 221801},
  [\href{https://arxiv.org/abs/1903.06146}{{\ttfamily 1903.06146}}].

\bibitem{Blasi:2020ktl}
S.~Blasi, C.~Cs{\'a}ki and F.~Goertz, \emph{{A Natural Composite Higgs via
  Universal Boundary Conditions}},
  \href{https://arxiv.org/abs/2004.06120}{{\ttfamily 2004.06120}}.

\bibitem{Csaki:2017cep}
C.~Cs{\'a}ki, T.~Ma and J.~Shu, \emph{{Maximally Symmetric Composite Higgs
  Models}},
  \href{https://doi.org/10.1103/PhysRevLett.119.131803}{\emph{Physical Review
  Letters} {\bfseries 119} (2017) 131803},
  [\href{https://arxiv.org/abs/1702.00405}{{\ttfamily 1702.00405}}].

\bibitem{skilling2006}
J.~Skilling, \emph{Nested sampling for general bayesian computation},
  \href{https://doi.org/10.1214/06-BA127}{\emph{Bayesian Analysis} {\bfseries
  1} (2006) 833--859}.

\bibitem{Handley:2015fda}
W.~Handley, M.~Hobson and A.~Lasenby, \emph{{PolyChord: nested sampling for
  cosmology}}, \href{https://doi.org/10.1093/mnrasl/slv047}{\emph{Monthly
  Notices of the Royal Astronomical Society} {\bfseries 450} (2015) L61--L65},
  [\href{https://arxiv.org/abs/1502.01856}{{\ttfamily 1502.01856}}].

\bibitem{Handley_2015}
W.~J. Handley, M.~P. Hobson and A.~N. Lasenby, \emph{{PolyChord:
  next-generation nested sampling}},
  \href{https://doi.org/10.1093/mnras/stv1911}{\emph{Monthly Notices of the
  Royal Astronomical Society} {\bfseries 453} (2015) 4385--4399}.

\bibitem{kullback1951}
S.~Kullback and R.~A. Leibler, \emph{On information and sufficiency},
  \href{https://doi.org/10.1214/aoms/1177729694}{\emph{Annals of Mathematical
  Statistics} {\bfseries 22} (1951) 79--86}.

\bibitem{2020arXiv200715632H}
C.~{Heymans}, T.~{Tr{\"o}ster}, M.~{Asgari}, C.~{Blake}, H.~{Hildebrandt},
  B.~{Joachimi} et~al., \emph{{KiDS-1000 Cosmology: Multi-probe weak
  gravitational lensing and spectroscopic galaxy clustering constraints}},
  \href{https://arxiv.org/abs/2007.15632}{{\ttfamily 2007.15632}}.

\bibitem{hergt}
L.~T. {Hergt}, W.~J. {Handley}, M.~P. {Hobson} and A.~N. {Lasenby},
  \emph{{Bayesian evidence for the tensor-to-scalar ratio r and neutrino masses
  m$_\nu$: Effects of uniform vs logarithmic priors}}, {\emph{(Awaiting
  submission)} }.

\bibitem{PhysRevD.98.030001}
{\scshape Particle Data Group} collaboration, M.~Tanabashi, K.~Hagiwara,
  K.~Hikasa, K.~Nakamura, Y.~Sumino, F.~Takahashi et~al., \emph{Review of
  particle physics},
  \href{https://doi.org/10.1103/PhysRevD.98.030001}{\emph{Physical Review D}
  {\bfseries 98} (2018) 030001}.

\bibitem{Hoang:2014oea}
A.~H. Hoang, \emph{{The Top Mass: Interpretation and Theoretical
  Uncertainties}},  in \emph{{Proceedings, 7th International Workshop on Top
  Quark Physics (TOP2014)}}, 2014,
  \href{https://arxiv.org/abs/1412.3649}{{\ttfamily 1412.3649}}.

\bibitem{ALEPH:2005ab}
{\scshape ALEPH, DELPHI, L3, OPAL, SLD, LEP Electroweak Working Group, SLD
  Electroweak Group, SLD Heavy Flavour Group} collaboration, S.~Schael et~al.,
  \emph{{Precision electroweak measurements on the $Z$ resonance}},
  \href{https://doi.org/10.1016/j.physrep.2005.12.006}{\emph{Physics Reports}
  {\bfseries 427} (2006) 257--454},
  [\href{https://arxiv.org/abs/hep-ex/0509008}{{\ttfamily hep-ex/0509008}}].

\bibitem{Khachatryan:2016vau}
{\scshape ATLAS, CMS} collaboration, \emph{{Measurements of the Higgs boson
  production and decay rates and constraints on its couplings from a combined
  ATLAS and CMS analysis of the LHC pp collision data at $\sqrt{s}=7 $ and 8
  TeV}}, \href{https://doi.org/10.1007/JHEP08(2016)045}{\emph{Journal of High
  Energy Physics} {\bfseries 08} (2016) 045},
  [\href{https://arxiv.org/abs/1606.02266}{{\ttfamily 1606.02266}}].

\bibitem{Sirunyan:2018koj}
{\scshape CMS} collaboration, \emph{{Combined measurements of Higgs boson
  couplings in proton--proton collisions at $\sqrt{s}=13\,\text {Te}\text {V}
  $}}, \href{https://doi.org/10.1140/epjc/s10052-019-6909-y}{\emph{European
  Physical Journal} {\bfseries C79} (2019) 421},
  [\href{https://arxiv.org/abs/1809.10733}{{\ttfamily 1809.10733}}].

\bibitem{ATLAS-CONF-2018-031}
{\scshape ATLAS} collaboration, \emph{{Combined measurements of Higgs boson
  production and decay using up to 80 fb$^{-1}$ of proton--proton collision
  data at $\sqrt{s}=$ 13 TeV collected with the ATLAS experiment}},  Tech. Rep.
  ATLAS-CONF-2018-031, CERN, Geneva, 2018.

\bibitem{CMS:1900lgv}
{\scshape CMS} collaboration, \emph{{Measurements of Higgs boson production via
  gluon fusion and vector boson fusion in the diphoton decay channel at
  $\sqrt{s} = 13$ TeV}},  Tech. Rep. CMS-PAS-HIG-18-029, Geneva, 2019.

\bibitem{Peskin:1991sw}
M.~E. Peskin and T.~Takeuchi, \emph{{Estimation of oblique electroweak
  corrections}}, \href{https://doi.org/10.1103/PhysRevD.46.381}{\emph{Physical
  Review} {\bfseries D46} (1992) 381--409}.

\bibitem{Barbieri:2004qk}
R.~Barbieri, A.~Pomarol, R.~Rattazzi and A.~Strumia, \emph{{Electroweak
  symmetry breaking after LEP-1 and LEP-2}},
  \href{https://doi.org/10.1016/j.nuclphysb.2004.10.014}{\emph{Nuclear Physics}
  {\bfseries B703} (2004) 127--146},
  [\href{https://arxiv.org/abs/hep-ph/0405040}{{\ttfamily hep-ph/0405040}}].

\bibitem{Banerjee:2017wmg}
A.~Banerjee, G.~Bhattacharyya, N.~Kumar and T.~S. Ray, \emph{{Constraining
  Composite Higgs Models using LHC data}},
  \href{https://doi.org/10.1007/JHEP03(2018)062}{\emph{Journal of High Energy
  Physics} {\bfseries 03} (2018) 062},
  [\href{https://arxiv.org/abs/1712.07494}{{\ttfamily 1712.07494}}].

\bibitem{Banerjee:2020tqc}
A.~Banerjee and G.~Bhattacharyya, \emph{{Probing the Higgs boson through Yukawa
  force}}, \href{https://doi.org/10.1016/j.nuclphysb.2020.115261}{\emph{Nuclear
  Physics B} {\bfseries 961} (2020) 115261},
  [\href{https://arxiv.org/abs/2006.01164}{{\ttfamily 2006.01164}}].

\bibitem{anesthetic}
W.~Handley, \emph{anesthetic: nested sampling visualisation},
  \href{https://doi.org/10.21105/joss.01414}{\emph{The Journal of Open Source
  Software} {\bfseries 4} (Jun, 2019) 1414}.

\bibitem{Murnane:2018ynd}
D.~Murnane, M.~J. White and A.~G. Williams, \emph{{Exploring Fine-tuning of the
  Next-to-Minimal Composite Higgs Model}},
  \href{https://doi.org/10.1007/JHEP04(2019)076}{\emph{Journal of High Energy
  Physics} {\bfseries 04} (2019) 076},
  [\href{https://arxiv.org/abs/1810.08355}{{\ttfamily 1810.08355}}].

\bibitem{Handley:2019pqx}
W.~Handley and P.~Lemos, \emph{{Quantifying dimensionality: Bayesian
  cosmological model complexities}},
  \href{https://doi.org/10.1103/PhysRevD.100.023512}{\emph{Physical Review D}
  {\bfseries 100} (2019) 023512},
  [\href{https://arxiv.org/abs/1903.06682}{{\ttfamily 1903.06682}}].

\bibitem{Barducci:2012kk}
D.~Barducci, A.~Belyaev, S.~De~Curtis, S.~Moretti and G.~M. Pruna,
  \emph{{Exploring Drell-Yan signals from the 4D Composite Higgs Model at the
  LHC}}, \href{https://doi.org/10.1007/JHEP04(2013)152}{\emph{Journal of High
  Energy Physics} {\bfseries 04} (2013) 152},
  [\href{https://arxiv.org/abs/1210.2927}{{\ttfamily 1210.2927}}].

\bibitem{Barducci:2012sk}
D.~Barducci, S.~De~Curtis, K.~Mimasu and S.~Moretti, \emph{{Multiple Z$^\prime
  \to t \bar{t}$ signals in a 4D Composite Higgs Model}},
  \href{https://doi.org/10.1103/PhysRevD.88.074024}{\emph{Physical Review D}
  {\bfseries 88} (2013) 074024},
  [\href{https://arxiv.org/abs/1212.5948}{{\ttfamily 1212.5948}}].

\bibitem{Low:2010mr}
I.~Low and A.~Vichi, \emph{{On the production of a composite Higgs boson}},
  \href{https://doi.org/10.1103/PhysRevD.84.045019}{\emph{Physical Review D}
  {\bfseries 84} (2011) 045019},
  [\href{https://arxiv.org/abs/1010.2753}{{\ttfamily 1010.2753}}].

\bibitem{Barducci:2013wjc}
D.~Barducci, A.~Belyaev, M.~Brown, S.~De~Curtis, S.~Moretti and G.~Pruna,
  \emph{{The 4-Dimensional Composite Higgs Model (4DCHM) and the 125 GeV
  Higgs-like signals at the LHC}},
  \href{https://doi.org/10.1007/JHEP09(2013)047}{\emph{Journal of High Energy
  Physics} {\bfseries 09} (2013) 047},
  [\href{https://arxiv.org/abs/1302.2371}{{\ttfamily 1302.2371}}].

\bibitem{CMS-DP-2016-064}
{\scshape CMS} collaboration, \emph{{Updates on Projections of Physics Reach
  with the Upgraded CMS Detector for High Luminosity LHC}},  Tech. Rep.
  CMS-DP-2016-064, 2016.

\bibitem{Savin:2015gda}
{\scshape ATLAS, CMS} collaboration, A.~A. Savin, \emph{{Prospects for Higgs
  and SM measurements at the HL-LHC}},
  \href{https://doi.org/10.1051/epjconf/20149504060}{\emph{European Physical
  Journal Web of Conferences} {\bfseries 95} (2016) 04060}.

\bibitem{Montull:2013mla}
M.~Montull, F.~Riva, E.~Salvioni and R.~Torre, \emph{{Higgs Couplings in
  Composite Models}},
  \href{https://doi.org/10.1103/PhysRevD.88.095006}{\emph{Physical Review}
  {\bfseries D88} (2013) 095006},
  [\href{https://arxiv.org/abs/1308.0559}{{\ttfamily 1308.0559}}].

\bibitem{Aad:2012dm}
{\scshape ATLAS} collaboration, \emph{{ATLAS search for a heavy gauge boson
  decaying to a charged lepton and a neutrino in $pp$ collisions at
  $\sqrt{s}=7$ TeV}},
  \href{https://doi.org/10.1140/epjc/s10052-012-2241-5}{\emph{European Physical
  Journal C} {\bfseries 72} (2012) 2241},
  [\href{https://arxiv.org/abs/1209.4446}{{\ttfamily 1209.4446}}].

\bibitem{CMS:2015kjy}
{\scshape CMS} collaboration, \emph{{Search for SSM W' production, in the
  lepton+MET final state at a center-of-mass energy of 13 TeV}},  Tech. Rep.
  CMS-PAS-EXO-15-006, 2015.

\bibitem{ATLAS:2016ecs}
{\scshape ATLAS} collaboration, \emph{{Search for new resonances decaying to a
  charged lepton and a neutrino in pp collisions at $\sqrt{s} = 13$ TeV with
  the ATLAS detector}},  Tech. Rep. ATLAS-CONF-2016-061, 2016.

\bibitem{ATLAS:2018lcz}
{\scshape ATLAS} collaboration, \emph{{Search for a new heavy gauge boson
  resonance decaying into a lepton and missing transverse momentum in 79.8
  fb$^{-1}$ of $pp$ collisions at $\sqrt{s} = 13$ TeV with the ATLAS
  experiment}},  Tech. Rep. ATLAS-CONF-2018-017, 2018.

\bibitem{Khachatryan:2015pua}
{\scshape CMS} collaboration, \emph{{Search for W' decaying to tau lepton and
  neutrino in proton-proton collisions at $\sqrt{s} =$ 8 TeV}},
  \href{https://doi.org/10.1016/j.physletb.2016.02.002}{\emph{Physics Letters}
  {\bfseries B755} (2016) 196--216},
  [\href{https://arxiv.org/abs/1508.04308}{{\ttfamily 1508.04308}}].

\bibitem{CMS:2016ppa}
{\scshape CMS} collaboration, \emph{{Search for W' decaying to tau lepton and
  neutrino in proton-proton collisions at $\sqrt{s} =$ 13 TeV}},  Tech. Rep.
  CMS-PAS-EXO-16-006, 2016.

\bibitem{Aad:2014cka}
{\scshape ATLAS} collaboration, \emph{{Search for high-mass dilepton resonances
  in pp collisions at $\sqrt{s}=8$ TeV with the ATLAS detector}},
  \href{https://doi.org/10.1103/PhysRevD.90.052005}{\emph{Physical Review}
  {\bfseries D90} (2014) 052005},
  [\href{https://arxiv.org/abs/1405.4123}{{\ttfamily 1405.4123}}].

\bibitem{Khachatryan:2014fba}
{\scshape CMS} collaboration, \emph{{Search for physics beyond the standard
  model in dilepton mass spectra in proton-proton collisions at $\sqrt{s}=8 $
  TeV}}, \href{https://doi.org/10.1007/JHEP04(2015)025}{\emph{Journal of High
  Energy Physics} {\bfseries 04} (2015) 025},
  [\href{https://arxiv.org/abs/1412.6302}{{\ttfamily 1412.6302}}].

\bibitem{CMS:2018wsn}
{\scshape CMS} collaboration, \emph{{Search for high mass resonances in
  dielectron final state}},  Tech. Rep. CMS-PAS-EXO-18-006, 2018.

\bibitem{CMS:2016abv}
{\scshape CMS} collaboration, \emph{{Search for a high-mass resonance decaying
  into a dilepton final state in 13 fb$^{-1}$ of pp collisions at
  $\sqrt{s}=13~\mathrm{TeV}$}},  Tech. Rep. CMS-PAS-EXO-16-031, 2016.

\bibitem{ATLAS:2016cyf}
{\scshape ATLAS} collaboration, \emph{{Search for new high-mass resonances in
  the dilepton final state using proton-proton collisions at $\sqrt{s}$ = 13
  TeV with the ATLAS detector}},  Tech. Rep. ATLAS-CONF-2016-045, 2016.

\bibitem{Aaboud:2017buh}
{\scshape ATLAS} collaboration, \emph{{Search for new high-mass phenomena in
  the dilepton final state using 36 fb$^{-1}$ of proton-proton collision data
  at $\sqrt{s}=13 $ TeV with the ATLAS detector}},
  \href{https://doi.org/10.1007/JHEP10(2017)182}{\emph{Journal of High Energy
  Physics} {\bfseries 10} (2017) 182},
  [\href{https://arxiv.org/abs/1707.02424}{{\ttfamily 1707.02424}}].

\bibitem{Aad:2019fac}
{\scshape ATLAS} collaboration, \emph{{Search for high-mass dilepton resonances
  using 139 fb$^{-1}$ of $pp$ collision data collected at $\sqrt{s}=$13 TeV
  with the ATLAS detector}},
  \href{https://doi.org/10.1016/j.physletb.2019.07.016}{\emph{Physics Letters}
  {\bfseries B796} (2019) 68--87},
  [\href{https://arxiv.org/abs/1903.06248}{{\ttfamily 1903.06248}}].

\bibitem{Sirunyan:2018exx}
{\scshape CMS} collaboration, \emph{{Search for high-mass resonances in
  dilepton final states in proton-proton collisions at $\sqrt{s}=$ 13 TeV}},
  \href{https://doi.org/10.1007/JHEP06(2018)120}{\emph{Journal of High Energy
  Physics} {\bfseries 06} (2018) 120},
  [\href{https://arxiv.org/abs/1803.06292}{{\ttfamily 1803.06292}}].

\bibitem{Aad:2015osa}
{\scshape ATLAS} collaboration, \emph{{A search for high-mass resonances
  decaying to $\tau^{+}\tau^{-}$ in $pp$ collisions at $\sqrt{s}=8$ TeV with
  the ATLAS detector}},
  \href{https://doi.org/10.1007/JHEP07(2015)157}{\emph{Journal of High Energy
  Physics} {\bfseries 07} (2015) 157},
  [\href{https://arxiv.org/abs/1502.07177}{{\ttfamily 1502.07177}}].

\bibitem{CMS:2015ufa}
{\scshape CMS} collaboration, \emph{{Z' to $\tau \tau - e\mu$ final state}},
  Tech. Rep. CMS-PAS-EXO-12-046, CERN, Geneva, 2015.

\bibitem{CMS:2016zxk}
{\scshape CMS} collaboration, \emph{{Search for new physics with high-mass tau
  lepton pairs in pp collisions at $\sqrt{s} =$ 13 TeV with the CMS detector}},
   Tech. Rep. CMS-PAS-EXO-16-008, 2016.

\bibitem{CMS:2016wpz}
{\scshape CMS} collaboration, \emph{{Searches for narrow resonances decaying to
  dijets in proton-proton collisions at 13 TeV using 12.9 inverse
  femtobarns.}},  Tech. Rep. CMS-PAS-EXO-16-032, 2016.

\bibitem{Khachatryan:2015dcf}
{\scshape CMS} collaboration, \emph{{Search for narrow resonances decaying to
  dijets in proton-proton collisions at $\sqrt{s} =$ 13 TeV}},
  \href{https://doi.org/10.1103/PhysRevLett.116.071801}{\emph{Physical Review
  Letters} {\bfseries 116} (2016) 071801},
  [\href{https://arxiv.org/abs/1512.01224}{{\ttfamily 1512.01224}}].

\bibitem{Sirunyan:2018xlo}
{\scshape CMS} collaboration, \emph{{Search for narrow and broad dijet
  resonances in proton-proton collisions at $\sqrt{s}=13 $ TeV and constraints
  on dark matter mediators and other new particles}},
  \href{https://doi.org/10.1007/JHEP08(2018)130}{\emph{Journal of High Energy
  Physics} {\bfseries 08} (2018) 130},
  [\href{https://arxiv.org/abs/1806.00843}{{\ttfamily 1806.00843}}].

\bibitem{ATLAS:2015nsi}
{\scshape ATLAS} collaboration, \emph{{Search for new phenomena in dijet mass
  and angular distributions from $pp$ collisions at $\sqrt{s}=$ 13 TeV with the
  ATLAS detector}},
  \href{https://doi.org/10.1016/j.physletb.2016.01.032}{\emph{Physics Letters}
  {\bfseries B754} (2016) 302--322},
  [\href{https://arxiv.org/abs/1512.01530}{{\ttfamily 1512.01530}}].

\bibitem{Aaboud:2019zxd}
{\scshape ATLAS} collaboration, \emph{{Search for low-mass resonances decaying
  into two jets and produced in association with a photon using $pp$ collisions
  at $\sqrt{s} = 13$ TeV with the ATLAS detector}},
  \href{https://doi.org/10.1016/j.physletb.2019.03.067}{\emph{Physics Letters}
  {\bfseries B795} (2019) 56--75},
  [\href{https://arxiv.org/abs/1901.10917}{{\ttfamily 1901.10917}}].

\bibitem{Chatrchyan:2014koa}
{\scshape CMS} collaboration, \emph{{Search for W' $\to $ tb decays in the
  lepton + jets final state in pp collisions at $\sqrt{s}$ = 8 TeV}},
  \href{https://doi.org/10.1007/JHEP05(2014)108}{\emph{Journal of High Energy
  Physics} {\bfseries 05} (2014) 108},
  [\href{https://arxiv.org/abs/1402.2176}{{\ttfamily 1402.2176}}].

\bibitem{Khachatryan:2015edz}
{\scshape CMS} collaboration, \emph{{Search for $W' \to tb$ in proton-proton
  collisions at $\sqrt{s} = $ 8 TeV}},
  \href{https://doi.org/10.1007/JHEP02(2016)122}{\emph{Journal of High Energy
  Physics} {\bfseries 02} (2016) 122},
  [\href{https://arxiv.org/abs/1509.06051}{{\ttfamily 1509.06051}}].

\bibitem{CMS:2016ude}
{\scshape CMS} collaboration, \emph{{Search for W' to tb in pp collisions at
  $\sqrt{s} =$ 13 TeV}},  Tech. Rep. CMS-PAS-B2G-16-009, 2016.

\bibitem{CMS:2016wqa}
{\scshape CMS} collaboration, \emph{{Search for W' boson resonances decaying
  into a top quark and a bottom quark in the leptonic final state at
  $\sqrt{s}=13$ TeV}},  Tech. Rep. CMS-PAS-B2G-16-017, 2016.

\bibitem{Aaboud:2018juj}
{\scshape ATLAS} collaboration, \emph{{Search for $W' \rightarrow tb$ decays in
  the hadronic final state using pp collisions at $\sqrt{s}=13$ TeV with the
  ATLAS detector}},
  \href{https://doi.org/10.1016/j.physletb.2018.03.036}{\emph{Physics Letters}
  {\bfseries B781} (2018) 327--348},
  [\href{https://arxiv.org/abs/1801.07893}{{\ttfamily 1801.07893}}].

\bibitem{ATLAS:2015aka}
{\scshape ATLAS} collaboration, \emph{{A search for $\mathbf{t\bar{t}}$
  resonances using lepton plus jets events in proton-proton collisions at
  $\sqrt{s}= 8$ TeV with the ATLAS detector}},  Tech. Rep. ATLAS-CONF-2015-009,
  2015.

\bibitem{Khachatryan:2015sma}
{\scshape CMS} collaboration, \emph{{Search for resonant $t \bar t$ production
  in proton-proton collisions at $\sqrt s=$ 8 TeV}},
  \href{https://doi.org/10.1103/PhysRevD.93.012001}{\emph{Physical Review}
  {\bfseries D93} (2016) 012001},
  [\href{https://arxiv.org/abs/1506.03062}{{\ttfamily 1506.03062}}].

\bibitem{CMS:2016ehh}
{\scshape CMS} collaboration, \emph{{Search for top quark-antiquark resonances
  in the all-hadronic final state at $\sqrt{s}=$13 TeV}},  Tech. Rep.
  CMS-PAS-B2G-15-003, 2016.

\bibitem{CMS:2016zte}
{\scshape CMS} collaboration, \emph{{Search for $\mathrm{t\bar{t}}$ resonances
  in boosted semileptonic final states in pp collisions at
  $\sqrt{s}=13~\mathrm{TeV}$}},  Tech. Rep. CMS-PAS-B2G-15-002, 2016.

\bibitem{Sirunyan:2018ryr}
{\scshape CMS} collaboration, \emph{{Search for resonant $
  \mathrm{t}\overline{\mathrm{t}} $ production in proton-proton collisions at
  $\sqrt{s}=13 $ TeV}},
  \href{https://doi.org/10.1007/JHEP04(2019)031}{\emph{Journal of High Energy
  Physics} {\bfseries 04} (2019) 031},
  [\href{https://arxiv.org/abs/1810.05905}{{\ttfamily 1810.05905}}].

\bibitem{Aaboud:2018mjh}
{\scshape ATLAS} collaboration, \emph{{Search for heavy particles decaying into
  top-quark pairs using lepton-plus-jets events in proton–proton collisions
  at $\sqrt{s} = 13$ $\text { TeV}$ with the ATLAS detector}},
  \href{https://doi.org/10.1140/epjc/s10052-018-5995-6}{\emph{European Physical
  Journal} {\bfseries C78} (2018) 565},
  [\href{https://arxiv.org/abs/1804.10823}{{\ttfamily 1804.10823}}].

\bibitem{Aaboud:2019roo}
{\scshape ATLAS} collaboration, \emph{{Search for heavy particles decaying into
  a top-quark pair in the fully hadronic final state in $pp$ collisions at
  $\sqrt{s} =$ 13 TeV with the ATLAS detector}},
  \href{https://doi.org/10.1103/PhysRevD.99.092004}{\emph{Physical Review}
  {\bfseries D99} (2019) 092004},
  [\href{https://arxiv.org/abs/1902.10077}{{\ttfamily 1902.10077}}].

\bibitem{Aad:2015yza}
{\scshape ATLAS} collaboration, \emph{{Search for a new resonance decaying to a
  W or Z boson and a Higgs boson in the $\ell \ell / \ell \nu / \nu \nu + b
  \bar{b}$ final states with the ATLAS detector}},
  \href{https://doi.org/10.1140/epjc/s10052-015-3474-x}{\emph{European Physical
  Journal} {\bfseries C75} (2015) 263},
  [\href{https://arxiv.org/abs/1503.08089}{{\ttfamily 1503.08089}}].

\bibitem{Khachatryan:2015ywa}
{\scshape CMS} collaboration, \emph{{Search for Narrow High-Mass Resonances in
  Proton–Proton Collisions at $\sqrt{s}$ = 8 TeV Decaying to a Z and a Higgs
  Boson}}, \href{https://doi.org/10.1016/j.physletb.2015.07.011}{\emph{Physics
  Letters} {\bfseries B748} (2015) 255--277},
  [\href{https://arxiv.org/abs/1502.04994}{{\ttfamily 1502.04994}}].

\bibitem{CMS:2016dzw}
{\scshape CMS} collaboration, \emph{{Search for heavy resonances decaying into
  a vector boson and a Higgs boson in the (ll, l$\nu$, $\nu\nu$) bb final
  state}},  Tech. Rep. CMS-PAS-B2G-16-003, 2016.

\bibitem{Sirunyan:2018fuh}
{\scshape CMS} collaboration, \emph{{Search for heavy resonances decaying into
  two Higgs bosons or into a Higgs boson and a W or Z boson in proton-proton
  collisions at 13 TeV}},
  \href{https://doi.org/10.1007/JHEP01(2019)051}{\emph{Journal of High Energy
  Physics} {\bfseries 01} (2019) 051},
  [\href{https://arxiv.org/abs/1808.01365}{{\ttfamily 1808.01365}}].

\bibitem{Sirunyan:2017wto}
{\scshape CMS} collaboration, \emph{{Search for heavy resonances that decay
  into a vector boson and a Higgs boson in hadronic final states at $\sqrt{s} =
  13$ $\,\text {TeV}$}},
  \href{https://doi.org/10.1140/epjc/s10052-017-5192-z}{\emph{European Physical
  Journal} {\bfseries C77} (2017) 636},
  [\href{https://arxiv.org/abs/1707.01303}{{\ttfamily 1707.01303}}].

\bibitem{Sirunyan:2018qob}
{\scshape CMS} collaboration, \emph{{Search for heavy resonances decaying into
  a vector boson and a Higgs boson in final states with charged leptons,
  neutrinos and b quarks at $\sqrt{s}=13 $ TeV}},
  \href{https://doi.org/10.1007/JHEP11(2018)172}{\emph{Journal of High Energy
  Physics} {\bfseries 11} (2018) 172},
  [\href{https://arxiv.org/abs/1807.02826}{{\ttfamily 1807.02826}}].

\bibitem{Aaboud:2016lwx}
{\scshape ATLAS} collaboration, \emph{{Search for new resonances decaying to a
  $W$ or $Z$ boson and a Higgs boson in the $\ell^+ \ell^- b\bar b$, $\ell \nu
  b\bar b$, and $\nu\bar{\nu} b\bar b$ channels with $pp$ collisions at $\sqrt
  s = 13$ TeV with the ATLAS detector}},
  \href{https://doi.org/10.1016/j.physletb.2016.11.045}{\emph{Physics Letters}
  {\bfseries B765} (2017) 32--52},
  [\href{https://arxiv.org/abs/1607.05621}{{\ttfamily 1607.05621}}].

\bibitem{TheATLAScollaboration:2015ulg}
{\scshape ATLAS} collaboration, \emph{{Search for new resonances decaying to a
  W or Z boson and a Higgs boson in the $\ell\ell b\bar b$, $\ell\nu b\bar b$,
  and $\nu\nu b\bar b$ channels in $pp$ collisions at $\sqrt s = 13$~TeV with
  the ATLAS detector}},  Tech. Rep. ATLAS-CONF-2015-074, 2015.

\bibitem{ATLAS:2016kxc}
{\scshape ATLAS} collaboration, \emph{{A Search for Resonances Decaying to a
  $W$ or $Z$ Boson and a Higgs Boson in the $q\bar{q}^{(\prime)}b\bar{b}$ Final
  State}},  Tech. Rep. ATLAS-CONF-2016-083, 2016.

\bibitem{Aad:2015owa}
{\scshape ATLAS} collaboration, \emph{{Search for high-mass diboson resonances
  with boson-tagged jets in proton-proton collisions at $ \sqrt{s}=8 $ TeV with
  the ATLAS detector}},
  \href{https://doi.org/10.1007/JHEP12(2015)055}{\emph{Journal of High Energy
  Physics} {\bfseries 12} (2015) 055},
  [\href{https://arxiv.org/abs/1506.00962}{{\ttfamily 1506.00962}}].

\bibitem{Aad:2015ufa}
{\scshape ATLAS} collaboration, \emph{{Search for production of $WW/WZ$
  resonances decaying to a lepton, neutrino and jets in $pp$ collisions at
  $\sqrt{s}=8$ TeV with the ATLAS detector}},
  \href{https://doi.org/10.1140/epjc/s10052-015-3593-4,
  10.1140/epjc/s10052-015-3425-6}{\emph{European Physical Journal} {\bfseries
  C75} (2015) 209}, [\href{https://arxiv.org/abs/1503.04677}{{\ttfamily
  1503.04677}}].

\bibitem{Aad:2014pha}
{\scshape ATLAS} collaboration, \emph{{Search for WZ resonances in the fully
  leptonic channel using pp collisions at $\sqrt{s} =$ 8 TeV with the ATLAS
  detector}},
  \href{https://doi.org/10.1016/j.physletb.2014.08.039}{\emph{Physics Letters}
  {\bfseries B737} (2014) 223--243},
  [\href{https://arxiv.org/abs/1406.4456}{{\ttfamily 1406.4456}}].

\bibitem{Khachatryan:2014hpa}
{\scshape CMS} collaboration, \emph{{Search for massive resonances in dijet
  systems containing jets tagged as W or Z boson decays in pp collisions at
  $\sqrt{s} $ = 8 TeV}},
  \href{https://doi.org/10.1007/JHEP08(2014)173}{\emph{Journal of High Energy
  Physics} {\bfseries 08} (2014) 173},
  [\href{https://arxiv.org/abs/1405.1994}{{\ttfamily 1405.1994}}].

\bibitem{CMS:2015nmz}
{\scshape CMS} collaboration, \emph{{Search for massive resonances decaying
  into pairs of boosted W and Z bosons at $\sqrt{s}$ = 13 TeV}},  Tech. Rep.
  CMS-PAS-EXO-15-002, 2015.

\bibitem{CMS:2016pfl}
{\scshape CMS} collaboration, \emph{{Search for new resonances decaying to
  $\mathrm{WW}/\mathrm{WZ} \to \ell\nu \mathrm{qq}$}},  Tech. Rep.
  CMS-PAS-B2G-16-020, 2016.

\bibitem{Sirunyan:2018iff}
{\scshape CMS} collaboration, \emph{{Search for a heavy resonance decaying to a
  pair of vector bosons in the lepton plus merged jet final state at
  $\sqrt{s}=13 $ TeV}},
  \href{https://doi.org/10.1007/JHEP05(2018)088}{\emph{Journal of High Energy
  Physics} {\bfseries 05} (2018) 088},
  [\href{https://arxiv.org/abs/1802.09407}{{\ttfamily 1802.09407}}].

\bibitem{Sirunyan:2017acf}
{\scshape CMS} collaboration, \emph{{Search for massive resonances decaying
  into $WW$, $WZ$, $ZZ$, $qW$, and $qZ$ with dijet final states at
  $\sqrt{s}=13\text{ }\text{ }\mathrm{TeV}$}},
  \href{https://doi.org/10.1103/PhysRevD.97.072006}{\emph{Physical Review}
  {\bfseries D97} (2018) 072006},
  [\href{https://arxiv.org/abs/1708.05379}{{\ttfamily 1708.05379}}].

\bibitem{Sirunyan:2018ivv}
{\scshape CMS} collaboration, \emph{{Search for a heavy resonance decaying into
  a Z boson and a vector boson in the $ \nu
  \overline{\nu}\mathrm{q}\overline{\mathrm{q}} $ final state}},
  \href{https://doi.org/10.1007/JHEP07(2018)075}{\emph{Journal of High Energy
  Physics} {\bfseries 07} (2018) 075},
  [\href{https://arxiv.org/abs/1803.03838}{{\ttfamily 1803.03838}}].

\bibitem{Sirunyan:2018hsl}
{\scshape CMS} collaboration, \emph{{Search for a heavy resonance decaying into
  a Z boson and a Z or W boson in 2$\ell$2q final states at $\sqrt{s}=13 $
  TeV}}, \href{https://doi.org/10.1007/JHEP09(2018)101}{\emph{Journal of High
  Energy Physics} {\bfseries 09} (2018) 101},
  [\href{https://arxiv.org/abs/1803.10093}{{\ttfamily 1803.10093}}].

\bibitem{Sirunyan:2019jbg}
{\scshape CMS} collaboration, \emph{{A multi-dimensional search for new heavy
  resonances decaying to boosted WW, WZ, or ZZ boson pairs in the dijet final
  state at 13 TeV}},
  \href{https://doi.org/10.1140/epjc/s10052-020-7773-5}{\emph{European Physical
  Journal C} {\bfseries 80} (2019) 237},
  [\href{https://arxiv.org/abs/1906.05977}{{\ttfamily 1906.05977}}].

\bibitem{ATLAS:2016yqq}
{\scshape ATLAS} collaboration, \emph{{Search for resonances with boson-tagged
  jets in 15.5 fb$^{-1}$ of $pp$ collisions at $\sqrt{s} = 13$ TeV collected
  with the ATLAS detector}},  Tech. Rep. ATLAS-CONF-2016-055, 2016.

\bibitem{ATLAS:2016cwq}
{\scshape ATLAS} collaboration, \emph{{Search for diboson resonance production
  in the $\ell\nu qq$ final state using $pp$ collisions at $\sqrt{s}=13$ TeV
  with the ATLAS detector at the LHC}},  Tech. Rep. ATLAS-CONF-2016-062, 2016.

\bibitem{ATLAS:2016npe}
{\scshape ATLAS} collaboration, \emph{{Searches for heavy ZZ and ZW resonances
  in the llqq and $\nu\nu$qq final states in pp collisions at $\sqrt{s} =$ 13
  TeV with the ATLAS detector}},  Tech. Rep. ATLAS-CONF-2016-082, 2016.

\bibitem{Khachatryan:2016yji}
{\scshape CMS} collaboration, \emph{{Search for massive WH resonances decaying
  into the $\ell \nu \mathrm{b} \overline{\mathrm{b}} $ final state at
  $\sqrt{s}=8$ $~\text {TeV}$}},
  \href{https://doi.org/10.1140/epjc/s10052-016-4067-z}{\emph{European Physical
  Journal} {\bfseries C76} (2016) 237},
  [\href{https://arxiv.org/abs/1601.06431}{{\ttfamily 1601.06431}}].

\bibitem{Khachatryan:2014gha}
{\scshape CMS} collaboration, \emph{{Search for massive resonances decaying
  into pairs of boosted bosons in semi-leptonic final states at $\sqrt{s}=$ 8
  TeV}}, \href{https://doi.org/10.1007/JHEP08(2014)174}{\emph{Journal of High
  Energy Physics} {\bfseries 08} (2014) 174},
  [\href{https://arxiv.org/abs/1405.3447}{{\ttfamily 1405.3447}}].

\bibitem{CMS:2016wev}
{\scshape CMS} collaboration, \emph{{Combination of diboson resonance searches
  at 8 and 13 TeV}},  Tech. Rep. CMS-PAS-B2G-16-007, 2016.

\bibitem{CDF:10110}
{\scshape CDF} collaboration, \emph{{Search for Heavy Top $t' \to Wq$ in Lepton
  Plus Jets Events in $\int \mathcal{L} dt = 4.6$ fb$^{-1}$}},  Tech. Rep.
  10110, 2010.

\bibitem{Aad:2012bt}
{\scshape ATLAS} collaboration, \emph{{Search for pair-produced heavy quarks
  decaying to Wq in the two-lepton channel at $\sqrt{s}=7$ TeV with the ATLAS
  detector}}, \href{https://doi.org/10.1103/PhysRevD.86.012007}{\emph{Physical
  Review} {\bfseries D86} (2012) 012007},
  [\href{https://arxiv.org/abs/1202.3389}{{\ttfamily 1202.3389}}].

\bibitem{Aad:2015tba}
{\scshape ATLAS} collaboration, \emph{{Search for pair production of a new
  heavy quark that decays into a $W$ boson and a light quark in $pp$ collisions
  at $\sqrt{s} = 8$ TeV with the ATLAS detector}},
  \href{https://doi.org/10.1103/PhysRevD.92.112007}{\emph{Physical Review}
  {\bfseries D92} (2015) 112007},
  [\href{https://arxiv.org/abs/1509.04261}{{\ttfamily 1509.04261}}].

\bibitem{CMS:2014dka}
{\scshape CMS} collaboration, \emph{{Search for vector-like quarks in final
  states with a single lepton and jets in pp collisions at $\sqrt{s} =$ 8
  TeV}},  Tech. Rep. CMS-PAS-B2G-12-017, 2014.

\bibitem{ATLAS:2012qe}
{\scshape ATLAS} collaboration, \emph{{Search for pair production of heavy
  top-like quarks decaying to a high-pT $W$ boson and a $b$ quark in the lepton
  plus jets final state at $\sqrt{s}$=7 TeV with the ATLAS detector}},
  \href{https://doi.org/10.1016/j.physletb.2012.11.071}{\emph{Physics Letters
  B} {\bfseries 718} (2013) 1284--1302},
  [\href{https://arxiv.org/abs/1210.5468}{{\ttfamily 1210.5468}}].

\bibitem{CMS:2012ab}
{\scshape CMS} collaboration, \emph{{Search for heavy, top-like quark pair
  production in the dilepton final state in $pp$ collisions at $\sqrt{s} = 7$
  TeV}}, \href{https://doi.org/10.1016/j.physletb.2012.07.059}{\emph{Physics
  Letters} {\bfseries B716} (2012) 103--121},
  [\href{https://arxiv.org/abs/1203.5410}{{\ttfamily 1203.5410}}].

\bibitem{Chatrchyan:2012vu}
{\scshape CMS} collaboration, \emph{{Search for Pair Produced Fourth-Generation
  Up-Type Quarks in $pp$ Collisions at $\sqrt{s}=7$ TeV with a Lepton in the
  Final State}},
  \href{https://doi.org/10.1016/j.physletb.2012.10.038}{\emph{Physics Letters}
  {\bfseries B718} (2012) 307--328},
  [\href{https://arxiv.org/abs/1209.0471}{{\ttfamily 1209.0471}}].

\bibitem{ATLAS:2015dka}
{\scshape ATLAS} collaboration, \emph{{Search for production of vector-like
  quark pairs and of four top quarks in the lepton plus jets final state in
  $pp$ collisions at $\sqrt{s}=8$ TeV with the ATLAS detector}},  Tech. Rep.
  ATLAS-CONF-2015-012, 2015.

\bibitem{Khachatryan:2015oba}
{\scshape CMS} collaboration, \emph{{Search for vector-like charge 2/3 T quarks
  in proton-proton collisions at $\sqrt{s} =$ 8 TeV}},
  \href{https://doi.org/10.1103/PhysRevD.93.012003}{\emph{Physical Review}
  {\bfseries D93} (2016) 012003},
  [\href{https://arxiv.org/abs/1509.04177}{{\ttfamily 1509.04177}}].

\bibitem{ATLAS:2016cuv}
{\scshape ATLAS} collaboration, \emph{{Search for pair production of heavy
  vector-like quarks decaying to high-$p_T$ $W$ bosons and b quarks in the
  lepton-plus-jets final state in pp collisions at $\sqrt{s}$=13 TeV with the
  ATLAS detector}},  Tech. Rep. ATLAS-CONF-2016-102, 2016.

\bibitem{Sirunyan:2017pks}
{\scshape CMS} collaboration, \emph{{Search for pair production of vector-like
  quarks in the bW$\overline{\mathrm{b}}$W channel from proton-proton
  collisions at $\sqrt{s} =$ 13 TeV}},
  \href{https://doi.org/10.1016/j.physletb.2018.01.077}{\emph{Physics Letters}
  {\bfseries B779} (2018) 82--106},
  [\href{https://arxiv.org/abs/1710.01539}{{\ttfamily 1710.01539}}].

\bibitem{Sirunyan:2017usq}
{\scshape CMS} collaboration, \emph{{Search for pair production of vector-like
  T and B quarks in single-lepton final states using boosted jet substructure
  in proton-proton collisions at $\sqrt{s}=13$ TeV}},
  \href{https://doi.org/10.1007/JHEP11(2017)085}{\emph{Journal of High Energy
  Physics} {\bfseries 11} (2017) 085},
  [\href{https://arxiv.org/abs/1706.03408}{{\ttfamily 1706.03408}}].

\bibitem{Aaltonen:2009nr}
{\scshape CDF} collaboration, \emph{{Search for New Bottomlike Quark Pair
  Decays $Q\bar{Q} \to (tW^{mp})(\bar{t}W^{\pm})$ in Same-Charge Dilepton
  Events}},
  \href{https://doi.org/10.1103/PhysRevLett.104.091801}{\emph{Physical Review
  Letters} {\bfseries 104} (2010) 091801},
  [\href{https://arxiv.org/abs/0912.1057}{{\ttfamily 0912.1057}}].

\bibitem{Chatrchyan:2012af}
{\scshape CMS} collaboration, \emph{{Search for Heavy Quarks Decaying into a
  Top Quark and a $W$ or $Z$ Boson using Lepton + Jets Events in $pp$
  Collisions at $\sqrt{s}$ = 7 TeV}},
  \href{https://doi.org/10.1007/JHEP01(2013)154}{\emph{Journal of High Energy
  Physics} {\bfseries 01} (2013) 154},
  [\href{https://arxiv.org/abs/1210.7471}{{\ttfamily 1210.7471}}].

\bibitem{Chatrchyan:2013wfa}
{\scshape CMS} collaboration, \emph{{Search for top-quark partners with charge
  5/3 in the same-sign dilepton final state}},
  \href{https://doi.org/10.1103/PhysRevLett.112.171801}{\emph{Physical Review
  Letters} {\bfseries 112} (2014) 171801},
  [\href{https://arxiv.org/abs/1312.2391}{{\ttfamily 1312.2391}}].

\bibitem{CMS:2013una}
{\scshape CMS} collaboration, \emph{{Search for Vector-Like b' Pair Production
  with Multilepton Final States in pp collisions at $\sqrt{s} =$ 8 TeV}},
  Tech. Rep. CMS-PAS-B2G-13-003, 2013.

\bibitem{Khachatryan:2015gza}
{\scshape CMS} collaboration, \emph{{Search for pair-produced vectorlike B
  quarks in proton-proton collisions at $\sqrt{s}$=8 TeV}},
  \href{https://doi.org/10.1103/PhysRevD.93.112009}{\emph{Physical Review}
  {\bfseries D93} (2016) 112009},
  [\href{https://arxiv.org/abs/1507.07129}{{\ttfamily 1507.07129}}].

\bibitem{Aad:2015gdg}
{\scshape ATLAS} collaboration, \emph{{Analysis of events with $b$-jets and a
  pair of leptons of the same charge in $pp$ collisions at $\sqrt{s}=8$ TeV
  with the ATLAS detector}},
  \href{https://doi.org/10.1007/JHEP10(2015)150}{\emph{Journal of High Energy
  Physics} {\bfseries 10} (2015) 150},
  [\href{https://arxiv.org/abs/1504.04605}{{\ttfamily 1504.04605}}].

\bibitem{Aad:2015mba}
{\scshape ATLAS} collaboration, \emph{{Search for vector-like $B$ quarks in
  events with one isolated lepton, missing transverse momentum and jets at
  $\sqrt{s}=$ 8 TeV with the ATLAS detector}},
  \href{https://doi.org/10.1103/PhysRevD.91.112011}{\emph{Physical Review}
  {\bfseries D91} (2015) 112011},
  [\href{https://arxiv.org/abs/1503.05425}{{\ttfamily 1503.05425}}].

\bibitem{Aaboud:2018xpj}
{\scshape ATLAS} collaboration, \emph{{Search for new phenomena in events with
  same-charge leptons and $b$-jets in $pp$ collisions at $\sqrt{s}= 13$ TeV
  with the ATLAS detector}},
  \href{https://doi.org/10.1007/JHEP12(2018)039}{\emph{Journal of High Energy
  Physics} {\bfseries 12} (2018) 039},
  [\href{https://arxiv.org/abs/1807.11883}{{\ttfamily 1807.11883}}].

\bibitem{Aaboud:2018uek}
{\scshape ATLAS} collaboration, \emph{{Search for pair production of heavy
  vector-like quarks decaying into high-$p_T$ $W$ bosons and top quarks in the
  lepton-plus-jets final state in $pp$ collisions at $\sqrt{s}=13$ TeV with the
  ATLAS detector}},
  \href{https://doi.org/10.1007/JHEP08(2018)048}{\emph{Journal of High Energy
  Physics} {\bfseries 08} (2018) 048},
  [\href{https://arxiv.org/abs/1806.01762}{{\ttfamily 1806.01762}}].

\bibitem{CMS:2015alb}
{\scshape CMS} collaboration, \emph{{Search for top quark partners with charge
  5/3 at $\sqrt{s}=13$ TeV}},  Tech. Rep. CMS-PAS-B2G-15-006, 2015.

\bibitem{CMS:2017jfv}
{\scshape CMS} collaboration, \emph{{Search for heavy vector-like quarks
  decaying to same-sign dileptons}},  Tech. Rep. CMS-PAS-B2G-16-019, 2017.

\bibitem{Sirunyan:2018yun}
{\scshape CMS} collaboration, \emph{{Search for top quark partners with charge
  5/3 in the same-sign dilepton and single-lepton final states in proton-proton
  collisions at $\sqrt{s}=13 $ TeV}},
  \href{https://doi.org/10.1007/JHEP03(2019)082}{\emph{Journal of High Energy
  Physics} {\bfseries 03} (2019) 082},
  [\href{https://arxiv.org/abs/1810.03188}{{\ttfamily 1810.03188}}].

\bibitem{Aaltonen:2007je}
{\scshape CDF} collaboration, \emph{{Search for New Particles Leading to $Z +$
  jets Final States in $p \bar{p}$ Collisions at $\sqrt{s}$ = 1.96-TeV}},
  \href{https://doi.org/10.1103/PhysRevD.76.072006}{\emph{Physical Review}
  {\bfseries D76} (2007) 072006},
  [\href{https://arxiv.org/abs/0706.3264}{{\ttfamily 0706.3264}}].

\bibitem{CMS:2012jwa}
{\scshape CMS} collaboration, \emph{{Search B' to bZ}},  Tech. Rep.
  CMS-PAS-EXO-11-066, 2012.

\bibitem{Aaboud:2018saj}
{\scshape ATLAS} collaboration, \emph{{Search for pair- and single-production
  of vector-like quarks in final states with at least one $Z$ boson decaying
  into a pair of electrons or muons in $pp$ collision data collected with the
  ATLAS detector at $\sqrt{s} = 13$ TeV}},
  \href{https://doi.org/10.1103/PhysRevD.98.112010}{\emph{Physical Review}
  {\bfseries D98} (2018) 112010},
  [\href{https://arxiv.org/abs/1806.10555}{{\ttfamily 1806.10555}}].

\bibitem{Chatrchyan:2011ay}
{\scshape CMS} collaboration, \emph{{Search for a Vector-like Quark with Charge
  2/3 in $t$ + $Z$ Events from $pp$ Collisions at $\sqrt{s}=7$ TeV}},
  \href{https://doi.org/10.1103/PhysRevLett.107.271802}{\emph{Physical Review
  Letters} {\bfseries 107} (2011) 271802},
  [\href{https://arxiv.org/abs/1109.4985}{{\ttfamily 1109.4985}}].

\bibitem{ATLAS:2016qlg}
{\scshape ATLAS} collaboration, \emph{{Search for pair production of
  vector-like top partners in events with exactly one lepton and large missing
  transverse momentum in $\sqrt{s}=13$ TeV $pp$ collisions with the ATLAS
  detector}},  Tech. Rep. ATLAS-CONF-2016-101, 2016.

\bibitem{Aaboud:2017qpr}
{\scshape ATLAS} collaboration, \emph{{Search for pair production of
  vector-like top quarks in events with one lepton, jets, and missing
  transverse momentum in $\sqrt{s}=13 $ TeV $pp$ collisions with the ATLAS
  detector}}, \href{https://doi.org/10.1007/JHEP08(2017)052}{\emph{Journal of
  High Energy Physics} {\bfseries 08} (2017) 052},
  [\href{https://arxiv.org/abs/1705.10751}{{\ttfamily 1705.10751}}].

\bibitem{Aaboud:2018xuw}
{\scshape ATLAS} collaboration, \emph{{Search for pair production of up-type
  vector-like quarks and for four-top-quark events in final states with
  multiple $b$-jets with the ATLAS detector}},
  \href{https://doi.org/10.1007/JHEP07(2018)089}{\emph{Journal of High Energy
  Physics} {\bfseries 07} (2018) 089},
  [\href{https://arxiv.org/abs/1803.09678}{{\ttfamily 1803.09678}}].

\bibitem{CMS:2012hfa}
{\scshape CMS} collaboration, \emph{{Search for pair-produced vector-like
  quarks of charge -1/3 in lepton+jets final state in pp collisions at
  $\sqrt{s} =$ 8 TeV}},  Tech. Rep. CMS-PAS-B2G-12-019, 2012.

\bibitem{CMS:2014afa}
{\scshape CMS} collaboration, \emph{{Search for pair-produced vector-like
  quarks of charge -1/3 decaying to bH using boosted Higgs jet-tagging in pp
  collisions at $\sqrt{s} =$ 8 TeV}},  Tech. Rep. CMS-PAS-B2G-14-001, 2014.

\bibitem{CMS:2016dmr}
{\scshape CMS} collaboration, \emph{{Search for vector-like quark pair
  production in final states with leptons and boosted Higgs bosons at
  $\sqrt{s}=13~\mathrm{TeV}$}},  Tech. Rep. CMS-PAS-B2G-16-011, 2016.

\bibitem{TheATLAScollaboration:2016gxs}
{\scshape ATLAS} collaboration, \emph{{Search for production of vector-like top
  quark pairs and of four top quarks in the lepton-plus-jets final state in
  $pp$ collisions at $\sqrt{s}=13$ TeV with the ATLAS detector}},  Tech. Rep.
  ATLAS-CONF-2016-013, 2016.

\end{thebibliography}\endgroup

\end{document}